\documentclass[a4paper]{article}

\usepackage{amsmath, amssymb, amsthm, bbm, booktabs, color, enumitem, graphicx, multicol, tikz, url, xcolor}
\usepackage[a4paper]{geometry}
\usepackage[authoryear, round]{natbib}

\usepackage[title]{appendix}

\usepackage[colorlinks, allcolors = blue, breaklinks=true, hypertexnames=false]{hyperref}
\usepackage[ruled, vlined]{algorithm2e}

\newcommand{\rational}{\mathbb{Q}}
\newcommand{\real}{\mathbb{R}}

\newcommand{\myE}{\mathbb{E}}
\newcommand{\myQ}{\mathbb{Q}}

\newcommand{\cA}{\mathcal{A}}
\newcommand{\cB}{\mathcal{B}}
\newcommand{\cF}{\mathcal{F}}
\newcommand{\cG}{\mathcal{G}}
\newcommand{\cT}{\mathcal{T}}
\newcommand{\cL}{\mathcal{L}}
\newcommand{\cN}{\mathcal{N}}

\newcommand{\CD}{{\operatorname R}^\ast} 

\newcommand{\myR}{{\operatorname R}}
\newcommand{\myS}{{\operatorname S}}
\newcommand{\myT}{{\operatorname T}}

\newcommand{\cm}{$\checkmark$}
\newcommand{\dd}{\,\mathrm{d}}

\newcommand{\one}{\mathbbm{1}}

\newcommand{\DSC}{{\sf{DSC}}}
\newcommand{\MCB}{{\sf{MCB}}}
\newcommand{\UNC}{{\sf{UNC}}}

\newcommand{\mg}{{\rm{mg}}}
\newcommand{\rc}{{\rm{rc}}}
\newcommand{\skill}{{\rm{skill}}}
\newcommand{\urc}{{\rm{urc}}}

\newcommand{\cond}{{\sf{c}}}
\newcommand{\uncond}{{\sf{u}}}

\newcommand{\hsp}{\hspace{0.2mm}}

\theoremstyle{plain}  
\newtheorem{theorem}{Theorem}[section] 
\newtheorem{lemma}[theorem]{Lemma} 
\newtheorem{proposition}[theorem]{Proposition}

\theoremstyle{definition} 
\newtheorem{definition}[theorem]{Definition}

\newtheorem{example}[theorem]{Example}
\newtheorem{assumption}[theorem]{Assumption}

\theoremstyle{remark} 
\newtheorem{remark}[theorem]{Remark}

\setlist[enumerate,1]{label={(\alph*)}} % enumerate as (a), (b), ... by default

\usepackage[defaultcolor = red]{changes} % \deleted

\begin{document}

\title{Regression Diagnostics meets Forecast Evaluation: \\ 
       Conditional Calibration, Reliability Diagrams, and Coefficient of Determination}
\author{Tilmann Gneiting$^{1,2}$ and Johannes Resin$^{1,2}$ \\ \\
        $^1$Heidelberg Institute for Theoretical Studies \\ $^2$Karlsruhe Institute of Technology} 
\maketitle

\begin{abstract}
  Model diagnostics and forecast evaluation are two sides of the same
  coin.  A common principle is that fitted or predicted distributions
  ought to be calibrated or reliable, ideally in the sense of
  auto-calibration, where the outcome is a random draw from the
  posited distribution.  For binary responses, this is the
  universal concept of reliability.  For real-valued outcomes, a
  general theory of calibration has been elusive, despite a recent
  surge of interest in distributional regression and machine
  learning.  We develop a framework rooted in probability
  theory, which gives rise to hierarchies of calibration, and
  applies to both predictive distributions and stand-alone point
  forecasts.  In a nutshell, a prediction --- distributional
  or single-valued --- is conditionally $\myT$-calibrated if it can
  be taken at face value in terms of the functional $\myT$.  Whenever
  $\myT$ is defined via an identification function --- as in the cases
  of threshold (non) exceedance probabilities, quantiles, expectiles,
  and moments --- auto-calibration implies $\myT$-calibration.  We
  introduce population versions of $\myT$-reliability diagrams and
  revisit a score decomposition into measures of miscalibration
  (\MCB), discrimination (\DSC), and uncertainty (\UNC).  In empirical
  settings, stable and efficient estimators of $\myT$-reliability
  diagrams and score components arise via nonparametric isotonic
  regression and the pool-adjacent-violators algorithm.  For in-sample
  model diagnostics, we propose a universal coefficient of
  determination,
\[
\text{R}^\ast = \frac{\text{DSC} - \MCB}{\UNC},
\] 
that nests and reinterprets the classical $\myR^2$ in least squares
(mean) regression and its natural analogue $\myR^1$ in quantile
regression, yet applies to $\myT$-regression in general, with $\MCB
\geq 0$, $\DSC \geq 0$, and $\CD \in [0,1]$ under modest conditions.

\medskip

\end{abstract}

\section{Introduction}  \label{sec:introduction} 

Predictive distributions ought to be calibrated or reliable
\citep{Dawid1984, Gneiting2014}.  More generally, statistical models
ought to provide plausible probabilistic explanations of observations,
be it in-sample or out-of-sample, ideally in the sense of
auto-calibration, meaning that the outcomes are indistinguishable from
random draws from the posited distributions.  For binary outcomes,
auto-calibration is the universal standard of reliability.  In the
general case of linearly ordered, real-valued outcomes, weaker,
typically unconditional facets of calibration have been studied, with
probabilistic calibration, which corresponds to the uniformity of the
probability integral transform \citep[PIT;][]{Dawid1984, Diebold1998},
being the most popular notion.  Recently, conditional notions have
been proposed \citep{Mason2007, Bentzien2014, Strahl2017}, and there
has been a surge of attention to calibration in the machine learning
community, where the full conditional distribution of a response,
given a feature vector, is of increasing interest, as exemplified by
the work of \citet{Guo2017}, \citet{Kuleshov2018}, \citet{Song2019},
\citet{Gupta2020}, \citet{Zhao2020}, \citet{Sahoo2021} and
\citet{Roelofs2020}.  While in the literature on forecast evaluation
predictive performance is judged out-of-sample, calibration is
relevant in regression diagnostics as well, where in-sample
goodness-of-fit is assessed via test statistics or criteria of
$\myR^2$ type.  In many ways, the complementary perspectives of model
diagnostics and forecast evaluation are two sides of the same coin.

\begin{figure}[t]
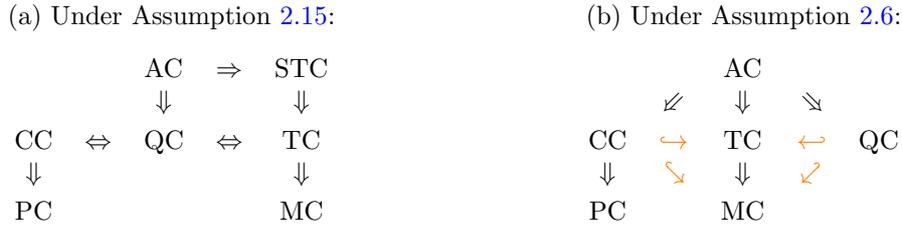
  
	\centering
	\newcommand\SEarrow{\mathrel{\rotatebox[origin=c]{-45}{$\Rightarrow$}}}
	\newcommand\hookSEarrow{\mathrel{\rotatebox[origin=c]{-45}{$\hookrightarrow$}}}
	\newcommand\SWarrow{\mathrel{\rotatebox[origin=c]{45}{$\Leftarrow$}}}
	\newcommand\hookSWarrow{\mathrel{\rotatebox[origin=c]{45}{$\hookleftarrow$}}}
	\begin{multicols}{2}
		(a) Under Assumption \ref{as:csi}:
		\begin{center}
			\begin{tabular}{ccccc}
				& & AC & $\Rightarrow$ & STC \\
				& & $\Downarrow$ & & $\Downarrow$ \\ [1mm]
				CC & $\Leftrightarrow$ & QC & $\Leftrightarrow$ & TC \\ 
				$\Downarrow$ & & & & $\Downarrow$ \\ [1mm]
				PC & & & & MC
			\end{tabular}
		\end{center}
		\columnbreak
		(b) Under Assumption \ref{as:T}:
		\begin{center}
			\begin{tabular}{ccccc}
				& & AC & & \\
				& $\SWarrow$ & $\Downarrow$ & $\SEarrow$ & \\ [1mm]
				CC & \color{orange} $\hookrightarrow$ & TC & \color{orange} $\hookleftarrow$ & QC \\
				$\Downarrow$ & \color{orange} $\hookSEarrow$ & $\Downarrow$ & \color{orange} $\hookSWarrow$ & 
				\\ [1mm]
				PC & & MC & &
			\end{tabular}
		\end{center}
	\end{multicols}
	\caption{Preview of key findings in
		Section~\ref{sec:conditional}: Hierarchies of calibration
		(a) for continuous, strictly increasing cumulative
		distribution functions (CDFs) with common support and (b)
		under minimal conditions, with auto-calibration (AC) being
		the strongest notion.  Conditional exceedance probability
		calibration (CC) is a conditional version of probabilistic
		calibration (PC), whereas threshold calibration (TC) is a
		conditional version of marginal calibration (MC).
		Quantile calibration (QC) differs from CC and TC in subtle
		ways. Strong threshold calibration (STC) is a
		stronger notion of threshold calibration introduced by
		\cite{Sahoo2021} for continuous CDFs.  Hook arrows show
		conjectured implications.  \label{fig:hierarchy}}
\end{figure}

In this paper, we strive to develop a theory of calibration for
real-valued outcomes that complements the aforementioned strands of
literature.  Starting from measure theoretic and probabilistic
foundations, we develop practical tools for visualizing, diagnosing
and testing calibration, for both in-sample and out-of-sample
settings, and applying both to full distributions and functionals
thereof.  Section \ref{sec:population} develops an overarching,
rigorous theoretical framework in a general population setting, where
we establish hierarchical relations between notions of unconditional
and conditional calibration, with Figure \ref{fig:hierarchy}
summarizing key results.  We reduce a posited distribution to a
typically single-valued statistical functional, $\myT$, and define
conditional calibration in terms of said functional.  While in general
auto-calibration fails to imply calibration in terms of a functional,
we prove this implication for functionals defined via an
identification function, such as event probabilities, means,
quantiles, and generalized quantiles.  We plot recalibrated values of
the functional against posited values to obtain $\myT$-reliability
diagrams and revisit extant score decompositions to define nonnegative
measures of miscalibration ($\MCB$), discrimination ($\DSC$) and
uncertainty ($\UNC$), for which the mean score satisfies
$\bar{\myS} = \MCB - \DSC + \UNC$.  These considerations continue to
apply when $\myT$-regression is studied as an end in itself, such as
in mean (least squares) and quantile regression.  In this setting,
Theorem \ref{th:optimal} establishes a general link between
unconditional calibration and canonical score optimization, which
nests classical results in least squares regression and the
partitioning inequalities of quantile regression \citep[Theorem
3.4]{Koenker1978}.

In Section \ref{sec:empirical}, we turn to empirical settings and
statistical inference.  We adopt and generalize the approach of
\citet{Dimitriadis2020a} that uses isotonic regression and the
pool-adjacent-violators (PAV) algorithm \citep{Ayer1955} to obtain
consistent, optimally binned, reproducible and PAV based (CORP)
estimates of $\myT$-reliability diagrams and score components, along
with uncertainty quantification via resampling.  As opposed to extant
estimators, the CORP approach yields non-decreasing reliability
diagrams and guarantees the nonnegativity of the estimated $\MCB$ and
$\DSC$ components.  The regularizing constraint of isotonicity avoids
artifacts and overfitting.  For in-sample model diagnostics, we
introduce a generalized coefficient of determination $\CD$ that links
to skill scores, and nests both the classical variance explained or
$\myR^2$ in least squares regression \citep{Kvalseth1985}, and its
natural analogue $\myR^1$ in quantile regression \citep{Koenker1999}.
Subject to modest conditions $\CD \in [0,1]$, with values of 0 and 1
indicating uninformative and immaculate fits, respectively.

\begin{figure}[t]
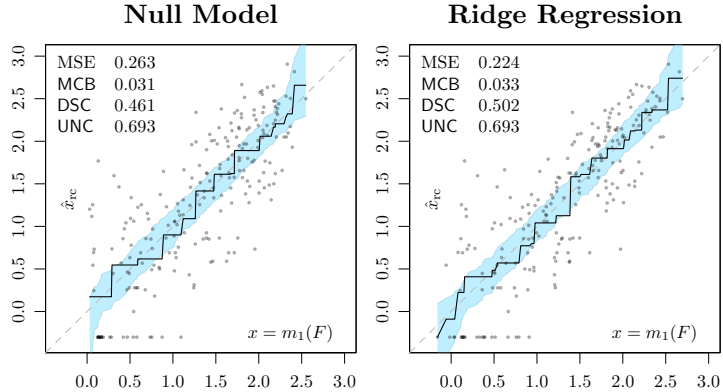
  
	\centering
	\scalebox{0.67}{\input{figs/tikz/relDiag_butterflies_null.tex}}
	\scalebox{0.67}{\input{figs/tikz/relDiag_butterflies_ridge.tex}} 
	\caption{CORP mean reliability diagrams for point forecasts of
		(log-transformed) butterfly population size from the null model
		(left) and ridge regression (right) of \citet{Tredennick2021}, along
		with 90\% consistency bands and miscalibration ($\MCB$),
		discrimination ($\DSC$) and uncertainty ($\UNC$) components of the
		mean squared error (MSE).  \label{fig:butterflies}}
\end{figure}

In forecast evaluation, reliability diagrams and score components
serve to diagnose and quantify performance on test samples.  The
most prominent case arises when $\myT$ is the mean functional and
performance is assessed by the mean squared error (MSE).  As a
preview of the diagnostic tools developed in this paper, we
assess point forecasts by \cite{Tredennick2021} of (log
transformed) butterfly population size from a ridge regression
and a null model.  The CORP mean reliability diagrams and
MSE decompositions in Figure \ref{fig:butterflies} show
that, while both models are reliable, ridge regression enjoys
considerably higher discrimination ability.

The paper closes in Section \ref{sec:discussion}, where we discuss our
findings and provide a roadmap for follow-up research.  While
\citet{Dimitriadis2020a} introduced the CORP approach in the nested
case of probability forecasts for binary outcomes, the setting of
real-valued outcomes treated in this paper is far more complex, as it
necessitates the consideration of statistical functionals in general.
Throughout, we link the traditional case of regression diagnostics and
(stand-alone) point forecast evaluation, where functionals such as
conditional means, moments, quantiles or expectiles are modeled and
predicted, to model diagnostics and forecast evaluation in the fully
distributional setting \citep{Gneiting2014, Hothorn2014}.
Appendices \ref{sec:examples}--\ref{sec:timeseries} include
material of more specialized or predominantly technical character.

\section{Notions of calibration, reliability diagrams, and score decompositions}  \label{sec:population} 

Generally, we use the symbol $\cL$ to denote a generic conditional or
unconditional law or distribution, and we identify distributions with
their cumulative distribution functions (CDFs).  We write
$\cN(m,c^2)$ to denote a normal distribution with mean $m$ and
variance $c^2$, and we let $\varphi$ and $\Phi$ denote the density and
the CDF, respectively, of a standard normal variable.

\subsection{Prediction spaces and prequential principle}  \label{sec:prequential} 

We consider the joint law of a posited distribution and the respective
outcome in the technical setting of \citet{Gneiting2013}.
Specifically, let $(\Omega, \cA, \myQ)$ be a \emph{prediction space},
i.e., a probability space where the elementary elements $\omega \in
\Omega$ correspond to realizations of the random triple
\[
(F,Y,U),  
\]
where $Y$ is the real-valued outcome, $F$ is a posited distribution
for $Y$ in the form of a CDF, and $U$ is uniformly distributed on the
unit interval.  Statements involving conditional or unconditional
distributions, expectations, or probabilities, generally refer to the
probability measure $\myQ$, which specifies the joint distribution of
the forecast $F$ and the outcome $Y$.  The uniform random variable $U$
allows for randomization.  Throughout, we assume that $U$ is
independent of the $\sigma$-algebra generated by the random variable
$Y$ and the random function $F$ in the technical sense detailed prior
to Definition 2.6 in \citet{Strahl2017}.

Let $\cA_0 \subseteq \cA$ denote the forecaster's \emph{information
	basis}, i.e., a sub-$\sigma$-algebra such that $F$ is measurable
with respect to $\cA_0$.  Then $F$ is \emph{ideal} relative to $\cA_0$
\citep{Gneiting2013} if
\[
F(y) = \myQ \left( Y \leq y \mid \cA_0 \right) 
\quad \text{almost surely}, \quad \text{for all} \quad y \in \real. 
\]
If $F$ is ideal relative to some sub $\sigma$-algebra
$\cA_0$, then it is \emph{auto-calibrated} \citep{Tsyplakov2013}
in the sense that
\[
F(y) = \myQ \left( Y \leq y \mid F \right) 
\quad \text{almost surely}, \quad \text{for all} \quad y \in \real,
\]
which is equivalent to being ideal relative to the information
basis $\sigma(F) \subseteq \cA_0$.  Extensions to prediction
spaces with tuples $(Y, F_1, \ldots, F_k, U)$ that allow for
multiple CDF-valued forecasts $F_1, \ldots, F_k$ with associated
information bases $\cA_1, \ldots, \cA_k \subset \cA$ are
straightforward.

\begin{example}[\citet{Gneiting2013, Pohle2020}]  \label{ex:perf.clim} 
	Conditionally on a standard normal variate $\mu$, let the outcome $Y$
	be normal with mean $\mu$ and variance 1.  Then the \emph{perfect}
	forecast $F_1 = \cN(\mu,1)$ is ideal relative to the information basis
	$\cA_1 = \sigma(\mu)$ generated by $\mu$.  The \emph{unconditional}
	forecast $F_2 = \cN(0,2)$ agrees with the marginal distribution of the
	outcome $Y$ and is ideal relative to the trivial $\sigma$-algebra
	$\cA_2 = \{ \emptyset, \Omega \}$.
\end{example} 

More elaborate notions of prediction spaces are feasible.  In
particular, one might include a covariate or feature vector $Z$
and consider random tuples of the form $(Z,F,Y,U)$.  Indeed,
the transdisciplinary scientific literature has considered
reliability relative to covariate information, under labels such
as \emph{strong} \citep{VanCalster2016} or \emph{individual}
\citep{Chung2020, Zhao2020} calibration.  We refrain from doing so, as
our simple setting adheres to the \emph{prequential principle} posited
by \citet{Dawid1984}, according to which predictive performance needs
to be evaluated on the basis of the tuple $(F,Y)$ only, without
consideration of the forecast-generating mechanism.  The
aforementioned extensions become critical in studies of
cross-calibration \citep{Strahl2017}, stratification \citep{Ehm2017,
	Ferro2020,Allen2022}, sensitivity \citep{Fissler2022b}, and fairness
\citep{Pleiss2017, Mitchell2021}.

\subsection{Traditional notions of unconditional calibration}  \label{sec:unconditional} 

Let us recall the classical notions of calibration of predictive
distributions for real-valued outcomes.  In order to do so, we define
the \emph{probability integral transform} (PIT)
\begin{equation}  \label{eq:PIT}
Z_F = F(Y-) + U \left( F(Y) - F(Y-) \right)
\end{equation}
of the CDF-valued random quantity $F$, where $F(y-) = \lim_{x \uparrow
	y} F(x)$ denotes the left-hand limit of $F$ at $y \in \real$, and
the random variable $U$ is standard uniform and independent of $F$ and
$Y$.  The PIT of a continuous CDF $F$ is simply $Z_F = F(Y)$.  The
predictive distribution $F$ is \emph{probabilistically calibrated} or
\emph{PIT calibrated} if $Z_F$ is uniformly distributed on the unit
interval.  The use of the probabilistic calibration criterion was
suggested by \citet{Dawid1984} and popularized by \citet{Diebold1998},
who proposed the use of PIT histograms as a diagnostic tool.
Importantly, in continuous settings probabilistic calibration implies
that prediction intervals bracketed by quantiles capture the outcomes
at the respective nominal level.

Furthermore, the predictive distribution $F$ is \emph{marginally
	calibrated} \citep{Gneiting2007a} if
\begin{equation*}  
\myE_\myQ [ F(y) ] = \myQ \left( Y \leq y \right) \quad \text{for all} \quad y \in \real.
\end{equation*}
Thus, for a marginally calibrated predictive distribution, the
frequency of (not) exceeding a threshold value matches the posited
unconditional probability.

\begin{example}  \label{ex:unf.lop} 
	In the setting of Example \ref{ex:perf.clim}, let $\eta$ attain the
	values $\pm \eta_0$ with equal probability, independently of $\mu$ and
	$Y$, where $\eta_0 > 0$. Then the \emph{unfocused} forecast with CDF
	\begin{equation}  \label{eq:unf} 
	F(y) = \frac{1}{2} \left( \Phi(y-\mu) + \Phi(y-\mu-\eta) \right)
	\end{equation} 
	is probabilistically calibrated, but fails to be marginally calibrated
	\citep{Gneiting2007a}.  Similarly, let $\delta$ take the values
	$\pm \delta_0$ with equal probability, independently of $\mu$ and $Y$,
	where $\delta_0 \in (0,1)$. Then the \emph{lopsided}\/ forecast $F$
	with density
	\begin{equation}  \label{eq:lop}  
	f(y) = (1-\delta) \hsp \varphi(y-\mu) \hsp \one(y < \mu) 
	+ (1+\delta) \hsp \varphi(y-\mu) \hsp \one(y > \mu)
	\end{equation} 
	is marginally calibrated, but fails to be probabilistically
	calibrated.  For details see Appendix \ref{sec:lop}.
\end{example} 

It is well known that an ideal forecast is both probabilistically
calibrated and marginally calibrated (\citealp[Theorem
2.8]{Gneiting2013}; \citealp[Theorem 1]{Song2019}).  Reformulated in
terms of auto-calibration the following holds.

\begin{theorem}  \label{th:AC->PMC} 
	Auto-calibration implies marginal and probabilistic calibration.
\end{theorem}

Auto-calibration thus is a stronger requirement than either marginal
or probabilistic calibration, and the latter are logically
independent.  However, in the special case of a binary outcome,
probabilistic calibration and auto-calibration are equivalent
\citep[Theorem 2.11]{Gneiting2013}, and auto-calibration serves as a
universal notion of calibration.  In the case of three or more
distinct outcomes, \citet{Gneiting2013} conjectured that
auto-calibration is stronger than simultaneous marginal and
probabilistic calibration.  We resolve and prove their conjecture
within the following example.

\begin{example}  \label{ex:p-u.tern} 
	We begin by considering continuous CDFs and then discuss a discrete
	example with three distinct outcomes only.
	\begin{enumerate} 
		\item
		Suppose that $\mu$ is normal with mean $0$ and variance $c^2$.
		Conditionally on $\mu$, let the \emph{piecewise uniform}\/ predictive
		distribution $F$ be a mixture of uniform measures on $[\mu, \mu + 1]$,
		$[\mu + 1, \mu + 2]$, and $[\mu + 2, \mu + 3]$ with weights $p_1$,
		$p_2$ and $p_3$, respectively, and let the outcome $Y$ be drawn from a
		mixture with weights $q_1$, $q_2$ and $q_3$ on these intervals.
		Finally, let the tuple $(p_1, p_2, p_3; q_1, q_2, q_3)$ attain each of
		the values
		\[\textstyle
		\left (\frac{1}{2}, \frac{1}{4}, \frac{1}{4}; \frac{5}{10}, \frac{1}{10}, \frac{4}{10} \right) \! ,
		\quad 
		\left( \frac{1}{4}, \frac{1}{2}, \frac{1}{4}; \frac{1}{10}, \frac{8}{10}, \frac{1}{10} \right) \! , 
		\quad \text{and} \quad 
		\left( \frac{1}{4}, \frac{1}{4}, \frac{1}{2}; \frac{4}{10}, \frac{1}{10}, \frac{5}{10} \right) \! , 
		\]
		with $\myQ$-probability $\frac{1}{3}$.  Evidently, $F$ fails to be
		auto-calibrated.  However, $F$ is marginally calibrated, as
		conditionally on $\mu$, it assigns the same mass $\frac{1}{3}$ to each
		of the intervals, in agreement with the conditional distribution of
		$Y$.  As for the PIT $Z_F$, conditionally on $\mu$ its CDF is
		piecewise linear on the partition induced by 0, $\frac{1}{4}$,
		$\frac{1}{2}$, $\frac{3}{4}$, and 1.  Thus, in order to establish
		probabilistic calibration it suffices to verify that $\myQ(Z_F \leq x)
		= x$ for $x \in \{ \frac{1}{4}, \frac{1}{2}, \frac{3}{4} \}$, as
		confirmed by elementary calculations.  Integration over $\mu$
		completes the argument.
		\item
		For a full resolution of the aforementioned conjecture by
		\citet{Gneiting2013}, we fix $\mu = 0$ and replace the intervals by
		fixed numbers $y_1 < y_2 < y_3$.  Thus, $F$ assigns mass $p_j$ to
		$y_j$, whereas the event $Y = y_j$ realizes with probability $q_j$ for
		$j = 1$, 2, and 3.  The forecast remains probabilistically and
		marginally calibrated, and fails to be auto-calibrated.
	\end{enumerate} 
\end{example} 

\subsection{Conditional calibration}  \label{sec:conditional} 

While checks for probabilistic calibration have become a cornerstone
of predictive distribution evaluation \citep{Dawid1984, Diebold1998,
	Gneiting2007a}, both marginal and probabilistic calibration concern
unconditional facets of predictive performance, which is increasingly
being considered insufficient \citep[e.g.,][]{Levi2019}.  Stronger
conditional notions of calibration, which condition on facets of the
predictive distribution, have emerged in various strands of the
scientific literature.  For example, \citet{Mason2007} used
conditional (non) exceedance probabilities (CEP) to assess the
calibration of ensemble weather forecasts.  These were used by
\citet{Held2010} and \citet{Strahl2017} to derive calibration tests,
which operate under the hypothesis that the forecast $F$ is \emph{CEP
	calibrated}\/ in the sense that
\begin{equation}  \label{eq:CEP}  
\myQ \left( Z_F \leq \alpha \mid q_\alpha^-(F) \right) = \alpha 
\quad \text{almost surely,} \quad \text{for all} \quad \alpha \in (0,1), 
\end{equation}
where $q_\alpha^-(F) = \inf \{ x \in \real : F(x) \geq \alpha \}$
denotes the (lower) $\alpha$-quantile of $F$.  Similarly,
\citet{Henzi2019} introduced the notion of a \emph{threshold
	calibrated} forecast $F$, which stipulates that
\begin{equation}  \label{eq:threshold}  
\myQ \left( Y \leq t \mid F(t) \right) = F(t) 
\quad \text{almost surely,} \quad \text{for all} \quad t \in \real.
\end{equation}

Essentially, CEP calibration is a conditional version of probabilistic
calibration, and threshold calibration is conditional marginal
calibration.

\begin{theorem}  \label{th:CEP->PIT,thresh->marg}
	CEP calibration implies probabilistic calibration, and
	threshold calibration implies marginal calibration.
\end{theorem}

\begin{proof}
	Immediate by taking unconditional expectations, as noted by
	\citet{Henzi2019}.
\end{proof}

Variants of these concepts can be found scattered in the
literature.  Notably, \citet{Sahoo2021} introduce a notion of
calibration for continuous predictive distributions, which requires
that
\begin{equation}  \label{eq:strongthreshold}
\myQ(Z_F \leq \alpha
\mid F(t)) = \alpha \quad \text{almost surely,} \quad \text{for
	all} \quad \alpha \in (0,1), \; t \in \real.
\end{equation}
As in Figure \ref{fig:hierarchy}, we refer to this property as
\emph{strong threshold calibration}.  The notion is weaker than
auto-calibration, but implies both CEP
calibration and threshold calibration, subject to conditions that we
discuss below.

We proceed to the general notion of conditional $\myT$-calibration in
terms of a statistical functional $\myT$ as introduced by
\citet{Arnold2020} and \citet{Ferro2020}. Other authors
\citep{Pohle2020, Krueger2020} refer to this notion or special cases
thereof as auto-calibration with respect to $\myT$.  A statistical
functional on some class $\cF$ of probability measures is a measurable
function $\myT \colon \cF \to \cT$ into a (typically,
finite-dimensional) space $\cT$ with Borel-$\sigma$-algebra
$\cB(\cT)$.  Technically, we work in the prediction space setting
under a natural measurability condition that is not restrictive
\citep{Fissler2022a}.

\begin{assumption}  \label{as:T}
	The class $\cF$ and the functional $\myT$ are such that $F \in \cF$,
	the mapping $\myT(F)\colon (\Omega,\cA) \rightarrow (\cT,\cB(\cT))$ is
	measurable, and $\cL \left( Y \mid \myT(F) \right) \in \cF$ almost
	surely.
\end{assumption}

\begin{definition}  \label{def:T} 
	Under Assumption \ref{as:T}, the predictive distribution $F$ is
	\emph{conditionally} $\myT$-\emph{calibrated}, or simply
	$\myT$-\emph{calibrated}, if
	\begin{equation*}  
	\myT \left( \cL(Y \mid \myT(F)) \right) = \myT(F) \quad \text{almost surely}.
	\end{equation*}
\end{definition}

Essentially, under a $\myT$-calibrated predictive distribution $F$, we
can take $\myT(F)$ at face value.  Perhaps surprisingly, an
auto-calibrated forecast is not necessarily $\myT$-calibrated, as
noted by \citet[Section 3.2]{Arnold2020}.  For a simple
counterexample, consider the perfect forecast from Example
\ref{ex:perf.clim}, which fails to be $\myT$-calibrated when $\myT$ is
the variance, the standard deviation, the interquartile range, or a
related measure of dispersion.

We proceed to show that this startling issue does not occur with
\emph{identifiable}\/ functionals, i.e., functionals induced by an
identification function (Theorem \ref{th:auto->T}).  Similar to
the classical procedure in $M$-estimation \citep{Huber1964}, an
identification function weighs negative values in the case of
underprediction against positive values in the case of overprediction,
and the corresponding functional maps to the possibly set-valued
argument at which an associated expectation changes sign.
Following \cite{Jordan2019}, a measurable function
$V \colon \real \times \real \to \real$ is an \emph{identification
	function}\/ if $V(\cdot,y)$ is increasing and left-continuous for
all $y \in \real$.  We operate under Assumption \ref{as:T} with the
implicit understanding that $V(x,\cdot)$ is quasi-integrable with
respect to all $F \in \cF$ for all $x \in \real$.  Then, for any
probability measure $F$ in the class $\cF$, the functional $\myT(F)$
\emph{induced}\/ by $V$ is defined as
\[
\myT(F) = [\myT^-(F), \myT^+(F)] \subseteq [-\infty, +\infty] = \bar{\real},
\]
where the lower and upper bounds are given by the random variables
\begin{equation}\label{eq:T}
	\myT^-(F) = \sup \left\{ x : \int V(x,y) \dd F(y) < 0 \right\} \quad \text{and} \quad
	\myT^+(F) = \inf \left\{ x : \int V(x,y) \dd F(y) > 0 \right\}.
\end{equation}
An identifiable functional $\myT$ is of \emph{singleton type}\/
if $\myT(F)$ is a singleton for every $F \in \cF$.  Otherwise, $\myT$
is of \emph{interval type}.  Table \ref{tab:functionals} lists key
examples, such as threshold-defined event probabilities, quantiles,
expectiles, and moments.  The definition of the Huber functional
involves the clipping function $\kappa_{a,b}(t) = \max(\min(t,b),-a)$
with parameters $a, b > 0$ \citep{Taggart2020}.  In the limiting cases
as $a = b \to 0$ and $a = b \to \infty$, the Huber functional recovers
the $\alpha$-quantile ($q_\alpha$) and the $\alpha$-expectile
($e_\alpha$), respectively.

\begin{table}[t]  
	\caption{Key examples of identifiable functionals with associated
		parameters, identification function, and generic type.  For a
		similar listing see Table 1 in
		\citet{Jordan2019}.  \label{tab:functionals}}
	\vspace{2mm}
	\centering 
	\footnotesize
	\begin{tabular}{llll} 
		\toprule
		Functional & Parameters & Identification function & Type \\ 
		\toprule
		Threshold (non) exceedance & $t \in \real$ & $V(x,y) = x - \one \{ y \leq t \}$ & singleton \\   
		\midrule
		Mean & & $V(x,y) = x - y$ & singleton \\
		Median & & $V(x,y) = \one \{ y < x \} - \frac{1}{2}$ & interval \\
		\midrule 
		Moment of order $n$ ($m_n$) 
		& $n = 1, 2, \ldots$ & $V(x,y) = x - y^n$ & singleton \\
		$\alpha$-Expectile $(e_\alpha)$ & $\alpha \in (0,1)$ & $V(x,y) = \left| \one \{ y < x \} - \alpha \right| \left( x - y \right)$ & singleton \\
		$\alpha$-Quantile $(q_\alpha)$ & $\alpha \in (0,1)$ & $V(x,y) = \one \{ y < x \} - \alpha$ & interval \\ 
		\midrule 
		Huber & $\alpha \in (0,1)$, $a, b > 0$
		& $V(x,y)= \left| \one \{ y < x \} - \alpha \right| \kappa_{a,b}(x-y)$ & interval \\
		\bottomrule
	\end{tabular}
\end{table}

For identifiable functionals we can define an unconditional notion of
$\myT$-ca\-li\-bra\-tion as well.  Note that in contrast to traditional
settings, where $F$ is fixed, we work in the prediction space setting,
where $F$ is a random CDF.  In principle, the subsequent
Definitions \ref{def:uncondT} and \ref{def:canonical} depend on the choice of
the identification function.  However, as we demonstrate in
Appendix \ref{sec:identification}, the following
condition ensures that the identification function is unique, up to
a positive constant, so that ambiguities are avoided.

\begin{assumption} \label{as:V} The identification function $V$
	induces the functional $\myT$ on a convex class
	$\cF_0 \supseteq \cF$ of probability measures, which contains the
	Dirac measures $\delta_y$ for all $y \in \real$.  The identification
	function $V$ is
	\begin{enumerate}[label = (\roman*)]
		\item of \emph{prediction error form}, i.e., there exists an
		increasing, left-continuous function $v\colon \real\rightarrow\real$
		such that $V(x,y) = v(x-y)$ with $v(-r) < 0$ and $v(r) > 0$ for some
		$r > 0$, \emph{or}
		\item of the form $V(x,y) = x - \myT(\delta_y)$ for a functional
		$\myT$ of singleton type.
	\end{enumerate}
\end{assumption}

The examples in Table \ref{tab:functionals} all satisfy Assumption
\ref{as:V}.

\begin{definition} \label{def:uncondT} Suppose that the functional
	$\myT$ is generated by an identification function $V$, and
	let Assumptions \ref{as:T} and \ref{as:V} hold.  Then the
	predictive distribution $F$ is \emph{unconditionally}\/
	$\myT$\emph{-calibrated}\/ if 
	\begin{equation}  \label{eq:uncondTcal}
	\myE_\myQ [V(\myT^+(F) - \varepsilon, Y)] \leq 0 \quad \text{and} \quad 
	\myE_\myQ [ V(\myT^-(F) + \varepsilon, Y)] \geq 0 \quad \text{for all} \quad \varepsilon > 0.		
	\end{equation}
\end{definition}

For the interval-valued $\alpha$-quantile functional,
$q_\alpha(F) = [q_\alpha^-(F),q_\alpha^+(F)]$, condition
\eqref{eq:uncondTcal} reduces to the traditional unconditional
coverage condition
\begin{equation}  \label{eq:uncondqcal}
\myQ \left( Y \leq q_\alpha^-(F) \right) \geq \alpha \quad \text{and} \quad 
\myQ \left( Y \geq q_\alpha^+(F) \right) \geq 1 - \alpha, 
\end{equation}
with the latter being equivalent to $\myQ(Y < q_\alpha^+(F)) \leq
\alpha$.  Probabilistic calibration implies unconditional
$\alpha$-quantile calibration at every level $\alpha \in (0,1)$, as
hinted at by \citet[Section 3.1]{Kuleshov2018}, and proved in our
Appendix \ref{sec:PCvsQC}.  Under technical assumptions,
condition \eqref{eq:uncondTcal} simplifies to
\begin{equation}  \label{eq:simple} 
\myE_\myQ [V(\myT(F), Y)] = 0, 
\end{equation} 
with the classical unbiasedness condition $\myE_\myQ [m_1(F)]
= \myE_\myQ [Y]$ arising in the case of the mean or expectation
functional.

\begin{example}  \label{ex:uncondT} 
	Let $\myT$ be the mean functional or a quantile.  Then the
	unfocused forecast from Example \ref{ex:unf.lop} is unconditionally
	$\myT$-calibrated, but fails to be conditionally $\myT$-calibrated.
	For details see Figure \ref{fig:reldiag} and Appendix
	\ref{sec:unf}.
\end{example}

Importantly, for any identifiable functional auto-calibration implies
both unconditional and conditional $\myT$-calibration, as we
demonstrate now.  Note that Assumption \ref{as:T} is a minimal
condition, as it is required to define conditional
$\myT$-calibra\-tion in the first place.

\begin{theorem}  \label{th:auto->T} 
	Suppose that the functional $\myT$ is generated by an identification
	function and Assumption \ref{as:T} holds.  Then auto-calibration
	implies conditional $\myT$-calibration, and subject to Assumption \ref{as:V} conditional
	$\myT$-calibration implies unconditional $\myT$-calibration.
\end{theorem} 

\begin{proof}
	The statements in this proof are understood to hold almost surely.  By
	Theorem 4.34 and Proposition 4.35 of \citet{Breiman1992} in concert
	with auto-calibration, $F$ is a regular conditional distribution of
	$Y$ given $F$, and we conclude that
	\[
	\myE \left[ V(x,Y) \mid F \right] = \int V(x,y) \dd F(y). 
	\]
	Furthermore, a regular conditional distribution $F_\myT = \cL( Y \mid
	\myT(F) )$ of $Y$ given $\myT(F)$ exists, and the tower property of
	conditional expectation implies that
	\begin{align*}
	\int \! V(x,y) \dd F_\myT(y) 
	&= \myE \left[ V(x,Y) \! \mid \! \myT(F) \right]
	= \myE \left[ \myE \left[ V(x,Y) \! \mid \! F \right] \! \mid \! \myT(F) \right] \\
	&= \myE \left[ \left. \int \! V(x,y) \dd F(y) \, \right| \myT(F) \right].
	\end{align*}
	Let $\myT(F) = [\myT^-(F), \myT^+(F)]$ and $\myT(F_\myT) =
	[\myT^-(F_\myT), \myT^+(F_\myT)]$, where the boundaries are random
	variables.  The proof of the first part is complete if we can show
	that $\myT^-(F_\myT) = \myT^-(F)$ and $\myT^+(F_\myT) = \myT^+(F)$.
	
	Let $\varepsilon > 0$.  By the definition of
	$\myT^+(F)$, we know that $\int V(\myT^+(F),y)\dd F(y) \leq 0$
	and $\int V(\myT^+(F) + \varepsilon,y)\dd F(y) > 0$.  Using
	nested conditional expectations as above, the same
	inequalities hold almost surely when integrating with respect
	to $F_\myT$.  Hence, by the definition of $\myT^+(F_\myT)$, we
	obtain
	$\myT^+(F) \leq \myT^+(F_\myT) < \myT^+(F) + \varepsilon$.  An
	analogous argument shows that
	$\myT^-(F) - \varepsilon \leq \myT^-(F_\myT) \leq \myT^-(F) +
	\varepsilon$.  This completes the proof of the first part and
	shows that $F$ is conditionally $\myT$-calibrated.
	
	Finally, if $F$ is conditionally $\myT$-calibrated, unconditional
	$\myT$-calibration follows by taking nested expectations in the terms
	in the defining inequalities.
\end{proof} 

An analogous result is easily derived for CEP calibration.

\begin{theorem} \label{th:AC->CEP}
	Under Assumption \ref{as:T} for quantiles, auto-calibration implies
	CEP calibration.
\end{theorem}

\begin{proof}
	It holds that
	\[
	\myQ (Z_F \leq \alpha \mid q_\alpha^-(F)) 
	= \myE \left[ \one \{ Z_F \leq \alpha \} \mid q_\alpha^-(F) \right] 
	= \myE \left[ \myE \left[ \one\{ Z_F \leq \alpha \} \mid F \right] \mid q_\alpha^-(F) \right]
	\]
	almost surely for $\alpha \in (0,1)$. As $F$ is a version of
	$\cL(Y \mid F)$, the nested expectation equals $\alpha$ almost
	surely by Proposition 2.1 of \citet{Rueschendorf2009}, which
	implies CEP calibration.
\end{proof}

When evaluating full predictive distributions, it is natural to
consider families of functionals as in the subsequent definition,
where part (a) is compatible with the extant notion in
\eqref{eq:threshold}.

\begin{definition}  \label{def:condcal}
	A predictive distribution $F$ is
	\begin{enumerate}
		\item \emph{threshold calibrated} if it is conditionally
		$F(t)$-calibrated for all $t \in \real$;
		\item \emph{quantile calibrated} if it is conditionally
		$q_\alpha$-calibrated for all $\alpha \in (0,1)$;
		\item \emph{expectile calibrated} if it is conditionally
		$e_\alpha$-calibrated for all $\alpha \in (0,1)$;
		\item \emph{moment calibrated} if it is conditionally $n$-th
		moment calibrated for all integers $n = 1, 2, \ldots$
	\end{enumerate}
\end{definition}

While CEP, quantile, and threshold calibration are closely related
notions, they generally are not equivalent.  For illustration, we
consider predictive CDFs in the spirit of Example \ref{ex:p-u.tern}.

\newpage

\begin{example}  \label{ex:CEP.q.diff}
	\mbox{}
	\begin{enumerate}
		\item Let $\mu \sim \cN(0,c^2)$.  Conditionally on $\mu$, let $F$ be a
		mixture of uniform distributions on the intervals
		$[\mu,\mu+1],[\mu+1,\mu+2],[\mu+2,\mu+3]$, and $[\mu+3,\mu+4]$ with
		weights $p_1, p_2, p_3$, and $p_4$, respectively, and let $Y$ be
		from a mixture with weights $q_1, q_2, q_3$, and $q_4$.
		Furthermore, let the tuple $(p_1, p_2, p_3, p_4; q_1, q_2, q_3,
		q_4)$ attain each of the values
		\begin{align*}
		&\textstyle
		\left( \frac{1}{2}, 0, \frac{1}{2}, 0; \frac{3}{4}, 0, \frac{1}{4}, 0 \right) \! , \quad
		\left( \frac{1}{2}, 0, 0, \frac{1}{2}; \frac{1}{4}, 0, 0, \frac{3}{4} \right) \! , \quad
		\left( 0, \frac{1}{2}, \frac{1}{2}, 0; 0, \frac{1}{4}, \frac{3}{4}, 0 \right) \! , \quad
		\left( 0, \frac{1}{2}, 0, \frac{1}{2}; 0, \frac{3}{4}, 0, \frac{1}{4} \right)
		\end{align*}
		with equal probability.  Then the \emph{continuous} forecast $F$ is
		threshold calibrated and CEP calibrated, but fails to be quantile calibrated.
		\item Let the tuple $(p_1, p_2, p_3; q_1, q_2, q_3)$ attain each of the values
		\[\textstyle
		\left (\frac{1}{2}, \frac{1}{4}, \frac{1}{4}; \frac{5}{10}, \frac{4}{10}, \frac{1}{10} \right) \! , \quad 
		\left( \frac{1}{4}, \frac{1}{2}, \frac{1}{4}; \frac{1}{10}, \frac{5}{10}, \frac{4}{10} \right) \! , \quad \text{and} \quad 
		\left( \frac{1}{4}, \frac{1}{4}, \frac{1}{2}; \frac{4}{10}, \frac{1}{10}, \frac{5}{10} \right)
		\]
		with equal probability.  Let $F$ assign mass $p_j$ to numbers $y_j$
		for $j = 1, 2, 3$, where $y_1 < y_2 < y_3$, and let the event $Y =
		y_j$ realize with probability $q_j$.  The resulting \emph{discrete}
		forecast is quantile and threshold calibrated.  However, it fails
		to be CEP calibrated or even PIT calibrated.
	\end{enumerate}
\end{example}

Under the following conditions CEP, quantile, and threshold calibration coincide.

\begin{assumption}  \label{as:csi} 
	In addition to Assumption \ref{as:T} for quantiles and threshold (non)
	exceedances at all levels and thresholds, respectively, let the
	following hold.
	\begin{enumerate}[label = (\roman*)]
		\item The CDFs in the class $\cF$ are continuous and
		strictly increasing on a shared support interval.
		\item There exists a countable set $\cG \subseteq \cF$
		such that $\myQ(F \in \cG) = 1$.
	\end{enumerate}
\end{assumption}

\begin{theorem}  \label{th:TC=QC} 
	\mbox{} 
	\begin{enumerate} 
		\item Under Assumption \ref{as:csi}(i) CEP and quantile calibration
		are equivalent and imply probabilistic calibration.
		\item Under Assumptions \ref{as:csi}(i)--(ii) CEP, quantile and
		threshold calibration are equivalent and imply both probabilistic
		and marginal calibration.
	\end{enumerate} 
\end{theorem} 

\begin{proof} 
	By Assumption \ref{as:csi}(i) the CDFs $F \in \cF$ are invertible on
	the common support with the quantile function $\alpha \mapsto
	q_\alpha^-(F)$ as inverse.  Hence, for every $\alpha \in (0,1)$ the
	functional $q_\alpha$ is of singleton-type and $q_\alpha(F) = \{
	q_\alpha^-(F) \}$.  In this light, the almost sure identity
	\[
	\myQ( Z_F \leq \alpha \mid q_\alpha^-(F)) = \myQ( Y \leq q_\alpha^-(F) \mid q_\alpha(F) )
	\]
	implies part (a).  To prove part (b), let $\cG$ be as in Assumption
	\ref{as:csi}(ii) and assume without loss of generality that $\myQ(F =
	G) > 0$ for all $G \in \cG$.  If $\alpha \in (0,1)$ and $t \in \real$
	are such that $\myQ(F(t) = \alpha) > 0$, then
	\[
	\myQ( Y \leq t \mid F(t) = \alpha) = \myQ(Y \leq q_\alpha^-(F) \mid q_\alpha^-(F) = t),
	\]
	where Assumption \ref{as:csi}(ii) ensures that the events
	conditioned on have positive probability.  Hence, quantile and
	threshold calibration are equivalent.  The remaining implications are
	immediate from Theorem \ref{th:CEP->PIT,thresh->marg}.
\end{proof} 

We conjecture that the statement of part (b) holds under Assumption
\ref{as:csi}(i) alone, but are unaware of a measure theoretic argument
that serves to generalize the discrete reasoning in our proof.  As
indicated in panel (b) of Figure \ref{fig:hierarchy}, we also
conjecture that CEP or quantile calibration imply threshold
calibration in general, though we have not been able to prove this,
nor can we show that CEP or quantile calibration imply marginal
calibration in general.  Strong threshold calibration as
defined in \eqref{eq:strongthreshold} implies both CEP and threshold
calibration under Assumption \ref{as:csi}, by arguments similar to
those in the above proof.  The following result thus demonstrates
that the hierarchies in panel (a) and, with the aforementioned
exceptions, in panel (b) of Figure \ref{fig:hierarchy} are complete,
with the caveat that hierarchies may collapse if the class $\cF$ is
sufficiently small, as exemplified by Theorem 2.11 of
\citet{Gneiting2013}.

\begin{proposition}  \label{prop:relations}
	Under Assumption \ref{as:csi}(i)--(ii) the following hold:
	\begin{enumerate}
		\item Strong threshold calibration does not imply auto-calibration.
		\item Joint CEP, quantile, and threshold calibration does not
		imply strong threshold calibration.
		\item Joint probabilistic and marginal calibration does not imply
		threshold calibration.
		\item Probabilistic calibration does not imply marginal calibration.
		\item Marginal calibration does not imply probabilistic calibration.
	\end{enumerate}
\end{proposition}

\begin{figure}[t]
	\centering
	\scalebox{0.67}{% Created by tikzDevice version 0.12.3.1 on 2021-07-01 12:10:56
% !TEX encoding = UTF-8 Unicode
\begin{tikzpicture}[x=1pt,y=1pt]
\definecolor{fillColor}{RGB}{255,255,255}
\path[use as bounding box,fill=fillColor,fill opacity=0.00] (0,0) rectangle (379.42,195.13);
\begin{scope}
\path[clip] ( 21.68, 21.68) rectangle (379.42,195.13);
\definecolor{drawColor}{RGB}{255,0,0}

\path[draw=drawColor,line width= 0.4pt,line join=round,line cap=round] (  0.00, 31.22) --
	( 34.93, 31.22) --
	(117.74, 70.20) --
	(200.55,109.18) --
	(283.36,124.78) --
	(366.17,187.15) --
	(379.42,187.15);
\end{scope}
\begin{scope}
\path[clip] (  0.00,  0.00) rectangle (379.42,195.13);
\definecolor{drawColor}{RGB}{0,0,0}

\path[draw=drawColor,line width= 0.4pt,line join=round,line cap=round] ( 34.93, 21.68) -- (366.17, 21.68);

\path[draw=drawColor,line width= 0.4pt,line join=round,line cap=round] ( 34.93, 21.68) -- ( 34.93, 15.68);

\path[draw=drawColor,line width= 0.4pt,line join=round,line cap=round] (117.74, 21.68) -- (117.74, 15.68);

\path[draw=drawColor,line width= 0.4pt,line join=round,line cap=round] (200.55, 21.68) -- (200.55, 15.68);

\path[draw=drawColor,line width= 0.4pt,line join=round,line cap=round] (283.36, 21.68) -- (283.36, 15.68);

\path[draw=drawColor,line width= 0.4pt,line join=round,line cap=round] (366.17, 21.68) -- (366.17, 15.68);

\node[text=drawColor,anchor=base,inner sep=0pt, outer sep=0pt, scale=  1.00] at ( 34.93,  2.48) {0};

\node[text=drawColor,anchor=base,inner sep=0pt, outer sep=0pt, scale=  1.00] at (117.74,  2.48) {1};

\node[text=drawColor,anchor=base,inner sep=0pt, outer sep=0pt, scale=  1.00] at (200.55,  2.48) {2};

\node[text=drawColor,anchor=base,inner sep=0pt, outer sep=0pt, scale=  1.00] at (283.36,  2.48) {3};

\node[text=drawColor,anchor=base,inner sep=0pt, outer sep=0pt, scale=  1.00] at (366.17,  2.48) {4};

\path[draw=drawColor,line width= 0.4pt,line join=round,line cap=round] ( 21.68, 30.44) -- ( 21.68,186.37);

\path[draw=drawColor,line width= 0.4pt,line join=round,line cap=round] ( 21.68, 30.44) -- ( 15.68, 30.44);

\path[draw=drawColor,line width= 0.4pt,line join=round,line cap=round] ( 21.68, 61.63) -- ( 15.68, 61.63);

\path[draw=drawColor,line width= 0.4pt,line join=round,line cap=round] ( 21.68, 92.81) -- ( 15.68, 92.81);

\path[draw=drawColor,line width= 0.4pt,line join=round,line cap=round] ( 21.68,124.00) -- ( 15.68,124.00);

\path[draw=drawColor,line width= 0.4pt,line join=round,line cap=round] ( 21.68,155.18) -- ( 15.68,155.18);

\path[draw=drawColor,line width= 0.4pt,line join=round,line cap=round] ( 21.68,186.37) -- ( 15.68,186.37);

\node[text=drawColor,rotate= 90.00,anchor=base,inner sep=0pt, outer sep=0pt, scale=  1.00] at (  9.68, 30.44) {0.0};

\node[text=drawColor,rotate= 90.00,anchor=base,inner sep=0pt, outer sep=0pt, scale=  1.00] at (  9.68, 61.63) {0.2};

\node[text=drawColor,rotate= 90.00,anchor=base,inner sep=0pt, outer sep=0pt, scale=  1.00] at (  9.68, 92.81) {0.4};

\node[text=drawColor,rotate= 90.00,anchor=base,inner sep=0pt, outer sep=0pt, scale=  1.00] at (  9.68,124.00) {0.6};

\node[text=drawColor,rotate= 90.00,anchor=base,inner sep=0pt, outer sep=0pt, scale=  1.00] at (  9.68,155.18) {0.8};

\node[text=drawColor,rotate= 90.00,anchor=base,inner sep=0pt, outer sep=0pt, scale=  1.00] at (  9.68,186.37) {1.0};

\path[draw=drawColor,line width= 0.6pt,line join=round,line cap=round] ( 21.68, 21.68) --
	(379.42, 21.68) --
	(379.42,195.13) --
	( 21.68,195.13) --
	( 21.68, 21.68);
\end{scope}
\begin{scope}
\path[clip] (  0.00,  0.00) rectangle (379.42,195.13);
\definecolor{drawColor}{RGB}{0,0,0}

\node[text=drawColor,anchor=base,inner sep=0pt, outer sep=0pt, scale=  1.00] at (200.55, 30.08) {$y$};

\node[text=drawColor,rotate= 90.00,anchor=base,inner sep=0pt, outer sep=0pt, scale=  1.00] at ( 37.28,108.41) {$F(y)$};
\end{scope}
\begin{scope}
\path[clip] ( 21.68, 21.68) rectangle (379.42,195.13);
\definecolor{drawColor}{RGB}{0,0,255}

\path[draw=drawColor,line width= 0.4pt,line join=round,line cap=round] (  0.00, 29.66) --
	( 34.93, 29.66) --
	(117.74, 68.64) --
	(200.55,107.63) --
	(283.36,169.99) --
	(366.17,185.59) --
	(379.42,185.59);
\definecolor{drawColor}{RGB}{255,0,0}

\path[draw=drawColor,line width= 0.4pt,dash pattern=on 4pt off 4pt ,line join=round,line cap=round] (  0.00, 32.78) --
	( 34.93, 32.78) --
	(117.74, 95.15) --
	(200.55,110.74) --
	(283.36,126.34) --
	(366.17,188.70) --
	(379.42,188.70);
\definecolor{drawColor}{RGB}{0,0,255}

\path[draw=drawColor,line width= 0.4pt,dash pattern=on 4pt off 4pt ,line join=round,line cap=round] (  0.00, 28.11) --
	( 34.93, 28.11) --
	(117.74, 43.70) --
	(200.55,106.07) --
	(283.36,168.44) --
	(366.17,184.03) --
	(379.42,184.03);
\definecolor{drawColor}{RGB}{0,0,0}

\path[draw=drawColor,line width= 0.6pt,line join=round,line cap=round] (256.34, 81.68) rectangle (379.42, 21.68);
\definecolor{drawColor}{RGB}{255,0,0}

\path[draw=drawColor,line width= 0.6pt,line join=round,line cap=round] (265.34, 69.68) -- (283.34, 69.68);
\definecolor{drawColor}{RGB}{0,0,255}

\path[draw=drawColor,line width= 0.6pt,line join=round,line cap=round] (265.34, 57.68) -- (283.34, 57.68);
\definecolor{drawColor}{RGB}{255,0,0}

\path[draw=drawColor,line width= 0.6pt,dash pattern=on 4pt off 4pt ,line join=round,line cap=round] (265.34, 45.68) -- (283.34, 45.68);
\definecolor{drawColor}{RGB}{0,0,255}

\path[draw=drawColor,line width= 0.6pt,dash pattern=on 4pt off 4pt ,line join=round,line cap=round] (265.34, 33.68) -- (283.34, 33.68);
\definecolor{drawColor}{RGB}{0,0,0}

\node[text=drawColor,anchor=base west,inner sep=0pt, outer sep=0pt, scale=  1.00] at (292.34, 66.24) {$F_1$};

\node[text=drawColor,anchor=base west,inner sep=0pt, outer sep=0pt, scale=  1.00] at (292.34, 54.24) {$F_2$};

\node[text=drawColor,anchor=base west,inner sep=0pt, outer sep=0pt, scale=  1.00] at (292.34, 42.24) {$\mathbb{Q}(Y \leq y \mid F = F_1)$};

\node[text=drawColor,anchor=base west,inner sep=0pt, outer sep=0pt, scale=  1.00] at (292.34, 30.24) {$\mathbb{Q}(Y \leq y \mid F = F_2)$};
\end{scope}
\end{tikzpicture}} 
	\caption{The \emph{equiprobable} predictive distribution $F$ picks the
		piecewise linear, partially (namely, for $y \leq 2$) identical CDFs
		$F_1$ and $F_2$ with equal probability.  It is jointly CEP,
		quantile, and threshold calibrated, but fails to be
		auto-calibrated.  \label{fig:counter_auto}}
\end{figure}
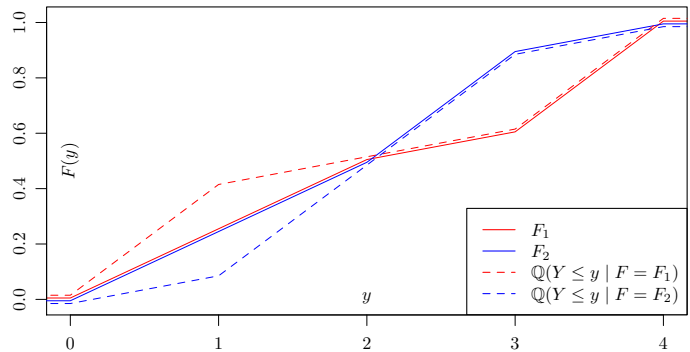

\begin{proof} 
	We establish the claims in a series of (counter) examples, starting
	with part (b), where we present an example based on two
	equiprobable, partially overlapping CDFs in Figure
	\ref{fig:counter_auto}. A similar example based on four equiprobable, partially overlapping CDFs in Appendix \ref{sec:STCnotAC} yields part (a).  As for part (c), we return to the piecewise
	uniform forecast in Example \ref{ex:p-u.tern}, where for simplicity
	we fix $\mu = 0$.  This forecast is probabilistically and marginally
	calibrated, but fails to be threshold calibrated, because
	\[
	\textstyle
	\myQ \left( Y \leq \frac{3}{2} \mid F(\frac{3}{2}) = \frac{5}{8} \right) 
	= \frac{5}{10} + \frac{1}{2} \cdot \frac{1}{10} 
	= \frac{11}{20} \not= \frac{5}{8}.  
	\]
	As for parts (d) and (e), we refer to the unfocused and lopsided
	forecasts from Example \ref{ex:unf.lop}.
\end{proof} 

\begin{table}[t]
	\caption{Properties of the forecasts in our examples.  We note whether
		they are auto-calibrated (AC), CEP calibrated (CC),
		quantile calibrated (QC), threshold calibrated (TC),
		probabilistically calibrated (PC), or marginally calibrated (MC),
		and whether the involved distributions are continuous and strictly
		increasing on a common support (CSI) as in Assumption
		\ref{as:csi}(i).  Except for the auto-calibrated cases, the
		forecasts fail to be moment calibrated.}
	\label{tab:hierarchy} 
	\vspace{4mm}
	\centering 
	\footnotesize
	\begin{tabular}{llccccccc} 
		\toprule
		Source                        & Forecast Type     & CSI          & AC  & CC  & QC  & TC  & PC  & MC \\ 
		\toprule
		Example \ref{ex:perf.clim}    & Perfect           & \cm          & \cm & \cm & \cm & \cm & \cm & \cm \\
		Example \ref{ex:perf.clim}    & Unconditional     & \cm          & \cm & \cm & \cm & \cm & \cm & \cm \\
		Figure \ref{fig:counter_auto} & Equiprobable      & \cm          &     & \cm & \cm & \cm & \cm & \cm \\   
		Example \ref{ex:p-u.tern}     & Piecewise uniform & as $c \to 0$ &     &     &     &     & \cm & \cm \\   
		Example \ref{ex:unf.lop}      & Unfocused         & \cm          &     &     &     &     & \cm &     \\   
		Example \ref{ex:unf.lop}      & Lopsided          & \cm          &     &     &     &     &     & \cm \\
		\midrule
		Example \ref{ex:CEP.q.diff}   & Continuous        &              &     & \cm &     & \cm & \cm & \cm \\
		Example \ref{ex:CEP.q.diff}   & Discrete          &              &     &     & \cm & \cm &     & \cm \\   
		\bottomrule
	\end{tabular}
\end{table}

Clearly, further hierarchical relations are immediate.  For example,
given that probabilistic calibration does not imply marginal
calibration, it does not imply threshold calibration nor
auto-calibration.  We leave further discussion to future work, but
note that moment calibration does not imply probabilistic nor marginal
calibration, as follows easily from classical results on the moment
problem \citep[e.g.,][]{Stoyanov2000}.  For an overview of
calibration properties in our examples, see Table \ref{tab:hierarchy}.

\subsection{Reliability diagrams}  \label{sec:reldiag} 

\begin{figure}[!t]
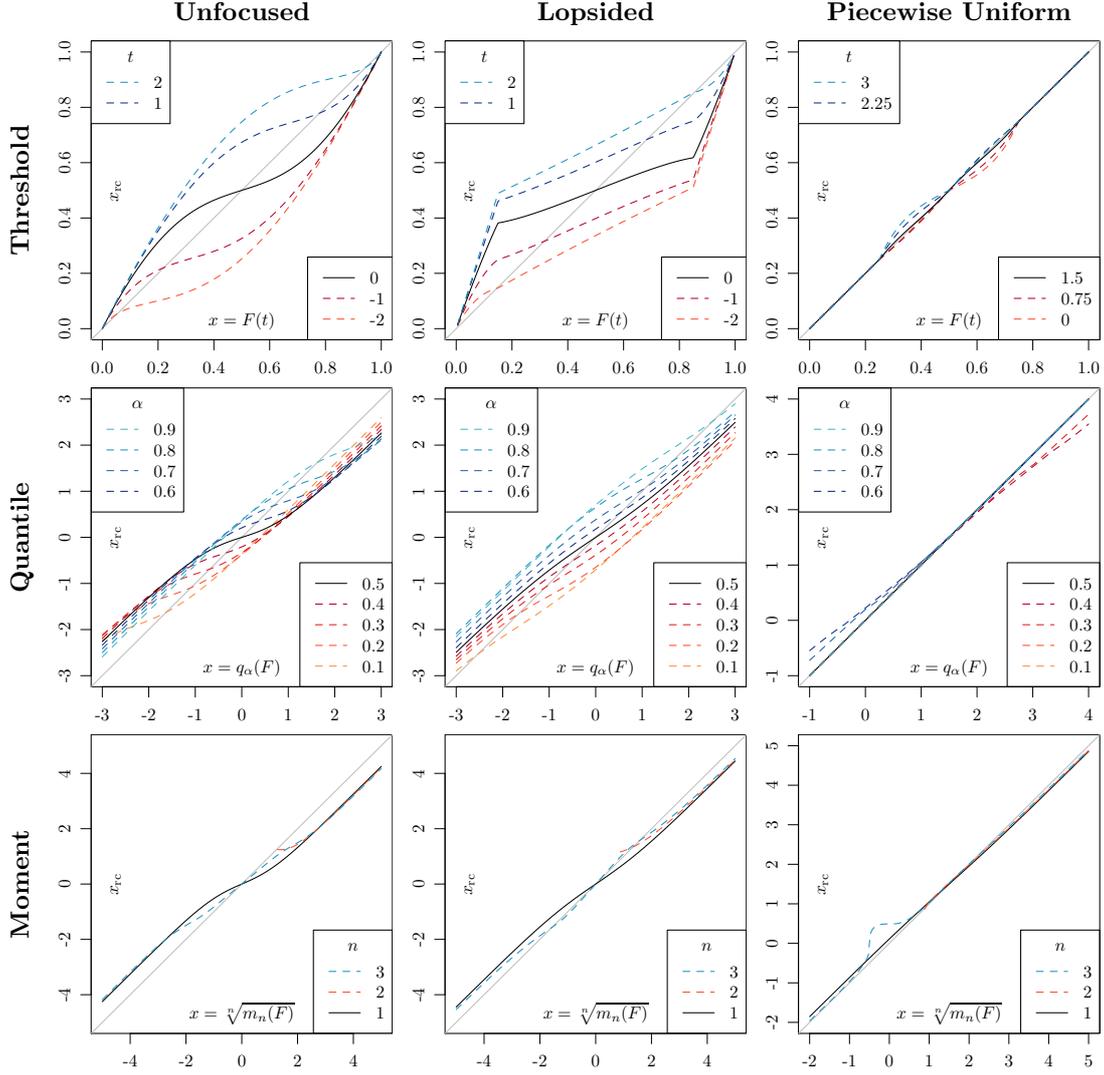

	\centering
	\scalebox{0.65}{\input{figs/tikz/theo_thresh_unf.tex}}
	\scalebox{0.65}{\input{figs/tikz/theo_thresh_lop.tex}}
	\scalebox{0.65}{\input{figs/tikz/theo_thresh_pwu.tex}} \\ \smallskip
	\scalebox{0.65}{\input{figs/tikz/theo_quant_unf.tex}}
	\scalebox{0.65}{\input{figs/tikz/theo_quant_lop.tex}}
	\scalebox{0.65}{\input{figs/tikz/theo_quant_pwu.tex}} \\ \smallskip
	\scalebox{0.65}{\input{figs/tikz/theo_moment_unf.tex}}
	\scalebox{0.65}{\input{figs/tikz/theo_moment_lop.tex}}
	\scalebox{0.65}{\input{figs/tikz/theo_moment_pwu.tex}}
	\caption{Threshold (top), quantile (middle) and moment (bottom)
		reliability diagrams for point forecasts induced by (left) the
		unfocused forecast with $\eta_0 = 1.5$ and (middle) the lopsided
		forecast with $\delta_0 = 0.7$ from Example \ref{ex:unf.lop}, and
		(right) the piecewise uniform forecast with $c = 0.5$ from Example
		\ref{ex:p-u.tern}.  Each display plots recalibrated against original
		values.  Deviations from the diagonal indicate violations of
		$\myT$-calibration.  For details see Appendix
		\ref{sec:examples}.  \label{fig:reldiag}}
\end{figure}

As we proceed to define reliability diagrams, it is useful to restrict
attention to single-valued functionals.  To this end, if an
identifiable functional $\myT$ is of interval type, we instead
consider its single-valued lower or upper bound, $\myT^-(F)$ or
$\myT^+(F)$, which we call the \emph{lower}\/ and \emph{upper
	version}\/ of $\myT$, or simply the \emph{lower}\/ and \emph{upper
	functional}, without explicit reference to the original functional
$\myT$.

The following result demonstrates that $\myT$-calibration implies
calibration of the upper and lower functional.

\begin{proposition} \label{prop:set2singleton} Suppose that the
	functional\/ $\myT$ is generated by an identification function $V$,
	and let Assumption \ref{as:T} hold.  Then
	conditional\/ $\myT$-calibration implies conditional\/ $\myT^-$- and\/
	$\myT^+$-calibration, and subject to Assumption \ref{as:V} unconditional\/ $\myT$-calibration implies
	unconditional\/ $\myT^-$- and\/ $\myT^+$-calibration.
\end{proposition}

\begin{proof}
	Suppose that $\myT^*$ is the lower or upper version of a
	functional $\myT$ generated by the identification function $V$.  As
	$\sigma(\myT^*(F)) \subseteq \sigma(\myT(F))$, we find that
	\[
	\myE \left[ V(x,Y) \mid \myT^*(F) \right] 
	= \myE \left[ \myE \left[ V(x,Y) \mid \myT(F) \right] \mid \myT^*(F) \right] 
	\]
	is almost surely $\leq 0$ if $x < \myT^*(F)$, and almost surely $\geq
	0$ if $x > \myT^*(F)$. Hence, $\myT^*(F) \in
	\myT(\cL(Y\mid \myT^*(F)))$. If $\myT^*$ is the lower functional, the
	former inequality is strict and hence $\myT^-(F) = \min \myT(\cL(Y\mid \myT^-(F)))$, whereas if $\myT^*$ is the upper
	functional, the latter is strict and hence $\myT^+(F) = \max \myT(\cL(Y\mid \myT^+(F)))$.
	
	Unconditional $\myT^*$-calibration is an immediate consequence of unconditional $\myT$-calibration.
\end{proof}

In this light, we restrict attention to single-valued functionals that
are lower or upper versions of identifiable functionals, or
identifiable functionals of singleton type.  Any such functional can
be associated with a random variable $X = \myT(F)$, and we call any
random variable $X_\rc$, for which
\begin{equation}  \label{eq:rc} 
X_\rc = \myT \left( \cL \left( Y \mid X \right) \right)
\end{equation} 
almost surely, a \emph{recalibrated}\/ version of $X$.  Clearly, we
can also define $X_\rc$ for a stand-alone point forecast $X$, based on
conceptualized distributions, by resorting to the joint distribution
of the random tuple $(X,Y)$, provided the right-hand side of
\eqref{eq:rc} is well defined and finite almost surely.  The point
forecast $X$ is \emph{conditionally}\/ $\myT$-\emph{calibrated}, or
simply $\myT$-\emph{calibrated}, if $X = X_\rc$ almost surely.
Subject to Assumption \ref{as:V}, $X$ is \emph{unconditionally}\/ $\myT$-\emph{calibrated}\/
if
\begin{equation}  \label{eq:XuncT}
\myE[ V(X - \varepsilon, Y) ] \leq 0 \quad \text{and} \quad 
\myE[ V(X + \varepsilon, Y) ] \geq 0 \quad \text{for all} \quad \varepsilon > 0.		
\end{equation}  
For recent discussions of the particular cases of the mean or
expectation and quantile functionals see, e.g., \citet[Sections
2.1--2.2]{Nolde2017}.  \citet[][Proposition 2]{Patton2020},
\citet[][Definition 3.1]{Krueger2020} and \cite[][Section
2]{Satopaa2021}.

To compare the posited functional $X$ with its recalibrated version
$X_\rc$, we introduce the $\myT$-reliability diagram.

\begin{assumption}  \label{as:X} 
	The functional $\myT$ is a lower
	or upper version of an identifiable functional, or an identifiable
	functional of singleton type.  The point forecast $X$ is a random variable, and the recalibrated forecast
	$X_\rc = \myT(\cL(Y \mid X))$ is well defined and finite almost
	surely.
\end{assumption} 

\begin{definition}  \label{def:reldiag} 
	Under Assumption \ref{as:X}, the $\myT$-\emph{reliability diagram} is
	the graph of a mapping $x \mapsto \myT \left( \cL(Y \mid X = x)
	\right)$ on the support of $X$.
\end{definition}

While technically the $\myT$-reliability diagram depends on the choice
of a regular conditional distribution for the outcome $Y$, this is not
a matter of practical relevance.  Evidently, for a $\myT$-calibrated
forecast the $\myT$-reliability diagram is concentrated on the
diagonal.  Conversely, deviations from the diagonal indicate
violations of $\myT$-calibration and can be interpreted
diagnostically, as illustrated in Figure \ref{fig:reldiag} for
threshold, quantile, and moment calibration.  For a similar display
in the specific case of mean calibration see Figure 1 of
\citet{Pohle2020}.

In the setting of fully specified predictive distributions, the
distinction between unconditional and conditional $\myT$-calibration
is natural.  Perhaps surprisingly, the distinction vanishes in the
setting of stand-alone point forecasts if the associated
identification function is of prediction error form and the
forecast and the residual are independent.

\begin{theorem}  \label{th:cond=uncond} 
	Let Assumption\/ \ref{as:X} hold, and suppose that the underlying
	identification function\/ $V$ satisfies Assumption \ref{as:V}.
	Suppose furthermore that the point forecast\/ $X$ and the generalized residual\/
	$\myT(\delta_Y) - X$ are independent.  Then $X$ is conditionally\/
	$\myT$-calibrated if, and only if, it is unconditionally\/
	$\myT$-calibrated.
\end{theorem}

\begin{proof}
	Given any constant $c \in \real$ it holds that
	\[
	\myE[ V(X+c,Y) \mid X] 
	= \myE[v(\myT(\delta_Y)-X-c) \mid X] 
	= \myE[v(\myT(\delta_Y)-X-c)].
	\]
	In view of \eqref{eq:rc} and \eqref{eq:XuncT}, conditional and
	unconditional $\myT$-calibration are equivalent.
\end{proof}

For quantiles, expectiles, and Huber functionals $V$
is of prediction error form and the generalized
residual reduces to the standard residual, $X-Y$.  In particular,
this applies in the case of least squares regression, where
$\myT$ is the mean functional, and the forecast and the residual have
typically been assumed to be independent in the literature.  We
discuss the statistical implications of Theorem
\ref{th:cond=uncond} in Appendix \ref{sec:UQ}.

\subsection{Score decompositions}  \label{sec:dcm} 

We now revisit a score decomposition into measures of miscalibration
(\MCB), discrimination (\DSC), and uncertainty (\UNC) based on
consistent scoring functions.  Specifically, suppose that $\myS$ is a
\emph{consistent}\/ loss or scoring function for the functional $\myT$
on the class $\cF$ in the sense that
\[
\myE_F \hsp [ \myS(t,Y) ] \leq \myE_F \hsp [ \myS(x,Y) ] 
\]
for all $F \in \cF$, all $t \in \myT(F) = [\myT^-(F),\myT^+(F)]$ and
all $x \in \real$ \citep{Savage1971, Gneiting2011}.  If the inequality
is strict unless $x \in \myT(F)$, then $\myS$ is \emph{strictly
	consistent}.  Consistent scoring functions serve as all-purpose
performance measures that elicit fair and honest assessments and
reward the utilization of broad information bases
\citep{Holzmann2014}.  If the functional $\myT$ is of interval type, a
consistent scoring function $\myS$ is consistent for both $\myT^-$ and
$\myT^+$, but strict consistency is lost when $\myT$ is replaced by
its lower or upper version and $\myS$ is strictly consistent for
$\myT$.  For prominent examples of consistent scoring functions, see
Table \ref{tab:loss}.

A functional is \emph{elicitable}\/ if it admits a strictly
consistent scoring function \citep{Gneiting2011}.  Under general
conditions, elicitability is equivalent to identifiability
\citep[][Theorem 5]{Steinwart2014}.  The respective functionals
allow for both principled relative forecast evaluation through the
use of consistent scoring functions, and principled absolute
forecast evaluation via $\myT$-reliability diagrams and score
decompositions, as discussed in what follows.

Let $\cL(Y)$ denote the unconditional distribution of the outcome and
suppose that $x_0 = \myT(\cL(Y))$ is well defined.  As before, we
operate under Assumption \ref{as:X} and work with $X = \myT(F)$, its
recalibrated version $X_\rc$, and the reference forecast $x_0$.
Again, the simplified notation accommodates stand-alone point
forecasts, and it suffices to consider the joint distribution of the
tuple $(X,Y)$.  Following the lead of \citet{Dawid1986} in the case of
binary outcomes, and \citet{Ehm2017} and \citet{Pohle2020} in the
setting of point forecasts for real-valued outcomes, we consider the
expected scores
\begin{equation}  \label{eq:Sexpected} 
\bar{\myS} = \myE_\myQ \hsp [ \myS(X,Y) ], \quad 
\bar{\myS}_\rc = \myE_\myQ \hsp [ \myS(X_\rc,Y) ], \quad \text{and} \quad 
\bar{\myS}_\mg = \myE_\myQ \hsp [ \myS(x_0,Y) ]
\end{equation} 
for the forecast at hand, its recalibrated version, and the
marginal reference forecast $x_0$, respectively.

\begin{definition}  \label{def:dcm} 
	Let Assumption \ref{as:X} hold, and let $x_0 = \myT(\cL(Y))$ and the
	expectations $\bar{\myS}$, $\bar{\myS}_\rc$, and $\bar{\myS}_\mg$ in
	\eqref{eq:Sexpected} be well defined and finite.  Then we refer to
	\begin{equation*}  
	\MCB_\myS = \bar{\myS} - \bar{\myS}_\rc, \quad  
	\DSC_\myS = \bar{\myS}_\mg - \bar{\myS}_\rc, \quad \text{and} \quad 
	\UNC_\myS = \bar{\myS}_\mg, 
	\end{equation*} 
	as \emph{miscalibration}, \emph{discrimination}, and
	\emph{uncertainty}, respectively.
\end{definition} 

The following result decomposes the expected score $\bar{\myS}$ for
the forecast at hand into miscalibration ($\MCB_\myS$), discrimination
($\DSC_\myS$), and uncertainty ($\UNC_\myS$) components.

\begin{theorem}[\citet{Dawid1986}, \citet{Pohle2020}]  \label{th:dcm} 
	In the setting of Definition \ref{def:dcm}, suppose that the scoring
	function $\myS$ is consistent for the functional $\myT$.  Then it
	holds that
	\begin{equation}  \label{eq:dcm}  
	\bar{\myS} = \MCB_\myS - \DSC_\myS + \UNC_\myS, 
	\end{equation} 
	where\/ $\MCB_\myS \geq 0$ with equality if\/ $X$ is conditionally\/
	$\myT$-calibrated, and\/ $\DSC_\myS \geq 0$ with equality if\/ $X_\rc
	= x_0$ almost surely.  If\/ $\myS$ is strictly consistent then\/
	$\MCB_\myS = 0$ only if\/ $X$ is conditionally\/ $\myT$-calibrated,
	and\/ $\DSC_\myS = 0$ only if $X_\rc = x_0$ almost surely.
\end{theorem}

A remaining question is what consistent scoring function $\myS$ ought
to be used in practice.  To address this issue, we resort to mixture
or Choquet representations of consistent loss functions, as introduced
by \citet{Ehm2016} for quantiles and expectiles and developed in full
generality by \citet{Dawid2016}, \citet{Ziegel2016} and
\citet{Jordan2019}.  Specifically, we rely on an obvious
generalization of Proposition 2.6 of \citet{Jordan2019}, as noted at
the start of their Section 2.  Let $\myT$ be identifiable with
identification function $V$ satisfying Assumption \ref{as:V},
and let $\eta \in \real$.  Then the \emph{elementary}\/ loss function
$\myS_\eta$, given by
\begin{equation}  \label{eq:elementary} 
\myS_\eta(x,y) 
= \left( \one \{ \eta \leq x \} - \one \{ \eta \leq y \} \right) V(\eta,y),
\end{equation}
is consistent for $\myT$.  As an immediate consequence, any well
defined function of the form
\begin{equation}  \label{eq:mixture}
\myS(x,y) = \int_\real \myS_\eta(x,y) \dd H(\eta), 
\end{equation}
where $H$ is a locally finite measure on $\real$, is consistent for
$\myT$.  If $\myT$ is a quantile, an expectile, an event probability
or a moment, then the construction includes all consistent scoring
functions, subject to standard conditions, and agrees with suitably
adapted classes of generalized piecewise linear (GPL) and Bregman
functions, respectively \citep{Gneiting2011, Ehm2016}.

\begin{table}[t]
	\caption{Canonical loss functions in the sense of Definition
		\ref{def:canonical}.  \label{tab:loss}}
	\vspace{2mm}
	\centering 
	\footnotesize
	\begin{tabular}{lll} 
		\toprule
		Functional          & Parameter          & Canonical Loss \\ 
		\midrule
		Moment of order $n$ & $n = 1, 2, \ldots$ & $\myS(x,y) = \left( x - y^n \right)^2$ \\
		$\alpha$-Expectile  & $\alpha \in (0,1)$ & $\myS(x,y) = 2 \left| \one \{ x \geq y \} - \alpha \right| \left( x - y \right)^2$ \rule{0mm}{3.5mm} \\ 
		$\alpha$-Quantile   & $\alpha \in (0,1)$ & $\myS(x,y) = 2 \left( \one \{ x \geq y \} - \alpha \right) \left( x - y \right)$ \rule{0mm}{3.5mm} \\ 
		\bottomrule
	\end{tabular}
\end{table}

We now formalize what \citet[p.~477]{Ehm2017} call the ``most
prominent'' choice, namely, scoring functions for which the mixing
measure $H$ in the representation \eqref{eq:mixture} is uniform.

\begin{definition} \label{def:canonical} Suppose that the functional
	$\myT$ is generated by an identification function $V$
	satisfying Assumption \ref{as:V}, with elementary loss functions
	$\myS_\eta$ as defined in \eqref{eq:elementary}.  Then a loss
	function $\myS$ is \emph{canonical}\/ for $\myT$ if it is
	nonnegative and admits a representation of the form
	\begin{equation}  \label{eq:canonical}
	\myS(x,y) = a \int_\real \myS_\eta(x,y) \dd \lambda(\eta) + b(y), 
	\end{equation}
	where $\lambda$ is the Lebesgue measure, $a > 0$ is a constant, and $b$ is 
	a measurable function. 
\end{definition} 

Clearly, any canonical loss function is a consistent scoring function
for $\myT$.  Furthermore, if the identification function is of the
prediction error form, then any canonical loss function has score
differentials that are invariant under translation in the sense that
$\myS(x_1+c, y+c) - \myS(x_2+c, y+c) = \myS(x_1, y) - \myS(x_2, y)$.  
Conversely, we
note from Section 5.1 of \citet{Savage1971} that for the mean
functional the canonical loss functions are the only consistent
scoring functions of this type.

\begin{table}[t]  
	\caption{Components of the decomposition \eqref{eq:dcm} for the mean
		squared error (MSE) under mean-forecasts induced by the predictive
		distributions in Examples \ref{ex:perf.clim} and
		\ref{ex:unf.lop}. Uncertainty ($\UNC$) equals 2
		irrespective of the forecast at hand.
		The term $I(\eta_0)$ is in integral form and can be evaluated
		numerically.  For details see Appendix
		\ref{sec:examples}.  \label{tab:dcm}}
	\vspace{2mm} \centering \footnotesize
	\begin{tabular}{lcccc} 
		\toprule
		Predictive Distribution	& Mean-Forecast & MSE & \MCB & \DSC \\ 
		\midrule
		Perfect       & $\mu$ & 1                             & 0    & 1 \rule{0mm}{0mm} \\  
		Unconditional & 0     & 2                             & 0    & 0 \rule{0mm}{3mm} \\   
		Unfocused     & $\mu + \frac{1}{2} \eta$ & $1 + \frac{1}{4} \eta_0^2$ 
		& $(\frac14 - I(\eta_0))\eta_0^2$ & $1 - I(\eta_0)\eta_0^2$ \rule{0mm}{3.5mm} \\  
		Lopsided      & $\mu + \frac{\sqrt{2}}{\sqrt{\pi}} \delta$ 
		& $1 + \frac{2}{\pi} \delta_0^2$ 
		& $(\frac{1}{4} - I(\sqrt{\frac{8}{\pi}}\delta_0)) \frac{8}{\pi}\delta_0^2$ 
		& $1 - I(\sqrt{\frac{8}{\pi}}\delta_0) \frac{8}{\pi} \delta_0^2$ \rule{0mm}{4.5mm} \\ 
		\bottomrule
	\end{tabular}
\end{table}  

\begin{figure}[t]  
	\centering
	\scalebox{0.67}{\input{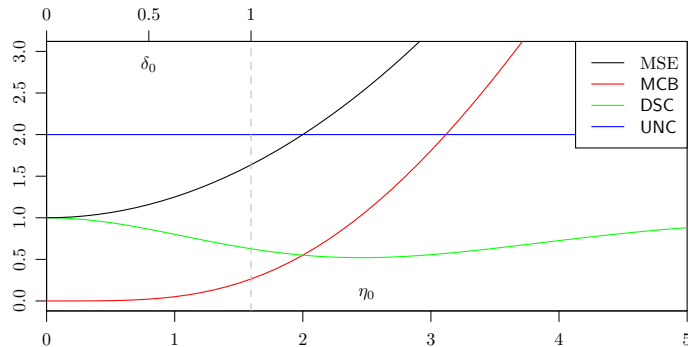}} 
	\caption{Components of the decomposition \eqref{eq:dcm} for the mean
		squared error (MSE) under mean-forecasts induced by the unfocused
		and lopsided predictive distributions from Example \ref{ex:unf.lop}
		and Table \ref{tab:dcm}, as functions of $\eta_0 \geq 0$ and
		$\delta_0 \in (0,1)$, respectively.  \label{fig:dcm}}
\end{figure}

Typically, one chooses the constant $a > 0$ and the measurable
function $b(y)$ in \eqref{eq:canonical} such that the canonical loss
admits a concise closed form, as exemplified in Table \ref{tab:loss}.
Since any selection incurs the same point forecast ranking, we refer
to the choice in Table \ref{tab:loss} as \emph{the}\/ canonical loss
function.  The most prominent example arises when $\myT$ is the
mean functional, where the ubiquitous \emph{quadratic}\/ or
\emph{squared error}\/ scoring function,
\begin{equation}  \label{eq:quadratic} 
\myS(x,y) = (x-y)^2, 
\end{equation} 
is canonical.  In this case, the $\UNC$ component equals the
unconditional variance of $Y$, as $x_0$ is simply the marginal mean
$\mu_Y$ of $Y$, and the $\MCB$ and $\DSC$ components of the general
score decomposition \eqref{eq:dcm} are
\begin{equation*}  
\MCB = \myE \left( X - X_\rc \right)^2    
\quad \text{and} \quad  
\DSC = \myE \left( X_\rc - \mu_Y \right)^2,    
\end{equation*} 
respectively.  Note that here and in the following, we drop the
subscript $\myS$ whenever we use a canonical loss.  Table
\ref{tab:dcm} and Figure \ref{fig:dcm} provide explicit
examples.

In the nested case of a binary outcome $Y$, where $X$ and $X_\rc$
specify event probabilities, the quadratic loss function reduces to
the Brier score \citep{Gneiting2007b}, and we refer to
\citet{Dimitriadis2020a} and references therein for details on score
decompositions.  In the case of threshold calibration, the point
forecast $x = F(t)$ is induced by a predictive distribution, and the
Brier score can be written as
\begin{equation}  \label{eq:Brier}  
\myS(x,y) = \left( F(t) - \one \{ y \leq t \} \right)^2.   
\end{equation}
For both real-valued and binary outcomes, it is often preferable to
use the square root of the miscalibration component ($\MCB^{1/2}$) as
a measure of calibration error that can be interpreted on natural
scales \citep[e.g.,][]{Roelofs2020}.

A canonical loss function for the Huber functional (Table
\ref{tab:functionals}) is given by
\[
\myS(x,y) = 
2 \hsp \left| \one \{ x \geq y \} - \alpha \right| 
\left\{ 
\begin{array}{ll} 
2 a |x-y| - a^2, & x-y < - a, \\ 
(x-y)^2,         & - a \leq x-y \leq b, \rule{0mm}{4mm} \\
2 b |x-y| - b^2, & x-y > b; \rule{0mm}{4mm} \\ 
\end{array}
\right.
\] 
cf.\ \citet[Definition 4.2]{Taggart2020}.  In the limiting case as $a
= b \to \infty$, we recover the canonical loss functions for the
$\alpha$-expectile, which include the quadratic loss in
\eqref{eq:quadratic}.  Similarly, if we rescale suitably and take the
limit as $a = b \to 0$, we recover the asymmetric \emph{piecewise
	linear}\/ or \emph{pinball}\/ loss, as listed in Table
\ref{tab:loss}, which lies at the heart of quantile regression.

We move on to a remarkable property of canonical loss functions.  In a
nutshell, the point forecast $X$ is unconditionally $\myT$-calibrated
if, and only if, the expected canonical loss deteriorates under
translation.  This property, which nests classical results in
regression theory, as we demonstrate at the end of Section
\ref{sec:dcm_emp}, does not hold under consistent scoring functions in
general.  For a counterexample see Appendix
\ref{sec:counterexample}.

\begin{assumption} \label{as:new} The point forecast $X$, the
	functional $\myT$ and the identification function $V$ satisfy
	Assumptions \ref{as:V} and \ref{as:X}, and $\myS$
	is a canonical loss for $\myT$.  Furthermore, $\myE[\myS(X+\eta,Y)]$
	and $\myE[V(X+\eta,Y)]$ are well defined and locally bounded as
	functions of $\eta \in \real$.
\end{assumption} 

\begin{theorem}  \label{th:optimal} 
	Under Assumption \ref{as:new}, the point forecast $X$ is
	unconditionally\/ $\myT$-calibrated if, and only if,
	\begin{equation*}  
	\myE \left[ \myS(X+c,Y) \right] \geq \myE \left[ \myS(X,Y) \right]
	\quad \text{for all} \quad c \in \real. 
	\end{equation*}
\end{theorem}

\begin{proof}
	If $X$ is unconditionally $\myT$-calibrated and $c > 0$, then
	\begin{align} \label{eq:SDiff}
	\myE \left[ \myS(X+c,Y) \right] - \myE \left[ \myS(X,Y) \right]
	&= \myE \left[ \int \left( \one \{ \eta \leq X + c \} - \one \{ \eta \leq X \} \right) V(\eta,Y) \dd\eta \right] \\ 
	&= \myE \left[ \int_{(0,c]} V(X+\eta,Y) \dd\eta \right]
	= \int_{(0,c]} \myE \left[ V(X+\eta,Y) \right] \dd\eta \notag
	\end{align} 
	is nonnegative by the second part of the unconditional
	$\myT$-calibration criterion \eqref{eq:XuncT}.  Conversely, if the
	score difference in \eqref{eq:SDiff} is nonnegative for all $c > 0$,
	then so is
	\[
	\myE[V(X+c,Y)] = \frac1c \int_{(0,c]} \myE[V(X+c,Y)] \dd\eta \geq \frac1c \int_{(0,c]} \myE[V(X+\eta,Y)] \dd\eta.
	\]
	Hence, the second part of \eqref{eq:XuncT} is satisfied.
	
	An analogous argument shows that the score difference \eqref{eq:SDiff}
	is nonnegative for all $c < 0$ if, and only if, the first inequality
	in \eqref{eq:XuncT} is satisfied.
\end{proof}

As a consequence, under a canonical loss function the $\MCB$ component
in the score decomposition \eqref{eq:dcm} of Theorem \ref{th:dcm}
decomposes into nonnegative \emph{unconditional}\/ and
\emph{conditional}\/ components $\MCB_\uncond$ and $\MCB_\cond$,
respectively, subject to the mild condition that unconditional
recalibration via translation is feasible.

\begin{theorem}  \label{th:MCB}
	Let Assumption \ref{as:new} hold, and suppose there is a constant $c$
	such that $X + c$ is unconditionally $\myT$-calibrated.  Let $X_\urc =
	X + c$ and $\bar{\myS}_\urc = \myE[\myS(X_\urc,Y)]$, and define
	\[
	\MCB_\uncond = \bar{\myS} - \bar{\myS}_\urc 
	\quad \text{and} \quad
	\MCB_\cond = \bar{\myS}_\urc - \bar{\myS}_\rc.
	\]
	Then
	\[
	\MCB = \MCB_\uncond + \MCB_\cond,
	\]
	where $\MCB_\uncond \geq 0$ with equality if\/ $X$ is
	unconditionally\/ $\myT$-calibrated, and $\MCB_\cond \geq 0$ with
	equality if\/ $X_\rc = X_\urc$ almost surely.  If\/ $\myS$ is strictly
	consistent, then\/ $\MCB_\uncond = 0$ only if\/ $X$ is
	unconditionally\/ $\myT$-calibrated, and\/ $\MCB_\cond = 0$ only if\/
	$X_\rc = X_\urc$ almost surely.
\end{theorem}

\begin{proof} 
	Immediate from Theorems \ref{th:dcm} and \ref{th:optimal},
	and the fact that conditional recalibration of $X$ and $X + c$ yields
	the same $X_\rc$.
\end{proof} 

In settings that are equivariant under translation, such as for
expectiles, quantiles and Huber functionals when both $X$ and $Y$ are
supported on the real line, $X$ can always be unconditionally
recalibrated by adding a constant.  Under any canonical loss function
$\myS$, the basic decomposition \eqref{eq:dcm} then extends to
\begin{equation}  \label{eq:dcm_ext}  
\bar{\myS} = \MCB_\uncond + \MCB_\cond - \DSC + \UNC.    
\end{equation}

For instance, when $\myS(x,y) = (x-y)^2$ is the canonical loss for the
mean functional, $\MCB_\uncond = c^2$ is the squared
unconditional bias.  The forecasts in Figure
\ref{fig:dcm} and Table \ref{tab:dcm} are free of unconditional
bias, so $\MCB_\uncond = 0$ and $\MCB_\cond = \MCB$.

In all cases studied thus far, canonical loss functions are strictly
consistent \citep{Ehm2016}, and so $\MCB_\uncond = 0$ if and only if
the forecast is unconditionally $\myT$-calibrated, and $\MCB_\cond =
0$ if and only if $X_\urc = X_\rc$ almost surely.  While in other
settings, such as when the outcomes are bounded, unconditional
recalibration by translation might be counterintuitive (in principle)
or impossible (in practice), the statement of Theorem \ref{th:MCB}
continues to hold, and the above results can be refined to admit more
general forms of unconditional recalibration.  We leave these and
other ramifications to future work.

\section{Empirical reliability diagrams and score decompositions: The CORP approach}  \label{sec:empirical} 

We turn to empirical settings, where calibration checks, scores, and
score decompositions address critical practical problems in both model
diagnostics and forecast evaluation.  The most direct usage is in the
evaluation of out-of-sample predictive performance, where forecasts
may either take the form of fully specified predictive distributions,
or be single-valued point forecasts that arise, implicitly or
explicitly, as functionals of predictive distributions.  Similarly, in
model diagnostics, where in-sample goodness-of-fit is of interest, the
model might supply fully specified, parametric or non-parametric
conditional distributions, or single-valued regression output that is
interpreted as a functional of an underlying, implicit or explicit,
probability distribution.  Prominent examples for the latter setting
include ordinary least squares regression, where the mean or
expectation functional is sought, and quantile regression.

In the case of fully specified predictive distributions, we work with
tuples of the form
\begin{equation}  \label{eq:F.y} 
(F_1, y_1), \ldots, (F_n, y_n), 
\end{equation}    
where $F_i$ is a posited conditional CDF for the real-valued
observation $y_i$ for $i = 1, \ldots, n$, which we interpret as a
sample from an underlying population $\myQ$ in the prediction space
setting of Section \ref{sec:population}.  In the case of stand-alone
point forecasts or regression output, we assume throughout that the
functional $\myT$ is of the type stated in Assumption \ref{as:X} and
work with tuples of the form
\begin{equation}  \label{eq:x.y} 
(x_1, y_1), \ldots, (x_n, y_n), 
\end{equation}    
where $x_i = \myT(F_i) \in \real$ derives explicitly or implicitly
from a predictive distribution $F_i$ for $i = 1, \ldots, n$.

In the remainder of the section, we introduce empirical versions of
$\myT$-reliability diagrams (Definition \ref{def:reldiag}) and score
components (Definition \ref{def:dcm}) for samples of the form
\eqref{eq:F.y} or \eqref{eq:x.y}, which allow for both diagnostic
checks and inference about an underlying population $\myQ$.
While practitioners may think of our empirical versions exclusively
from diagnostic perspectives, we emphasize that they can be
interpreted as estimators of the population quantities and be analyzed
as such.  A key feature of our approach is the use of nonparametric
isotonic regression via the pool-adjacent-violators algorithm, as
proposed by \citet{Dimitriadis2020a} in the particular case of binary
outcomes.  The generalization that we discuss here is hinted at in the
discussion section of their paper.

\subsection{The $\myT$-pool-adjacent-violators ($\myT$-PAV) algorithm}  \label{sec:PAV} 

Our key tool and workhorse is a very general version of the classical
pool-adjacent-violators (PAV) algorithm for nonparametric isotonic
regression \citep{Ayer1955, vanEeden1958}.  Historically, work on the
PAV algorithm has focused on the mean functional, as reviewed
by \citet{Barlow1972}, \citet{Robertson1980}, and \citet{deLeeuw2009},
among others.  In contrast, \citet{Jordan2019} study the PAV algorithm
in very general terms that accommodate our setting.

We rely on their work and describe the $\myT$-\emph{pool-adjacent-violators}
algorithm based on tuples $(x_1,y_1), \ldots,
(x_n,y_n)$ of the form \eqref{eq:x.y}, where without loss of
generality we may assume that $x_1 \leq \cdots \leq x_n$.
Furthermore, we let $\delta_i$ denote the point measure in the outcome
$y_i$.  More generally, for $1 \leq k \leq l \leq n$ we let
\[
\delta_{k:l} = \frac{1}{l-k+1} \sum_{i=k}^l \delta_i 
\]
be the associated empirical measure.  Algorithm \ref{alg:PAV}
describes the generation of an increasing sequence
$\hat{x}_1 \leq \dots \leq \hat{x}_n$ of recalibrated
values, which by construction are conditionally $\myT$-calibrated
with respect to the empirical measure associated with
$(\hat{x}_1, y_1), \ldots, (\hat{x}_n, y_n)$.  The algorithm rests
on partitions of the index set $\{ 1, \ldots, n \}$ into groups
$G_{k:l} = \{ k, \ldots, l \}$ of consecutive integers.

\medskip

\begin{algorithm}[t]  
	\caption{General $\myT$-PAV algorithm based on data of the form \eqref{eq:x.y}}
	\label{alg:PAV} 
	\SetAlgoLined 
	\KwIn{$(x_1,y_1), \ldots, (x_n,y_n) \in \real^2$ where $x_1 \leq \cdots \leq x_n$}
	\KwOut{$\myT$-calibrated values $\hat{x}_1, \ldots, \hat{x}_n$} 
	partition into groups $G_{1:1}, \ldots, G_{n:n}$ and let $\hat{x}_i =
	\myT(\delta_i)$ for $i = 1, \ldots, n$ \\
	\While{there are groups $G_{k:i}$ and $G_{(i+1):l}$ such that
		$\hat{x}_1 \leq \cdots \leq \hat{x}_i$ and $\hat{x}_i >
		\hat{x}_{i+1}$}{merge $G_{k:i}$ and $G_{(i+1):l}$ into $G_{k:l}$ and
		let $\hat{x}_i = \myT(\delta_{k:l})$ for $i = k, \ldots, l$}
\end{algorithm}

\medskip 

The following result summarizes the remarkable properties of the
$\myT$-PAV algorithm, as proved in Section 3.2 of \citet{Jordan2019}.

\begin{theorem}[\citet{Jordan2019}]  \label{th:PAV} 
	Suppose that the functional\/ $\myT$ is as stated in Assumption
	\ref{as:X}.  Then Algorithm\/ \ref{alg:PAV} generates a sequence\/
	$\hat{x}_1, \ldots, \hat{x}_n$ such that the empirical measure
	associated with\/ $(\hat{x}_1,y_1), \ldots, (\hat{x}_n,y_n)$ is
	conditionally\/ $\myT$-calibrated.  This sequence is optimal with
	respect to any scoring function\/ $\myS$ of the form\/
	\eqref{eq:mixture}, in that
	\begin{equation}  \label{eq:optimal}  
	\frac{1}{n} \sum_{i=1}^n \myS(\hat{x}_i,y_i) \leq \frac{1}{n} \sum_{i=1}^n \myS(t_i,y_i) 
	\end{equation} 
	for any non-decreasing sequence\/ $t_1 \leq \cdots \leq t_n$.  
\end{theorem} 

We note that for a functional of interval type, the minimum on
the left-hand side of \eqref{eq:optimal} is the same under the lower
and upper version, respectively.  For customary functionals, such as threshold
(non) exceedance probabilities, quantiles, expectiles, and moments,
the optimality is universal, as functions of the form
\eqref{eq:mixture} exhaust the class of the $\myT$-consistent scoring
functions subject to mild conditions \citep{Ehm2016}.  While the PAV
algorithm has been used extensively for the recalibration of
probabilistic classifiers \citep[e.g.,][]{Flach2012}, we are unaware
of any extant work that uses Algorithm \ref{alg:PAV} for forecast
recalibration, forecast evaluation, or model diagnostics in non-binary
settings.

\subsection{Empirical $\myT$-reliability diagrams}  \label{sec:reldiag_emp} 

Recently, \citet{Dimitriadis2020a} introduced the CORP approach for
the estimation of reliability diagrams and score decompositions in the
case of probability forecasts for binary outcomes.  In a nutshell, the
acronym CORP refers to an estimator that is Consistent under the
assumption of isotonicity for the population recalibration function
and Optimal in both finite sample and asymptotic settings, while
facilitating Reproducibility, and being based on the PAV algorithm.
Here, we extend the CORP approach and employ nonparametric isotonic
$\myT$-regression via the $\myT$-PAV algorithm under Assumption
\ref{as:X}, where $\myT$ is the lower or upper version of an
identifiable functional, or an identifiable singleton functional.

We begin by defining the empirical $\myT$-reliability diagram, which
is a sample version of the population diagram in Definition
\ref{def:reldiag}.

\begin{definition}  \label{def:reldiag_emp} 
	Let the functional $\myT$ be as stated in Assumption \ref{as:X}, and
	suppose that $\hat{x}_1, \ldots, \hat{x}_n$ originate from tuples\/
	$(x_1,y_1), \ldots, (x_n,y_n)$ with $x_1 \leq \cdots \leq x_n$ via
	Algorithm \ref{alg:PAV}.  Then the CORP \emph{empirical}\/
	$\myT$-\emph{reliability diagram}\/ is the graph of the piecewise
	linear function that connects the points $(x_1,\hat{x}_1), \ldots,
	(x_n,\hat{x}_n)$ in the Euclidean plane.
\end{definition} 

A few scattered references in the literature on forecast evaluation
have proposed displays of recalibrated against original values for
functionals other than binary event probabilities: Figures 3
and 7 of \citet{Bentzien2014} and Figure 8 of \citet{Pohle2020}
consider quantiles, and Figures 2--5 of \citet{Satopaa2015} concern
the mean functional.  However, none of these papers employ the PAV
algorithm, and the resulting diagrams are subject to issues of
stability and efficiency, as illustrated by \citet{Dimitriadis2020a}
in the case of binary outcomes.

For the CORP empirical $\myT$-reliability diagram to be consistent in
the sense of large sample convergence to the population version of
Definition \ref{def:reldiag}, the assumption of isotonicity of the
population recalibration function needs to be invoked.  As argued by
\citet{Roelofs2020} and \citet{Dimitriadis2020a}, such an assumption
is natural, and practitioners tend to dismiss nonisotonic
recalibration functions as artifacts.  Evidently, these arguments
transfer to arbitrary functionals, and any violations of the
isotonicity assumption entail horizontal segments in CORP reliability
diagrams, thereby indicating a lack of reliability.  Large sample
theory for CORP estimates of the recalibration function and the
$\myT$-reliability diagram depends on the functional $\myT$, the type
--- discrete or continuous --- of the marginal distribution of the
point forecast $X$, and smoothness conditions.  \citet{Mosching2020}
establish rates of uniform convergence in the cases of threshold (non)
exceedance and quantile functionals that complement classical
theory \citep{Barlow1972, Casady1976, Wright1984, Robertson1988,
	ElBarmi2005, Guntuboyina2018}.

\begin{figure}[!t]
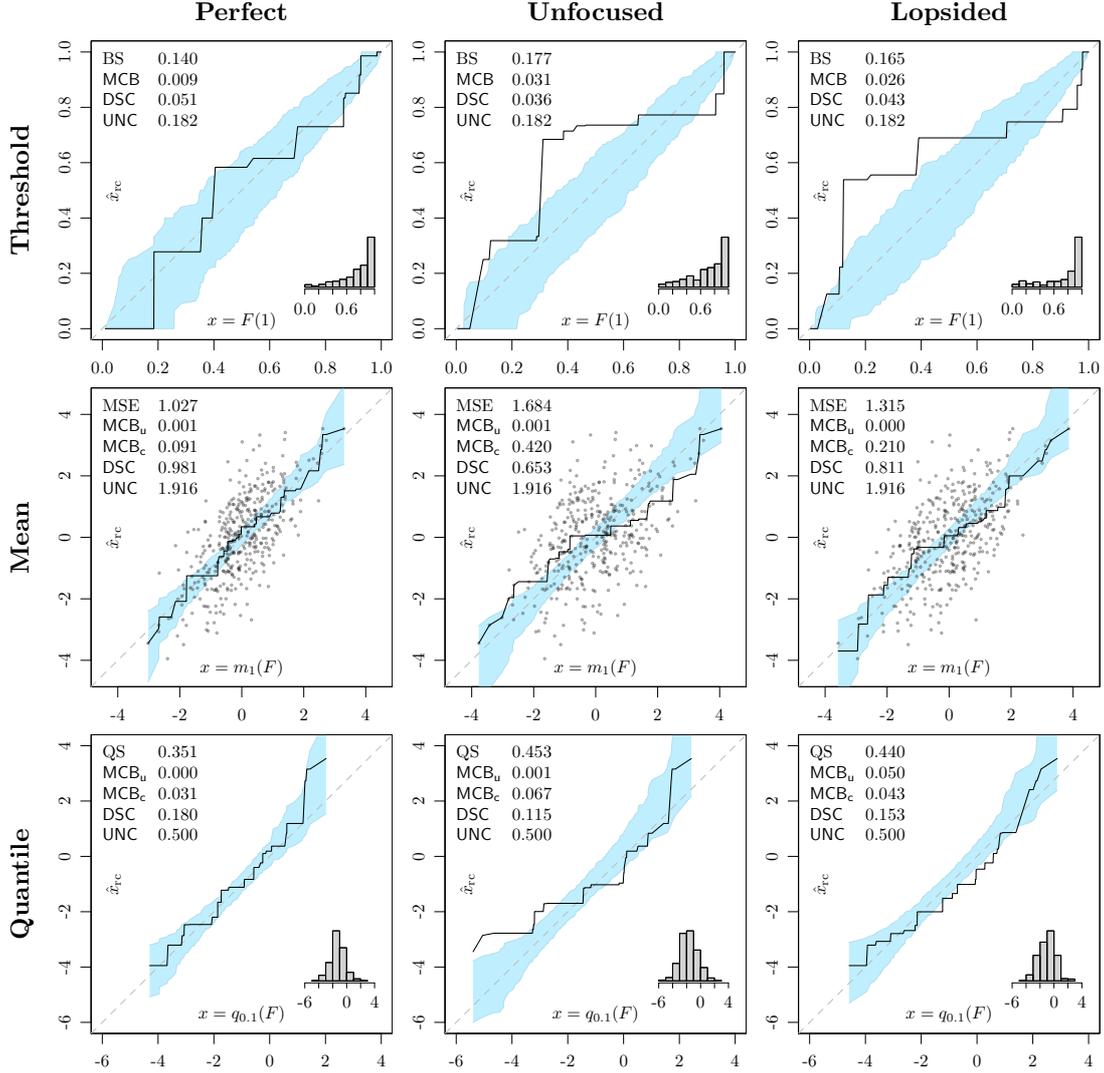

	\centering
	\scalebox{0.65}{\input{figs/tikz/relDiag_thresh_perf400.tex}}
	\scalebox{0.65}{\input{figs/tikz/relDiag_thresh_unf400.tex}}
	\scalebox{0.65}{\input{figs/tikz/relDiag_thresh_lop400.tex}} \\ \smallskip
	\scalebox{0.65}{\input{figs/tikz/relDiag_mean_perf400.tex}}
	\scalebox{0.65}{\input{figs/tikz/relDiag_mean_unf400.tex}}
	\scalebox{0.65}{\input{figs/tikz/relDiag_mean_lop400.tex}} \\ \smallskip
	\scalebox{0.65}{\input{figs/tikz/relDiag_quant_perf400.tex}}
	\scalebox{0.65}{\input{figs/tikz/relDiag_quant_unf400.tex}}
	\scalebox{0.65}{\input{figs/tikz/relDiag_quant_lop400.tex}}
	\caption{CORP empirical threshold (top, $t = 1$), mean (middle) and
		quantile (bottom, $\alpha = 0.10$) reliability diagrams for the
		perfect (left), unfocused (middle), and lopsided (right) forecast
		from Examples \ref{ex:perf.clim} and \ref{ex:unf.lop} with 90\%
		consistency bands and CORP score components under the associated
		canonical loss function based on samples of size
		400. \label{fig:reldiag_emp}}
\end{figure}

In the case of binary outcomes, \citet[p.~651]{Brocker2007} argue that
reliability diagrams ought to be supplemented by \emph{consistency
	bars}\/ for ``immediate visual evaluation as to just how likely the
observed relative frequencies are under the assumption that the
predicted probabilities are reliable.''  \citet{Dimitriadis2020a}
develop asymptotic and Monte Carlo based methods for the generation of
consistency bands to accompany a CORP reliability diagram for
dichotomous outcomes, and provide code in the form of the
\texttt{reliabilitydiag} package \citep{Dimitriadis2020b} for
\textsf{R} \citep{R}.  The consistency bands quantify and visualize
the variability of the empirical reliability diagram under the
respective null hypothesis, i.e., they show the pointwise range of the
CORP $\myT$-reliability diagram that we expect to see under a
calibrated forecast.  Algorithms \ref{alg:bands} and \ref{alg:auto} in
Appendix \ref{sec:UQ} generalize this approach to produce
consistency bands from data of the form \eqref{eq:F.y} under the
assumption of auto-calibration.  In the specific case of threshold
calibration, where the induced outcome is dichotomous, the assumptions
of auto-calibration (in the binary setting) and $\myT$-calibration
(for the non-exceedance functional) coincide \citep[Theorem
2.11]{Gneiting2013}, and we use the aforementioned
algorithms to generate consistency bands (Figure
\ref{fig:reldiag_emp}, top row).  Generally, auto-calibration is a
strictly stronger assumption than $\myT$-calibration, with ensuing
issues, which we discuss in Appendix \ref{sec:UQ_auto}.
Furthermore, to generate consistency bands from data of the form
\eqref{eq:x.y}, we cannot operate under the assumption of
auto-calibration.

As a crude yet viable alternative, we propose in Appendix
\ref{sec:UQ_T} a Monte Carlo technique for the generation of
consistency bands that is based on resampling residuals.  As in
traditional regression diagnostics, the approach depends on the
assumption of independence between point forecasts and residuals.
Figure \ref{fig:reldiag_emp} shows examples of $\myT$-reliability
diagrams with associated residual-based 90\% consistency bands for the
perfect, unfocused, and lopsided forecasts from Section
\ref{sec:population} for the mean functional (middle row) and the
lower quantile functional at level 0.10 (bottom row).  For further
discussion see Appendix \ref{sec:UQ_T}.  In the case of
the mean functional, we add the scatter diagram for the original data
of the form \eqref{eq:x.y}, whereas in the other two cases, inset
histograms visualize the marginal distribution of the point forecast.

We encourage follow-up work on both Monte Carlo and asymptotic methods
for the generation of consistency and confidence bands that are
tailored to specific functionals of interest, similar to the analysis
by \citet{Dimitriadis2020a} in the basic case of probability forecasts
for binary outcomes.

\subsection{Empirical score decompositions}  \label{sec:dcm_emp}

In this section, we consider data $(x_1,y_1), \ldots, \allowbreak (x_n,y_n)$ of
the form \eqref{eq:x.y}, where implicitly or explicitly
$x_i = \myT(F_i)$ for a single-valued functional $\myT$.  Let
$\hat{x}_1, \ldots, \hat{x}_n$ denote the respective $\myT$-PAV
recalibrated values, and let $\hat{x}_0 = \myT(\hat{F}_0)$, where
$\hat{F}_0$ is the empirical CDF of the outcomes $y_1, \ldots, y_n$.
Let
\begin{equation}  \label{eq:S_emp} 
\widehat{\myS} = \frac{1}{n} \sum_{i=1}^n \myS(x_i,y_i), \quad
\widehat{\myS}_\rc = \frac{1}{n} \sum_{i=1}^n \myS(\hat{x}_i,y_i), 
\quad \text{and} \quad 
\widehat{\myS}_\mg = \frac{1}{n} \sum_{i=1}^n \myS(\hat{x}_0,y_i) \quad  
\end{equation} 
denote the mean score of the point forecast at hand, the recalibrated
point forecast, and the functional $\myT$ applied to the
unconditional, marginal distribution of the outcome, respectively.  If
all quantities in \eqref{eq:S_emp} are finite, we refer to
\begin{equation}  \label{eq:components_emp} 
\widehat{\MCB}_\myS = \widehat{\myS} - \widehat{\myS}_\rc, \quad
\widehat{\DSC}_\myS = \widehat{\myS}_\mg - \widehat{\myS}_\rc, \quad \text{and} \quad 
\widehat{\UNC}_\myS = \widehat{\myS}_\mg
\end{equation} 
as the \emph{miscalibration}, \emph{discrimination}\/ and
\emph{uncertainty}\/ components of the mean score $\widehat{\myS}$.
Our next result generalizes Theorem 1 of \citet{Dimitriadis2020a} and
decomposes the mean score $\widehat{\myS}$ into a signed sum of
nonnegative, readily interpretable components.

\begin{theorem} \label{th:dcm_emp} 
	Suppose that the functional\/ $\myT$ satisfies the conditions in
	Assumption \ref{as:X}.  Let the scoring function\/ $\myS$ be of the
	form \eqref{eq:mixture}, suppose that\/ $\hat{x}_1, \ldots \hat{x}_n$
	originate from tuples $(x_1,y_1), \ldots, (x_n,y_n)$ via Algorithm
	\ref{alg:PAV}, and let all terms in \eqref{eq:S_emp} be finite.  Then
	\begin{equation}  \label{eq:dcm_emp}  
	\widehat{\myS} = \widehat{\MCB}_\myS - \widehat{\DSC}_\myS + \widehat{\UNC}_\myS, 
	\end{equation} 
	where\/ $\widehat{\MCB}_\myS \geq 0$ with equality if\/ $\hat{x}_i =
	x_i$ for\/ $i = 1, \ldots, n$, and\/ $\widehat{\DSC}_\myS \geq 0$ with
	equality if\/ $\hat{x}_i = \hat{x}_0$ for\/ $i = 1, \ldots, n$.
	
	If\/ $\myS$ is strictly consistent, then\/ $\widehat{\MCB}_\myS = 0$
	only if\/ $\hat{x}_i = x_i$ for\/ $i = 1, \ldots, n$ and\/
	$\widehat{\DSC}_\myS = 0$ only if\/ $\hat{x}_i = \hat{x}_0$ for\/
	$i = 1, \ldots, n$.
\end{theorem} 

\begin{proof} 
	Immediate from Theorem \ref{th:PAV}.
\end{proof} 

Thus, CORP estimates of score components enjoy the same properties as
the respective population quantities (Theorem \ref{th:dcm},
eq.~\eqref{eq:dcm}).  This is not to be taken for granted, as the
nonnegativity of the estimated components cannot be guaranteed if
approaches other than the $\myT$-PAV algorithm are used for
recalibration \citep[Supplementary Section S5]{Dimitriadis2020a}.

Recently, the estimation of calibration error has seen a surge of
interest in machine learning \citep{Guo2017, Kuleshov2018, Kumar2019,
	Nixon2019, Roelofs2020}.  Under the natural assumption of
isotonicity of the population recalibration function,
$\widehat{\MCB}_\myS$ is a consistent estimate of the population
quantity $\MCB_\myS$, with canonical loss functions being natural
choices for $\myS$.  As noted, it is often preferable to use the
square root of the miscalibration component under squared error as a
measure of calibration error that can be interpreted on natural
scales.  Asymptotic distributions for our estimators depend on the
functional $\myT$, the scoring function $\myS$, and regularity
conditions.  Large sample theory can leverage extant theory for
nonparametric isotonic regression, as hinted at in the previous
section, though score components might show distinct asymptotic
behavior.  Further development is beyond the scope of the present
paper and strongly encouraged.

In the remainder of the section, we assume that $\myS$ is a canonical
score and drop the subscript in the score components.  If there is a
constant $\hat{c} \in \real$ such that the empirical measure in $(x_1
+ \hat{c}, y_1), \ldots, (x_n + \hat{c}, y_n)$ is unconditionally
$\myT$-calibrated, let
\begin{equation}  \label{eq:S_emp_urc} 
\widehat{\myS}_\urc = \frac{1}{n} \sum_{i=1}^n \myS(x_i + \hat{c}, y_i). 
\end{equation} 
We then refer to
\begin{equation*}  
\widehat{\MCB}_\uncond = \widehat{\myS} - \widehat{\myS}_\urc \quad \text{and} \quad 
\widehat{\MCB}_\cond = \widehat{\myS}_\urc - \widehat{\myS}_\rc
\end{equation*}
as the CORP \emph{unconditional}\/ and \emph{conditional}\/
miscalibration components of the mean canonical score, respectively.
Under mild conditions, these estimates are nonnegative
and share properties of the respective population quantities
in Theorem \ref{th:MCB}.

\begin{theorem} \label{th:MCB_emp} 
	Let the conditions of Theorem\/ \ref{th:dcm_emp} hold, and let\/
	$\myS$ be a canonical loss function for\/ $\myT$.  Suppose there is a
	constant\/ $\hat{c} \in \real$ such that the empirical measure in\/
	$(x_1 + \hat{c}, y_1), \ldots, (x_n + \hat{c}, y_n)$ is
	unconditionally\/ $\myT$-calibrated, and suppose that all
	terms in \eqref{eq:S_emp_urc} are finite.  Then
	\begin{equation*}  
	\widehat{\MCB} = \widehat{\MCB}_\uncond + \widehat{\MCB}_\cond,
	\end{equation*} 
	where\/ $\widehat{\MCB}_\uncond \geq 0$ and\/ $\widehat{\MCB}_\cond
	\geq 0$.
\end{theorem} 

\begin{proof} 
	Immediate from Theorems \ref{th:optimal} and \ref{th:PAV},
	and the trivial fact that the addition of a constant is a special
	case of an isotonic mapping.
\end{proof} 

In the middle row of Figure \ref{fig:reldiag_emp}, the
extended CORP decomposition,
\begin{equation}  \label{eq:dcm_emp_ext}  
\widehat{\myS} = \widehat{\MCB}_\uncond + \widehat{\MCB}_\cond - \widehat{\DSC} + \widehat{\UNC}, 
\end{equation} 
which estimates the population decomposition \eqref{eq:dcm_ext}, is
applied to the mean squared error (MSE).  Likewise, the extended CORP
decomposition of the canonical score for quantiles, i.e., the
piecewise linear quantile score (QS) from Table \ref{tab:loss}, is
shown in the bottom row.  The top row concerns threshold calibration,
and we report the standard CORP decomposition \eqref{eq:dcm_emp} of
the Brier score (BS) from \eqref{eq:Brier}.  While the assumptions of
Theorem \ref{th:dcm_emp} are satisfied in this setting, the addition
of the constant $\hat{c}$ may yield forecast values outside the unit
interval, whence we refrain from considering the refined decomposition
in \eqref{eq:dcm_emp_ext}.

In this context, the distinction between out-of-sample forecast
evaluation and in-sample model diagnostics is critical.  When
evaluating out-of-sample forecasts, both unconditional and conditional
miscalibration are relevant.  In contrast, in-sample model fits
frequently enforce unconditional calibration.  For example, if we fit
a regression model with intercept by minimizing the canonical loss for
a functional $\myT$, Theorem \ref{th:optimal} applied to the
associated empirical measure guarantees in-sample unconditional
$\myT$-calibration.  As special cases, this line of reasoning yields
classical results in ordinary least squares regression, and the
partitioning inequalities of quantile regression in Theorem 3.4 of
\citet{Koenker1978}.

\subsection{Skill scores and a universal coefficient of determination}  \label{sec:CD} 

Let us revisit the mean scores in \eqref{eq:S_emp} under the natural
assumption that the terms in $\widehat{\myS}$ and $\widehat{\myS}_\mg$
are finite and that $\widehat{\myS}_\mg$ is strictly positive.  In
out-of-sample forecast evaluation, the quantity
\begin{equation}  \label{eq:skill} 
\widehat\myS_\skill
= 1 - \frac{\widehat{\myS}}{\widehat{\myS}_\mg} 
= \frac{\widehat{\myS}_\mg - \widehat{\myS}}{\widehat{\myS}_\mg} 
= \frac{\widehat{\DSC}_\myS - \widehat{\MCB}_\myS}{\widehat{\UNC}_\myS} 
\end{equation}
is known as \emph{skill score}\/ \citep{Murphy1989, Murphy1996,
	Gneiting2007b, Jolliffe2012} and may attain both positive and
negative values.  In particular, when $\myS(x,y) = (x-y)^2$ is the
canonical loss function for the mean functional, $\widehat\myS_\skill$
coincides with the popular Nash-Sutcliffe model efficiency coefficient
\citep[NSE;][]{Nash1970, Moriasi2007}.  A positive skill score
indicates predictive performance better than the simplistic
unconditional reference forecast, whereas a negative skill score
suggests that we are better off by using the simple reference
forecast.  Of course, it is possible, and frequently advisable, to
base skill scores on reference standards that are more sophisticated
than an unconditional, constant point forecast \citep{Hyndman2006}.

In contrast, if the goal is in-sample model diagnostics, the quantity
in \eqref{eq:skill} typically is nonnegative.  As we demonstrate now,
it constitutes a powerful generalization of the coefficient of
determination, $\myR^2$, or variance explained in least squares
regression, and its close cousin, the $\myR^1$ measure in quantile
regression \citep{Koenker1999}.  Specifically, we propose the use of
\begin{equation}  \label{eq:CD}  
\CD = \frac{\widehat{\DSC}_\myS - \widehat{\MCB}_\myS}{\widehat{\UNC}_\myS},      
\end{equation} 
as a universal \emph{coefficient of determination}.  In practice, one
takes $\myS$ to be a canonical loss for the functional $\myT$ at hand,
and we drop the subscripts in this case.  The classical $\myR^2$
measure arises when $\myS(x,y) = (x-y)^2$ is the canonical squared
error loss function for the mean functional, and the $\myR^1$ measure
of \citet{Koenker1999} emerges when $\myS(x,y) = 2 \left( \one \{ x
\geq y \} - \alpha \right) \left( x - y \right)$ is the canonical
piecewise linear loss under the $\alpha$-quantile functional.  Of
course, in the case $\alpha = \frac{1}{2}$ of the median, the
piecewise linear loss reduces to the absolute error.

\begin{figure}[t]  
	\centering % Please do not change the sizes of the boxes, this will mess up the font size!
	\scalebox{0.67}{% Created by tikzDevice version 0.12.3.1 on 2021-07-22 19:02:42
% !TEX encoding = UTF-8 Unicode
\begin{tikzpicture}[x=1pt,y=1pt]
\definecolor{fillColor}{RGB}{255,255,255}
\path[use as bounding box,fill=fillColor,fill opacity=0.00] (0,0) rectangle (361.35,195.13);
\begin{scope}
\path[clip] ( 25.29, 21.68) rectangle (198.74,195.13);
\definecolor{drawColor}{RGB}{0,0,0}

\path[draw=drawColor,line width= 0.6pt,line join=round,line cap=round] ( 42.43, 49.52) circle (  2.25);

\path[draw=drawColor,line width= 0.6pt,line join=round,line cap=round] ( 53.13, 60.22) circle (  2.25);

\path[draw=drawColor,line width= 0.6pt,line join=round,line cap=round] ( 74.55, 70.93) circle (  2.25);

\path[draw=drawColor,line width= 0.6pt,line join=round,line cap=round] ( 95.96,103.05) circle (  2.25);

\path[draw=drawColor,line width= 0.6pt,line join=round,line cap=round] (117.37,113.76) circle (  2.25);

\path[draw=drawColor,line width= 0.6pt,line join=round,line cap=round] (138.79,124.46) circle (  2.25);

\path[draw=drawColor,line width= 0.6pt,line join=round,line cap=round] (149.49,145.88) circle (  2.25);

\path[draw=drawColor,line width= 0.6pt,line join=round,line cap=round] (160.20, 92.34) circle (  2.25);

\path[draw=drawColor,line width= 0.6pt,line join=round,line cap=round] (181.61,167.29) circle (  2.25);
\end{scope}
\begin{scope}
\path[clip] (  0.00,  0.00) rectangle (361.35,195.13);
\definecolor{drawColor}{RGB}{0,0,0}

\path[draw=drawColor,line width= 0.4pt,line join=round,line cap=round] ( 31.72, 21.68) -- (192.32, 21.68);

\path[draw=drawColor,line width= 0.4pt,line join=round,line cap=round] ( 31.72, 21.68) -- ( 31.72, 15.68);

\path[draw=drawColor,line width= 0.4pt,line join=round,line cap=round] ( 85.25, 21.68) -- ( 85.25, 15.68);

\path[draw=drawColor,line width= 0.4pt,line join=round,line cap=round] (138.79, 21.68) -- (138.79, 15.68);

\path[draw=drawColor,line width= 0.4pt,line join=round,line cap=round] (192.32, 21.68) -- (192.32, 15.68);

\node[text=drawColor,anchor=base,inner sep=0pt, outer sep=0pt, scale=  1.00] at ( 31.72,  2.48) {0};

\node[text=drawColor,anchor=base,inner sep=0pt, outer sep=0pt, scale=  1.00] at ( 85.25,  2.48) {5};

\node[text=drawColor,anchor=base,inner sep=0pt, outer sep=0pt, scale=  1.00] at (138.79,  2.48) {10};

\node[text=drawColor,anchor=base,inner sep=0pt, outer sep=0pt, scale=  1.00] at (192.32,  2.48) {15};

\path[draw=drawColor,line width= 0.4pt,line join=round,line cap=round] ( 25.29, 60.22) -- ( 25.29,167.29);

\path[draw=drawColor,line width= 0.4pt,line join=round,line cap=round] ( 25.29, 60.22) -- ( 19.29, 60.22);

\path[draw=drawColor,line width= 0.4pt,line join=round,line cap=round] ( 25.29,113.76) -- ( 19.29,113.76);

\path[draw=drawColor,line width= 0.4pt,line join=round,line cap=round] ( 25.29,167.29) -- ( 19.29,167.29);

\node[text=drawColor,rotate= 90.00,anchor=base,inner sep=0pt, outer sep=0pt, scale=  1.00] at ( 13.29, 60.22) {5};

\node[text=drawColor,rotate= 90.00,anchor=base,inner sep=0pt, outer sep=0pt, scale=  1.00] at ( 13.29,113.76) {10};

\node[text=drawColor,rotate= 90.00,anchor=base,inner sep=0pt, outer sep=0pt, scale=  1.00] at ( 13.29,167.29) {15};

\path[draw=drawColor,line width= 0.6pt,line join=round,line cap=round] ( 25.29, 21.68) --
	(198.74, 21.68) --
	(198.74,195.13) --
	( 25.29,195.13) --
	( 25.29, 21.68);
\end{scope}
\begin{scope}
\path[clip] (  0.00,  0.00) rectangle (361.35,195.13);
\definecolor{drawColor}{RGB}{0,0,0}

\node[text=drawColor,anchor=base,inner sep=0pt, outer sep=0pt, scale=  1.00] at (112.02, 30.08) {$x$};

\node[text=drawColor,rotate= 90.00,anchor=base,inner sep=0pt, outer sep=0pt, scale=  1.00] at ( 40.89,108.41) {$y$};
\end{scope}
\begin{scope}
\path[clip] ( 25.29, 21.68) rectangle (198.74,195.13);
\definecolor{fillColor}{RGB}{0,255,0}

\path[fill=fillColor,fill opacity=0.15] (138.79,125.00) --
	(149.49,146.41) --
	(160.20,146.41) --
	(181.61,167.83) --
	(160.20,125.00) --
	(149.49,125.00) --
	cycle;
\definecolor{drawColor}{RGB}{0,0,255}

\path[draw=drawColor,line width= 0.6pt,line join=round,line cap=round] ( 25.29, 41.28) -- (198.74,163.98);
\definecolor{drawColor}{RGB}{0,255,0}

\path[draw=drawColor,line width= 0.6pt,line join=round,line cap=round] ( 25.29, 35.02) -- (198.74,181.79);

\path[draw=drawColor,line width= 0.6pt,dash pattern=on 4pt off 4pt ,line join=round,line cap=round] ( 42.43, 50.05) --
	( 53.13, 60.76) --
	( 74.55, 71.47) --
	( 95.96,103.59) --
	(117.37,114.29) --
	(138.79,125.00) --
	(149.49,125.00) --
	(160.20,125.00) --
	(181.61,167.83);

\path[draw=drawColor,line width= 0.6pt,dash pattern=on 4pt off 4pt ,line join=round,line cap=round] ( 42.43, 50.05) --
	( 53.13, 60.76) --
	( 74.55, 71.47) --
	( 95.96,103.59) --
	(117.37,114.29) --
	(138.79,125.00) --
	(149.49,146.41) --
	(160.20,146.41) --
	(181.61,167.83);
\definecolor{drawColor}{RGB}{0,0,255}

\path[draw=drawColor,line width= 0.6pt,dash pattern=on 4pt off 4pt ,line join=round,line cap=round] ( 42.43, 48.98) --
	( 53.13, 59.69) --
	( 74.55, 70.40) --
	( 95.96,102.52) --
	(117.37,113.22) --
	(138.79,120.36) --
	(149.49,120.36) --
	(160.20,120.36) --
	(181.61,166.76);
\end{scope}
\begin{scope}
\path[clip] (  0.00,  0.00) rectangle (361.35,195.13);
\definecolor{drawColor}{RGB}{0,0,0}
\definecolor{fillColor}{RGB}{255,255,255}

\path[draw=drawColor,line width= 0.6pt,line join=round,line cap=round,fill=fillColor] (198.87,105.68) rectangle (354.85, 21.68);
\definecolor{drawColor}{RGB}{0,0,255}

\path[draw=drawColor,line width= 0.6pt,line join=round,line cap=round] (207.87, 93.68) -- (225.87, 93.68);
\definecolor{drawColor}{RGB}{0,255,0}

\path[draw=drawColor,line width= 0.6pt,line join=round,line cap=round] (207.87, 69.68) -- (225.87, 69.68);
\definecolor{drawColor}{RGB}{0,0,255}

\path[draw=drawColor,line width= 0.6pt,dash pattern=on 4pt off 4pt ,line join=round,line cap=round] (207.87, 45.68) -- (225.87, 45.68);
\definecolor{drawColor}{RGB}{0,255,0}

\path[draw=drawColor,line width= 0.6pt,dash pattern=on 4pt off 4pt ,line join=round,line cap=round] (207.87, 33.68) -- (225.87, 33.68);
\definecolor{drawColor}{RGB}{0,0,0}

\node[text=drawColor,anchor=base west,inner sep=0pt, outer sep=0pt, scale=  1.00] at (234.87, 90.24) {Linear mean regression};

\node[text=drawColor,anchor=base west,inner sep=0pt, outer sep=0pt, scale=  1.00] at (234.87, 78.24) {$\operatorname{R}^2 = 0.779$, $\operatorname{R}^1 = 0.611$};

\node[text=drawColor,anchor=base west,inner sep=0pt, outer sep=0pt, scale=  1.00] at (234.87, 66.24) {Linear median regression};

\node[text=drawColor,anchor=base west,inner sep=0pt, outer sep=0pt, scale=  1.00] at (234.87, 54.24) {$\operatorname{R}^2 = 0.725$, $\operatorname{R}^1 = 0.692$};

\node[text=drawColor,anchor=base west,inner sep=0pt, outer sep=0pt, scale=  1.00] at (234.87, 42.24) {Isotonic mean regression};

\node[text=drawColor,anchor=base west,inner sep=0pt, outer sep=0pt, scale=  1.00] at (234.87, 30.24) {Isotonic median regression};
\end{scope}
\end{tikzpicture}}
	\caption{Linear mean and linear median regression lines for toy
		example from \citet[Figure 1]{Kvalseth1985}, along with
		nonparametric isotonic mean and median regression fits.  The
		isotonic median regression fit is not unique and framed by the
		respective lower and upper functional.  \label{fig:CD}}
\end{figure}
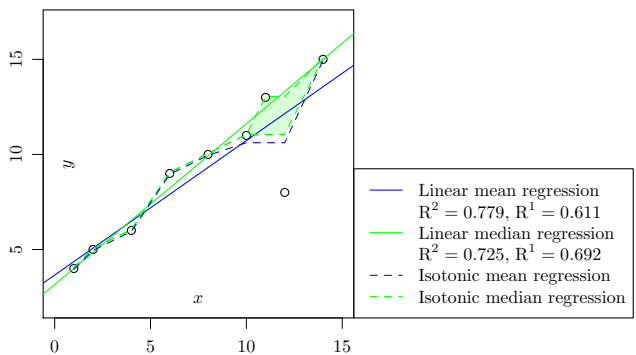

In Figure \ref{fig:CD}, we present a numerical example on the toy data
from Figure 1 in \citet{Kvalseth1985}.  The straight lines show the
linear (ordinary least squares) mean and linear (Laplace)
median regression fits, which \citet{Kvalseth1985} sought to compare.
The piecewise linear broken curves illustrate the nonparametric
isotonic regression fits, as realized by the $\myT$-PAV algorithm,
where $\myT$ is the mean and the lower and the upper median,
respectively.  As the linear regression fits induce the same ranking
of the point forecasts, they yield the same PAV-recalibrated values
that enter the terms in the score decomposition
\eqref{eq:components_emp}, and thus they have identical discrimination
components in \eqref{eq:dcm_emp}, which equal 10.593 under squared
error and 2.333 under absolute error, regardless of which isotonic
median is used.  The uncertainty components, which equal 12.000 under
squared error, and 2.889 under absolute error, are identical as well,
since they depend on the observations only.  Thus, the differences in
$\myR^2$ respectively $\myR^1$ in Figure \ref{fig:CD} stem from
distinct miscalibration components.  Of course, linear mean regression
is preferred under squared error, and linear median regression is
preferred under absolute error.

Various authors have discussed desiderata for a generally applicable
definition of a coefficient of determination \citep{Kvalseth1985,
	Nakagawa2013} for the assessment of in-sample fit.  In particular,
such a coefficient ought to be dimensionless and take values in the
unit interval, with a value of 1 indicating a perfect fit, and a value
of 0 representing a complete lack of fit.  The universal coefficient
of determination $\myR^*$ enjoys these properties under modest
conditions.

\begin{assumption}  \label{as:CD} 
	Suppose that the functional $\myT$ is as stated in Assumption
	\ref{as:X} with associated identification function $V$.  Let the
	scoring function $\myS$ be of the form \eqref{eq:mixture}, and suppose
	that\/ $\hat{x}_1, \ldots \hat{x}_n$ in \eqref{eq:S_emp} originate
	from tuples $(x_1,y_1), \ldots, (x_n,y_n)$ via Algorithm
	\ref{alg:PAV}.  Furthermore, let the following hold.
	\begin{enumerate} 
		\item[(i)] The terms contributing to\/ $\widehat{\myS}$ and\/
		$\widehat{\myS}_\mg$ in\/ \eqref{eq:S_emp} are finite, and\/
		$\widehat{\myS}_\mg > 0$.
		\item[(ii)] The values\/ $x_1, \ldots, x_n$ have been fitted to\/
		$y_1, \ldots, y_n$ by in-sample empirical loss minimization with
		respect to $\myS$, with any constant fit\/ $x_1 = \cdots = x_n$
		being admissible.
	\end{enumerate} 
\end{assumption} 

For example, suppose that $\myT$ is the mean functional and $\myS$ is
the canonical squared error scoring function.  Then condition (i) is
satisfied with the exception of the trivial case where $y_1 = \cdots =
y_n$, and condition (ii) is satisfied under linear (ordinary least
squares) mean regression with intercept.  Similarly, if $\myT$ is a
quantile and $\myS$ is the canonical piecewise linear loss function,
then (i) is satisfied except when $y_1 = \cdots = y_n$, and (ii) is
satisfied under linear quantile regression with intercept.  In this
light, the following theorem covers the classical settings for the
$\myR^2$ and $\myR^1$ measures.

\begin{theorem}  \label{th:CD} 
	Under Assumption\/ \ref{as:CD}  it holds that 
	\begin{equation*}  
	\CD \in [0,1]
	\end{equation*} 
	with\/ $\CD = 0$ if\/ $x_i = \hat{x}_0$ for\/ $i = 1, \ldots, n$,
	and\/ $\CD = 1$ if\/ $x_i = \myT(\delta_i)$ for\/
	$i = 1, \ldots, n$.
\end{theorem} 

\begin{proof} 
	The claim follows from Theorem \ref{th:PAV}, the trivial fact that a
	constant fit is a special case of an isotonic mapping, and the assumed
	form \eqref{eq:mixture} of the scoring function.
\end{proof} 

We emphasize that Assumption \ref{as:CD} and Theorem \ref{th:CD} are
concerned with, and tailored to, in-sample model diagnostics.  At the
expense of technicalities, the regularity conditions can be relaxed,
but the details are tedious and we leave them to subsequent work.
The condition that any constant fit $x_1 = \cdots = x_n$ be
admissible is critical and cannot be relaxed.

\subsection{Empirical examples}  \label{sec:empirical_ex} 

We now illustrate the use of reliability diagrams, score
decompositions, skill scores, and the coefficient of determination
$\myR^*$ for the purposes of forecast evaluation and model
diagnostics.

In the basic setting of tuples $(x_1, y_1), \ldots, (x_n, y_n)$
of the form \eqref{eq:x.y}, the point forecast $x_i$
represents the functional $\myT$ of a posited distribution for $y_i$.
The most prominent case of the mean functional and canonical
squared error loss \eqref{eq:quadratic} is illustrated in Figure
\ref{fig:butterflies}, where point forecasts by
\citet{Tredennick2021} of (log transformed) butterfly population
size are assessed. The CORP mean reliability diagram along with
90\% consistency bands under the hypothesis of mean calibration
complements the scatter plot provided by \citet[][Figure
6]{Tredennick2021}.  With a mean squared error (MSE) of 0.224,
ridge regression performs much better than the null model with
an MSE of 0.262.  The CORP score decomposition shown in Figure
\ref{fig:butterflies} refines and supports the analysis.

We move on to discuss the more complex setting of tuples $(F_1, y_1),
\ldots, (F_n, y_n)$ of the form \eqref{eq:x.y}, where $F_i$ is a
posited distribution for $y_i$ ($i = 1, \ldots, n$).  As discussed in
Section \ref{sec:population} and visualized in Figure
\ref{fig:hierarchy}, the traditional unconditional notions of
calibration, namely, probabilistic and marginal calibration,
constitute weak forms of reliability.  For this very reason, we
recommend that checks for probabilistic and marginal calibration are
given priority in this setting, much in line with current practice.
Typically, probabilistic calibration is checked by plotting histograms
of empirical probability integral transform (PIT) values
\citep{Diebold1998, Gneiting2007a}, though this is hindered by the
need for binning.  In Appendix \ref{sec:UQ_uncond}, we
discuss the \emph{PIT reliability diagram}, a rarely used alternative
that avoids binning and retains the spirit of our CORP approach by
plotting the CDF of the empirical PIT values.  Similarly, as we also
discuss in Appendix \ref{sec:UQ_uncond}, the
\emph{marginal reliability diagram}\/ can be used to assess marginal
calibration in the spirit of the CORP approach.  If the analysis
indicates gross violations of probabilistic or marginal calibration,
we note from Section \ref{sec:population} and Figure
\ref{fig:hierarchy} that key notions of conditional calibration must
be violated as well.  Otherwise, we might proceed to check stronger
conditional notions of calibration, such as threshold, mean, and
quantile calibration.

\begin{figure}[p]  
	\centering
	\scalebox{0.65}{\input{figs/tikz/relDiag_PIT_BoE1.tex}}
	\scalebox{0.65}{% Created by tikzDevice version 0.12.3.1 on 2021-08-06 11:48:27
% !TEX encoding = UTF-8 Unicode
\begin{tikzpicture}[x=1pt,y=1pt]
\definecolor{fillColor}{RGB}{255,255,255}
\path[use as bounding box,fill=fillColor,fill opacity=0.00] (0,0) rectangle (195.13,227.65);
\begin{scope}
\path[clip] ( 21.68, 21.68) rectangle (195.13,195.13);
\definecolor{drawColor}{RGB}{0,0,0}

\path[draw=drawColor,line width= 0.6pt,line join=round,line cap=round] ( 28.11, 56.43) -- ( 28.11,188.70);

\path[draw=drawColor,line width= 0.6pt,line join=round,line cap=round] ( 37.55, 56.43) -- ( 37.55, 99.52);

\path[draw=drawColor,line width= 0.6pt,line join=round,line cap=round] ( 47.00, 56.43) -- ( 47.00, 44.34);

\path[draw=drawColor,line width= 0.6pt,line join=round,line cap=round] ( 56.45, 56.43) -- ( 56.45, 68.75);

\path[draw=drawColor,line width= 0.6pt,line join=round,line cap=round] ( 65.89, 56.43) -- ( 65.89, 63.23);

\path[draw=drawColor,line width= 0.6pt,line join=round,line cap=round] ( 75.34, 56.43) -- ( 75.34, 76.71);

\path[draw=drawColor,line width= 0.6pt,line join=round,line cap=round] ( 84.79, 56.43) -- ( 84.79, 85.99);

\path[draw=drawColor,line width= 0.6pt,line join=round,line cap=round] ( 94.23, 56.43) -- ( 94.23, 71.42);

\path[draw=drawColor,line width= 0.6pt,line join=round,line cap=round] (103.68, 56.43) -- (103.68, 53.79);

\path[draw=drawColor,line width= 0.6pt,line join=round,line cap=round] (113.13, 56.43) -- (113.13, 54.87);

\path[draw=drawColor,line width= 0.6pt,line join=round,line cap=round] (122.58, 56.43) -- (122.58, 49.99);

\path[draw=drawColor,line width= 0.6pt,line join=round,line cap=round] (132.02, 56.43) -- (132.02, 62.58);

\path[draw=drawColor,line width= 0.6pt,line join=round,line cap=round] (141.47, 56.43) -- (141.47, 80.52);

\path[draw=drawColor,line width= 0.6pt,line join=round,line cap=round] (150.92, 56.43) -- (150.92, 50.55);

\path[draw=drawColor,line width= 0.6pt,line join=round,line cap=round] (160.36, 56.43) -- (160.36, 49.48);

\path[draw=drawColor,line width= 0.6pt,line join=round,line cap=round] (169.81, 56.43) -- (169.81, 44.65);

\path[draw=drawColor,line width= 0.6pt,line join=round,line cap=round] (179.26, 56.43) -- (179.26, 31.86);

\path[draw=drawColor,line width= 0.6pt,line join=round,line cap=round] (188.71, 56.43) -- (188.71, 49.19);
\end{scope}
\begin{scope}
\path[clip] (  0.00,  0.00) rectangle (195.13,227.65);
\definecolor{drawColor}{RGB}{0,0,0}

\path[draw=drawColor,line width= 0.4pt,line join=round,line cap=round] ( 28.11, 21.68) -- (169.81, 21.68);

\path[draw=drawColor,line width= 0.4pt,line join=round,line cap=round] ( 28.11, 21.68) -- ( 28.11, 15.68);

\path[draw=drawColor,line width= 0.4pt,line join=round,line cap=round] ( 75.34, 21.68) -- ( 75.34, 15.68);

\path[draw=drawColor,line width= 0.4pt,line join=round,line cap=round] (122.58, 21.68) -- (122.58, 15.68);

\path[draw=drawColor,line width= 0.4pt,line join=round,line cap=round] (169.81, 21.68) -- (169.81, 15.68);

\node[text=drawColor,anchor=base,inner sep=0pt, outer sep=0pt, scale=  1.00] at ( 28.11,  2.48) {0};

\node[text=drawColor,anchor=base,inner sep=0pt, outer sep=0pt, scale=  1.00] at ( 75.34,  2.48) {5};

\node[text=drawColor,anchor=base,inner sep=0pt, outer sep=0pt, scale=  1.00] at (122.58,  2.48) {10};

\node[text=drawColor,anchor=base,inner sep=0pt, outer sep=0pt, scale=  1.00] at (169.81,  2.48) {15};

\path[draw=drawColor,line width= 0.4pt,line join=round,line cap=round] ( 21.68, 29.98) -- ( 21.68,188.70);

\path[draw=drawColor,line width= 0.4pt,line join=round,line cap=round] ( 21.68, 29.98) -- ( 15.68, 29.98);

\path[draw=drawColor,line width= 0.4pt,line join=round,line cap=round] ( 21.68, 56.43) -- ( 15.68, 56.43);

\path[draw=drawColor,line width= 0.4pt,line join=round,line cap=round] ( 21.68, 82.89) -- ( 15.68, 82.89);

\path[draw=drawColor,line width= 0.4pt,line join=round,line cap=round] ( 21.68,109.34) -- ( 15.68,109.34);

\path[draw=drawColor,line width= 0.4pt,line join=round,line cap=round] ( 21.68,135.80) -- ( 15.68,135.80);

\path[draw=drawColor,line width= 0.4pt,line join=round,line cap=round] ( 21.68,162.25) -- ( 15.68,162.25);

\path[draw=drawColor,line width= 0.4pt,line join=round,line cap=round] ( 21.68,188.70) -- ( 15.68,188.70);

\node[text=drawColor,rotate= 90.00,anchor=base,inner sep=0pt, outer sep=0pt, scale=  1.00] at (  9.68, 29.98) {-0.2};

\node[text=drawColor,rotate= 90.00,anchor=base,inner sep=0pt, outer sep=0pt, scale=  1.00] at (  9.68, 56.43) {0.0};

\node[text=drawColor,rotate= 90.00,anchor=base,inner sep=0pt, outer sep=0pt, scale=  1.00] at (  9.68, 82.89) {0.2};

\node[text=drawColor,rotate= 90.00,anchor=base,inner sep=0pt, outer sep=0pt, scale=  1.00] at (  9.68,109.34) {0.4};

\node[text=drawColor,rotate= 90.00,anchor=base,inner sep=0pt, outer sep=0pt, scale=  1.00] at (  9.68,135.80) {0.6};

\node[text=drawColor,rotate= 90.00,anchor=base,inner sep=0pt, outer sep=0pt, scale=  1.00] at (  9.68,162.25) {0.8};

\node[text=drawColor,rotate= 90.00,anchor=base,inner sep=0pt, outer sep=0pt, scale=  1.00] at (  9.68,188.70) {1.0};

\path[draw=drawColor,line width= 0.6pt,line join=round,line cap=round] ( 21.68, 21.68) --
	(195.13, 21.68) --
	(195.13,195.13) --
	( 21.68,195.13) --
	( 21.68, 21.68);
\end{scope}
\begin{scope}
\path[clip] (  0.00,  0.00) rectangle (195.13,227.65);
\definecolor{drawColor}{RGB}{0,0,0}

\node[text=drawColor,anchor=base,inner sep=0pt, outer sep=0pt, scale=  1.00] at (108.41, 30.08) {lag};
\end{scope}
\begin{scope}
\path[clip] ( 21.68, 21.68) rectangle (195.13,195.13);
\definecolor{drawColor}{RGB}{0,0,0}

\path[draw=drawColor,line width= 0.6pt,line join=round,line cap=round] ( 21.68, 56.43) -- (195.13, 56.43);
\definecolor{drawColor}{RGB}{0,0,255}

\path[draw=drawColor,line width= 0.6pt,dash pattern=on 4pt off 4pt ,line join=round,line cap=round] ( 21.68, 84.76) -- (195.13, 84.76);

\path[draw=drawColor,line width= 0.6pt,dash pattern=on 4pt off 4pt ,line join=round,line cap=round] ( 21.68, 28.11) -- (195.13, 28.11);
\end{scope}
\begin{scope}
\path[clip] (  0.00,  0.00) rectangle (195.13,227.65);
\definecolor{drawColor}{RGB}{0,0,0}

\node[text=drawColor,anchor=base,inner sep=0pt, outer sep=0pt, scale=  1.50] at (108.41,207.13) {\bfseries (b) ACF of PIT};
\end{scope}
\end{tikzpicture}}
	\scalebox{0.65}{% Created by tikzDevice version 0.12.3.1 on 2021-08-06 11:48:27
% !TEX encoding = UTF-8 Unicode
\begin{tikzpicture}[x=1pt,y=1pt]
\definecolor{fillColor}{RGB}{255,255,255}
\path[use as bounding box,fill=fillColor,fill opacity=0.00] (0,0) rectangle (195.13,227.65);
\begin{scope}
\path[clip] ( 21.68, 21.68) rectangle (195.13,195.13);
\definecolor{drawColor}{RGB}{0,0,0}

\path[draw=drawColor,line width= 0.6pt,line join=round,line cap=round] ( 28.11, 56.43) -- ( 28.11,188.70);

\path[draw=drawColor,line width= 0.6pt,line join=round,line cap=round] ( 37.55, 56.43) -- ( 37.55,133.15);

\path[draw=drawColor,line width= 0.6pt,line join=round,line cap=round] ( 47.00, 56.43) -- ( 47.00,107.83);

\path[draw=drawColor,line width= 0.6pt,line join=round,line cap=round] ( 56.45, 56.43) -- ( 56.45,100.27);

\path[draw=drawColor,line width= 0.6pt,line join=round,line cap=round] ( 65.89, 56.43) -- ( 65.89, 95.56);

\path[draw=drawColor,line width= 0.6pt,line join=round,line cap=round] ( 75.34, 56.43) -- ( 75.34, 90.77);

\path[draw=drawColor,line width= 0.6pt,line join=round,line cap=round] ( 84.79, 56.43) -- ( 84.79, 69.52);

\path[draw=drawColor,line width= 0.6pt,line join=round,line cap=round] ( 94.23, 56.43) -- ( 94.23, 52.99);

\path[draw=drawColor,line width= 0.6pt,line join=round,line cap=round] (103.68, 56.43) -- (103.68, 65.82);

\path[draw=drawColor,line width= 0.6pt,line join=round,line cap=round] (113.13, 56.43) -- (113.13, 72.47);

\path[draw=drawColor,line width= 0.6pt,line join=round,line cap=round] (122.58, 56.43) -- (122.58, 67.06);

\path[draw=drawColor,line width= 0.6pt,line join=round,line cap=round] (132.02, 56.43) -- (132.02, 70.37);

\path[draw=drawColor,line width= 0.6pt,line join=round,line cap=round] (141.47, 56.43) -- (141.47, 62.04);

\path[draw=drawColor,line width= 0.6pt,line join=round,line cap=round] (150.92, 56.43) -- (150.92, 60.21);

\path[draw=drawColor,line width= 0.6pt,line join=round,line cap=round] (160.36, 56.43) -- (160.36, 70.97);

\path[draw=drawColor,line width= 0.6pt,line join=round,line cap=round] (169.81, 56.43) -- (169.81, 58.62);

\path[draw=drawColor,line width= 0.6pt,line join=round,line cap=round] (179.26, 56.43) -- (179.26, 65.58);

\path[draw=drawColor,line width= 0.6pt,line join=round,line cap=round] (188.71, 56.43) -- (188.71, 63.89);
\end{scope}
\begin{scope}
\path[clip] (  0.00,  0.00) rectangle (195.13,227.65);
\definecolor{drawColor}{RGB}{0,0,0}

\path[draw=drawColor,line width= 0.4pt,line join=round,line cap=round] ( 28.11, 21.68) -- (169.81, 21.68);

\path[draw=drawColor,line width= 0.4pt,line join=round,line cap=round] ( 28.11, 21.68) -- ( 28.11, 15.68);

\path[draw=drawColor,line width= 0.4pt,line join=round,line cap=round] ( 75.34, 21.68) -- ( 75.34, 15.68);

\path[draw=drawColor,line width= 0.4pt,line join=round,line cap=round] (122.58, 21.68) -- (122.58, 15.68);

\path[draw=drawColor,line width= 0.4pt,line join=round,line cap=round] (169.81, 21.68) -- (169.81, 15.68);

\node[text=drawColor,anchor=base,inner sep=0pt, outer sep=0pt, scale=  1.00] at ( 28.11,  2.48) {0};

\node[text=drawColor,anchor=base,inner sep=0pt, outer sep=0pt, scale=  1.00] at ( 75.34,  2.48) {5};

\node[text=drawColor,anchor=base,inner sep=0pt, outer sep=0pt, scale=  1.00] at (122.58,  2.48) {10};

\node[text=drawColor,anchor=base,inner sep=0pt, outer sep=0pt, scale=  1.00] at (169.81,  2.48) {15};

\path[draw=drawColor,line width= 0.4pt,line join=round,line cap=round] ( 21.68, 29.98) -- ( 21.68,188.70);

\path[draw=drawColor,line width= 0.4pt,line join=round,line cap=round] ( 21.68, 29.98) -- ( 15.68, 29.98);

\path[draw=drawColor,line width= 0.4pt,line join=round,line cap=round] ( 21.68, 56.43) -- ( 15.68, 56.43);

\path[draw=drawColor,line width= 0.4pt,line join=round,line cap=round] ( 21.68, 82.89) -- ( 15.68, 82.89);

\path[draw=drawColor,line width= 0.4pt,line join=round,line cap=round] ( 21.68,109.34) -- ( 15.68,109.34);

\path[draw=drawColor,line width= 0.4pt,line join=round,line cap=round] ( 21.68,135.80) -- ( 15.68,135.80);

\path[draw=drawColor,line width= 0.4pt,line join=round,line cap=round] ( 21.68,162.25) -- ( 15.68,162.25);

\path[draw=drawColor,line width= 0.4pt,line join=round,line cap=round] ( 21.68,188.70) -- ( 15.68,188.70);

\node[text=drawColor,rotate= 90.00,anchor=base,inner sep=0pt, outer sep=0pt, scale=  1.00] at (  9.68, 29.98) {-0.2};

\node[text=drawColor,rotate= 90.00,anchor=base,inner sep=0pt, outer sep=0pt, scale=  1.00] at (  9.68, 56.43) {0.0};

\node[text=drawColor,rotate= 90.00,anchor=base,inner sep=0pt, outer sep=0pt, scale=  1.00] at (  9.68, 82.89) {0.2};

\node[text=drawColor,rotate= 90.00,anchor=base,inner sep=0pt, outer sep=0pt, scale=  1.00] at (  9.68,109.34) {0.4};

\node[text=drawColor,rotate= 90.00,anchor=base,inner sep=0pt, outer sep=0pt, scale=  1.00] at (  9.68,135.80) {0.6};

\node[text=drawColor,rotate= 90.00,anchor=base,inner sep=0pt, outer sep=0pt, scale=  1.00] at (  9.68,162.25) {0.8};

\node[text=drawColor,rotate= 90.00,anchor=base,inner sep=0pt, outer sep=0pt, scale=  1.00] at (  9.68,188.70) {1.0};

\path[draw=drawColor,line width= 0.6pt,line join=round,line cap=round] ( 21.68, 21.68) --
	(195.13, 21.68) --
	(195.13,195.13) --
	( 21.68,195.13) --
	( 21.68, 21.68);
\end{scope}
\begin{scope}
\path[clip] (  0.00,  0.00) rectangle (195.13,227.65);
\definecolor{drawColor}{RGB}{0,0,0}

\node[text=drawColor,anchor=base,inner sep=0pt, outer sep=0pt, scale=  1.00] at (108.41, 30.08) {lag};
\end{scope}
\begin{scope}
\path[clip] ( 21.68, 21.68) rectangle (195.13,195.13);
\definecolor{drawColor}{RGB}{0,0,0}

\path[draw=drawColor,line width= 0.6pt,line join=round,line cap=round] ( 21.68, 56.43) -- (195.13, 56.43);
\definecolor{drawColor}{RGB}{0,0,255}

\path[draw=drawColor,line width= 0.6pt,dash pattern=on 4pt off 4pt ,line join=round,line cap=round] ( 21.68, 84.76) -- (195.13, 84.76);

\path[draw=drawColor,line width= 0.6pt,dash pattern=on 4pt off 4pt ,line join=round,line cap=round] ( 21.68, 28.11) -- (195.13, 28.11);
\end{scope}
\begin{scope}
\path[clip] (  0.00,  0.00) rectangle (195.13,227.65);
\definecolor{drawColor}{RGB}{0,0,0}

\node[text=drawColor,anchor=base,inner sep=0pt, outer sep=0pt, scale=  1.50] at (108.41,207.13) {\bfseries (c) ACF of $\protect\boldmath{(\text{PIT} - \frac12)^2}$};
\end{scope}
\end{tikzpicture}} \\ \smallskip
	\scalebox{0.65}{\input{figs/tikz/relDiag_marg_BoE1.tex}}
	\scalebox{0.65}{\input{figs/tikz/relDiag_thresh_BoE1.tex}}
	\scalebox{0.65}{\input{figs/tikz/relDiag_quant_BoE1.tex}} 
	\caption{Calibration diagnostics for Bank of England forecasts of CPI
		inflation at a prediction horizon of one quarter: (a) PIT
		reliability diagram, along with the empirical autocorrelation
		functions of (b) original and (c) squared, centered PIT values, (d)
		marginal, (e) threshold, and (f) 75\%-quantile reliability diagram.
		If applicable, we show 90\% consistency bands and CORP score
		components under the associated canonical loss function,
		namely, the Brier score (BS) and the piecewise linear quantile
		score (QS), respectively.  \label{fig:1}}
	\scalebox{0.65}{\input{figs/tikz/thresh_score_decomp_data.tex}}
	\scalebox{0.65}{\input{figs/tikz/quant_score_decomp_data.tex}}
	\caption{Score decomposition \eqref{eq:dcm_emp} respectively
		\eqref{eq:dcm_emp_ext} and skill score \eqref{eq:skill} for probability
		forecasts of not exceeding the 2\% inflation target (left) and
		75\%-quantile forecasts (right) induced by Bank of England fancharts
		for CPI inflation, under the associated canonical scoring
		function.  \label{fig:BoE_dcm}}
\end{figure}

To illustrate this process, we consider quarterly Bank of England
forecasts of consumer price index (CPI) inflation rates, as issued
since 2004.  The forecast distributions, for which we give details and
refer to extant analyses in Appendix \ref{sec:BoE}, are
two-piece normal distributions that are communicated to the public via
fancharts.  The forecasts are at prediction horizons up to six
quarters ahead in the time series setting, where $k$ step ahead
forecasts that are ideal with respect to the canonical filtration show
PIT values that are independent at lags $\geq k + 1$ in addition to
being uniformly distributed \citep{Diebold1998}. However, as discussed
in Appendix \ref{sec:iidPITs}, independent, uniformly
distributed PIT values do not imply auto-calibration, except in a
special case.  Thus, calibration diagnostics beyond checks of the
uniformity and independence of the PIT are warranted.

In Figure \ref{fig:1}, we consider forecasts one quarter ahead and show
PIT and marginal reliability diagrams, along with empirical
autocorrelation functions (ACFs) for the first two moments of the PIT.
In part, the PIT reliability diagram and the ACFs lie outside the
respective 90\% consistency bands.  For a closer look, we also plot
the threshold reliability diagram at the policy target of 2\% and the
lower $\alpha$-quantile reliability diagram for $\alpha = 0.75$.  The
deviations from reliability remain minor, in stark contrast to
calibration diagnostics at prediction horizons $k \geq 4$, for which
we refer to Appendix \ref{sec:BoE}.

Figure \ref{fig:BoE_dcm} shows the standard CORP decomposition
\eqref{eq:dcm_emp} of the Brier score (BS) for the induced probability
forecasts at the 2\% target and the extended CORP decomposition
\eqref{eq:dcm_emp_ext} of the piecewise linear quantile score for
$\alpha$-quantile forecasts at level $\alpha = 0.75$ and lead times up
to six quarters ahead.  In the latter case, the difference between
$\MCB$ and $\MCB_\uncond$ equals the $\MCB_\cond$ component.
Generally, the miscalibration components increase while the
discrimination components decrease with the lead time.  Related
results for the quantile functional can be found in \citet[Table 5,
Figures 7 and 8]{Pohle2020}, where there is a notable increase in
the discrimination (resolution) component at the largest two lead
times, which is caused by counterintuitive decays in the recalibration
functions.  In contrast, the regularizing constraint of isotonicity
prevents overfitting in the CORP approach.

The coefficient of determination or skill score $\myR^*$
decays with the prediction horizon and becomes negative at lead
times $k \geq 4$.  This suggests that forecasts remain informative
at lead times up to at most three quarters ahead, in line with the
substantive findings in \citet{Pohle2020} and other extant work, as
hinted at in Appendix \ref{sec:BoE}.

\section{Discussion}  \label{sec:discussion} 

We have developed a comprehensive theoretical and methodological
framework for the analysis of calibration and reliability, serving the
purposes of both (out-of-sample) forecast evaluation and (in-sample)
model diagnostics.  A common principle is that fitted or predicted
distributions ought to be calibrated or reliable, ideally in the sense
of auto-calibration, which stipulates that the outcomes are random
draws from the posited distributions.  For general real-valued
outcomes, we have seen that auto-calibration is stronger than
both classical unconditional and recently proposed
conditional notions of calibration.  We have developed
hierarchies of calibration in the spirit of
\citet{VanCalster2016}, as highlighted in Figure
\ref{fig:hierarchy}, and proposed a generic notion of
conditional calibration in terms of statistical
functionals.  Specifically, a posited distribution is
conditionally $\myT$-calibrated if the induced point forecast for the
functional $\myT$ can be taken at face value.  This concept continues
to apply when stand-alone point forecasts or regression output in
terms of the functional $\myT$ are to be evaluated, and can be
assessed via $\myT$-reliability diagrams and associated score
decompositions. Importantly, our tools apply regardless of how
forecasts are generated, be it through the use of
traditional statistical regression models, modern machine
learning techniques, or even subjective human judgment.

We have adopted and generalized the nonparametric approach of
\citet{Dimitriadis2020a}, who obtained consistent, optimally binned,
reproducible and PAV based (CORP) estimators of $\myT$-reliability
diagrams and score components in the case of probability forecasts
for binary outcomes.  While our tools apply in the much broader
setting of identifiable functionals and real-valued outcomes, the
arguments put forth by \citet{Dimitriadis2020a} continue to apply,
in that CORP estimators are bound to, simultaneously, improve
statistical efficiency, reproducibility \citep{Stodden2016}, and
stability \citep{Yu2020}.  In a nutshell, the CORP approach is
flexible, due to its use of nonparametric regression for
recalibration, and yet it avoids overfitting, owing to the
regularizing constraint of isotonicity.  Notably, the CORP score
decomposition yields a new, universal coefficient of determination,
$\myR^*$, that nests and generalizes the classical $\myR^2$ in
ordinary least squares (mean) regression, and its cousin $\myR^1$ in
quantile regression.  In independent work, \cite{Allen2022} also observes the link between skill scores, score decompositions and the coefficient of determination.
We have illustrated the CORP approach on Bank of
England forecasts of inflation, along with a brief ecological
example.  \cite{Gneiting2022} provide an in depth review of
the particular case of forecasts in the form of (one or multiple)
quantiles, accompanied by case studies.  Code in \textsf{R}
\citep{R} for reproducing our results is available at
\url{https://github.com/resinj/replication\_GR21}.

Follow-up work on the CORP approach for specific functionals $\myT$ is
essential, including but not limited to the ubiquitous cases of
quantiles and the mean functional, where the newly developed tools can
supplement classical approaches to regression diagnostics, as hinted
at in the ecological example.  In particular, we have applied a crude,
all-purpose, residual-based permutation approach to generate
consistency bands for $\myT$-reliability diagrams under the hypothesis
of $\myT$-calibration.  Clearly, this approach can be refined, and we
anticipate vigorous work on consistency and confidence bands, based on
either resampling or large sample theory, akin to the developments in
\citet{Dimitriadis2020a} for probability forecasts of binary outcomes.
Similarly, CORP estimates of miscalibration components under canonical
loss functions are natural candidates for the quantification of
calibration error in empirical work.  Reliability and discrimination
ability are complementary attributes of point forecasts and regression
output, and discrimination can be assessed quantitatively via
the respective score component.  When many forecasts are to
be compared with each other, scatter plots of CORP
miscalibration (\MCB) and discrimination (\DSC) components admit
succinct visual displays of predictive performance. In
this type of display, forecasts with the same score or,
equivalently, identical coefficient of determination, $\myR^*$,
gather on lines with unit slope, and multiple facets of forecast quality
can be assessed simultaneously, for a general alternative to the
widely used \citet{Taylor2001} diagram.

Formal tests of hypotheses of calibration are critical in both
specific applications, such as banking regulation
\citep[e.g.,][]{Nolde2017}, and in generic tasks, such as the
assessment of goodness of fit in regression \citep[Section
S2]{Dimitriadis2020a}.  In Appendix \ref{sec:tests}, we comment
on this problem from the perspective of the theoretical and
methodological advances presented here.  While specific developments
need to be deferred to future work, it is our belief that the progress
in our understanding of notions and hierarchies of calibration, paired
with the CORP approach to estimating reliability diagrams and score
components, can spur a wealth of new and fruitful developments in
these directions.

\section*{Acknowledgments}

We thank Sebastian Arnold, Fadoua
Balabdaoui--Mohr, Jonas Brehmer, Frank Diebold, Timo
Dimitriadis, Uwe Ehret, Andreas Fink, Tobias Fissler, Rafael
Frongillo, Alexander Henzi, Alexander I.\ Jordan, Kristof Kraus,
Fabian Kr\"uger, Sebastian Lerch, Michael Maier-Gerber, Anja
M\"uhlemann, Jim Pitman, Marc-Oliver Pohle, Roopesh Ranjan, Benedikt
Schulz, Ville Satop\"a\"a, Daniel Wolffram and Johanna F.\
Ziegel, as well as anonymous reviewers, for helpful comments
and discussion. Our research has been funded by the
Klaus Tschira Foundation.

\bibliography{manuscript_v3}
\bibliographystyle{abbrvnat}

\newpage

\begin{appendices}

\section{Supporting calculations for Section \ref{sec:population}}  \label{sec:examples} 

Here, we provide supporting computations and discussion for Examples
\ref{ex:unf.lop} and \ref{ex:p-u.tern}, Definitions
\ref{def:uncondT} and \ref{def:canonical}, Figures
\ref{fig:reldiag} and \ref{fig:dcm}, and Table \ref{tab:dcm}, along
with a discussion of the relation between probabilistic calibration
and unconditional quantile calibration, and a counterexample hinted at
in the main text.  For subsequent use, the first three (non-centered)
moments of the normal distribution $\cN(\mu,\sigma^2)$ are $\mu$,
$\mu^2 + \sigma^2$, and $\mu^3 + 3\mu\sigma^2$.  As in the main text,
we let $\varphi$ and $\Phi$ denote the density and the cumulative
distribution function (CDF), respectively, of a standard normal
variable.

\subsection{Unfocused forecast}  \label{sec:unf} 

For fixed $a, b \in \real$, the function 
\[ 
y \mapsto \Phi_a(y-b) = \frac{1}{2} \left( \Phi(y-b) + \Phi(y-a-b) \right)
\] 
is a CDF.  The random CDF \eqref{eq:unf} for the unfocused forecast in
Example \ref{ex:unf.lop} can be written as $F(y) = \Phi_\eta(y-\mu)$,
where $\eta$ and $\mu$ are independent random variables and $\eta =
\pm \eta_0$ for some constant $\eta_0 > 0$.  Then the conditional CDF
for the outcome $Y$ given the posited (non) exceedance probability
$F(t)$ at any fixed threshold $t \in \real$ or, equivalently, given
the quantile forecast $F^{-1}(\alpha)$ at any fixed level $\alpha \in
(0,1)$ is
\begin{align*}
\myQ( Y \leq y \mid F(t) = \alpha) 
&= \myQ( Y \leq y \mid F^{-1}(\alpha) = t)
= \myQ( Y \leq y \mid \mu = t - \Phi_\eta^{-1}(\alpha)) \\
& = \frac{1}{\sum_{s = \pm 1} \varphi(t - \Phi_{s\eta_0}^{-1}(\alpha))} \sum_{s = \pm 1}
\varphi(t - \Phi_{s\eta_0}^{-1}(\alpha)) \: \Phi(y - (t - \Phi_{s\eta_0}^{-1}(\alpha))).
\end{align*}
As $F$ is symmetric, conditioning on the mean is the same as
conditioning on the median.  The second moment is $m_2(F) = 1 + \mu^2
+ \mu\eta + \frac{1}{2} \eta^2 \geq 1 + \frac{1}{4} \eta^2$, so that
\[ 
\myQ( Y \leq y \mid m_2(F) = m) 
= \myQ \left( Y \leq y \mid \mu = -\frac{1}{2} \eta \pm \sqrt{m - 1 - \frac{1}{4}\eta^2} \: \right)  
\]
is a mixture of normal distributions.  Similarly, the third moment is
$m_3(F) = \mu^3 + \frac{3}{2} \eta \mu^2 + 3 \left( \frac{1}{2} \eta^2
+ 1 \right) \mu + \frac{1}{2} \eta \left( \eta^2 + 3 \right) =
f(\mu;\eta)$, so that $\myQ( Y \leq y \mid m_3(F) = m) = \myQ( Y \leq
y \mid f(\mu;\eta) = m)$ also is a mixture of normal distributions.
To compute the roots of the mapping $x \mapsto f(x;\eta)$, we fix
$\eta$ at $\pm \eta_0$ and use a numeric solver (\texttt{polyroot} in
\textsf{R}).

As regards the score decomposition \eqref{eq:dcm} with $\myS(x,y) =
(x-y)^2$ for the implied mean-forecast, $m_1(F) = \mu +
\frac{1}{2}\eta$, the expected score of the recalibrated mean-forecast
is
\begin{align*}
\bar\myS_\rc 
& = \myE\left[\frac{\sum_{s = \pm 1} \varphi \left( m_1(F) + \frac{s}{2} \eta_0 \right)
	\left( m_1(F) + \frac{s}{2} \eta_0 \right)}{\sum_{s = \pm 1} \varphi \left( m_1(F) + \frac{s}{2} \eta_0 \right)} 
- Y \right]^2 \\
& = \myE \left[ \frac{\sum_{s = \pm 1} \varphi \! \left( \mu + \frac{1}{2} \eta + \frac{s}{2} \eta_0 \right)
	\left(\frac{1}{2} \eta + \frac{s}{2} \eta_0\right)}{\sum_{s = \pm 1} \varphi \left( \mu + \frac{1}{2} \eta + \frac{s}{2} \eta_0 \right)} 
- (Y - \mu) \right]^2 \\
& = \eta_0^2 \,\myE \left[ \frac{ \varphi(\mu + \eta)}{\varphi(\mu) + \varphi(\mu + \eta)} \right]^2
+ \: \myE \hsp [Y - \mu]^2 \\
& = \eta_0^2 \, \myE \! \left[ \Psi_{\eta_0}^2(\mu) \right] + 1, \rule{0mm}{5mm}
\end{align*}
where we define $\Psi_a(x) = \varphi(x + a) / (\varphi(x) + \varphi(x
+ a))$ for $a \in \real$ and note that $\myE\hsp[\Psi_\eta^2(\mu)\mid
\eta] = \myE \hsp [\Psi_{\eta_0}^2(\mu)]$.  The associated integral
\[
I(\eta_0) = \myE \! \left[ \Psi_{\eta_0}^2(\mu) \right]  
= \int_{-\infty}^\infty \left( \frac{\varphi(x + \eta_0)}{\varphi(x) + \varphi(x + \eta_0)} \right)^2 \varphi(x) \dd x
\]
needs to be evaluated numerically.

\subsection{Lopsided forecast}  \label{sec:lop}

We proceed in analogy to the development for the unfocused forecast.
For fixed $a \in [0,1]$ and $b \in \real$, the function
\[
y \mapsto \Phi_a(y-b) 
= (1-a) \Phi(y-b) \one \{ y \leq b \} + \left( (1+a)\Phi(y-b) - a \right) \one \{ y > b \}
\]
is a CDF.  The CDF for the lopsided forecast with random density
\eqref{eq:lop} from Example \ref{ex:unf.lop} can be written as
$F(y) = \Phi_\delta(y-\mu)$, where $\delta$ and $\mu$ are independent
random variables and $\delta = \pm \delta_0$ for some
$\delta_0 \in (0,1)$.  As
$\myE[\Phi_\delta(y-\mu) \mid \mu] = \Phi(y-\mu)$, the
lopsided forecast is marginally calibrated.  It fails to be
probabilistically calibrated, since $Z_F = \Phi_\delta(Y-\mu)$
has CDF
\[
\myQ(Z_F \leq u)
= \frac{1}{2} \sum_{s = \pm 1} \left(
\frac{u}{1-s\delta_0} \one \left\{ \frac{u}{1-s\delta_0} \leq \frac{1}{2} \right\} +
\frac{u+s\delta_0}{1+s\delta_0} \one \left\{ \frac{u}{1-s\delta_0} > \frac{1}{2} \right\} \right)
\]
for $u \in (0,1)$ by the law of total probability.

The conditional CDF for the outcome $Y$ given the posited (non)
exceedance probability $F(t)$ at any fixed threshold $t \in \real$ or,
equivalently, given the quantile forecast $F^{-1}(\alpha)$ at any
fixed level $\alpha \in (0,1)$ is
\begin{align*}
\myQ(Y\leq y \mid F(t) = \alpha)
&= \myQ(Y\leq y \mid F^{-1}(\alpha) = t)
= \myQ(Y\leq y \mid \mu = t - \Phi_\delta^{-1}(\alpha)) \\
&= \frac{1}{\sum_{s = \pm 1} \varphi(t - \Phi_{s \delta_0}^{-1}(\alpha))} 
\sum_{s = \pm 1} \varphi(t - \Phi_{s\delta_0}^{-1}(\alpha)) \, \Phi(y - (t - \Phi_{s\delta_0}^{-1}(\alpha))), 
\end{align*}
where $\Phi_a^{-1}(\alpha) = \Phi^{-1}(\alpha/(1-a))$ if $\alpha \leq
\frac{1}{2} (1-a)$ and $\Phi_a^{-1}(\alpha) = \Phi^{-1}((a +
\alpha)/(a + 1))$ otherwise.

As $F$ is a mixture of truncated normal distributions, its moments are
mixtures of the component moments, for which we refer to
\cite{Orjebin2014}.  The first moment is $m_1(F) = \mu +
2\delta\varphi(0)$, so that
\begin{align*}
\myQ(Y \leq y \mid m_1(F) = m) 
&= \myQ(Y \leq y \mid \mu = m - 2\delta\varphi(0)) \\
& = \frac{1}{\sum_{s = \pm 1} \varphi(m - 2s\delta_0\varphi(0))} 
\sum_{s = \pm 1} \varphi(m - 2s\delta_0\varphi(0)) \, \Phi(y - (m - 2s\delta_0\varphi(0)))
\end{align*}
is a mixture of normal distributions.  Similarly, the second and third
moments are $m_2(F) = \mu^2 + 1 + 4\delta\varphi(0)\mu \geq 1 -
4\delta^2\varphi(0)^2$ and $m_3(F) = \mu^3 + 3\mu +
2\delta\varphi(0)(3\mu^2 + 2) = f(\mu;\delta)$, respectively, so that
\begin{align*}
\myQ(Y \leq y \mid m_2(F) = m) & = \myQ(Y \leq y\mid \mu = -2\delta\varphi(0) \pm \sqrt{4\delta^2\varphi(0)^2 - 1 + m}), \\
\myQ(Y \leq y \mid m_3(F) = m) & = \myQ(Y \leq y\mid f(\mu;\delta) = m)
\end{align*}
also admit expressions as mixtures of normal distributions.  Again, we
use a numeric solver to find the roots of $x \mapsto f(x;
\pm \delta_0)$.

As the implied mean-forecast, $m_1(F) = \mu + 2\delta\varphi(0)$,
agrees with the implied mean-forecast of the unfocused forecast with
$\eta = (8/\pi)^{1/2} \delta$, the terms in the score decomposition
\eqref{eq:dcm} with $\myS(x,y) = (x-y)^2$ derive from the respective
terms in the score decomposition for the unfocused forecast, as
illustrated in Figure \ref{fig:dcm}.

\subsection{Piecewise uniform forecast}  \label{sec:p-u} 

Given any fixed index $i \in \{ 1, 2, 3 \}$, let the tuple
$(p_1^{(i)},p_2^{(i)},p_3^{(i)};q_1^{(i)},q_2^{(i)},q_3^{(i)})$ attain
the value $(\frac{1}{2}, \frac{1}{4}, \frac{1}{4}; \frac{5}{10},
\frac{1}{10}, \frac{4}{10})$ if $i = 1$, the value $(\frac{1}{4},
\frac{1}{2}, \frac{1}{4}; \frac{1}{10}, \frac{8}{10}, \frac{1}{10})$
if $i = 2$, and the value $(\frac{1}{4}, \frac{1}{4}, \frac{1}{2};
\frac{4}{10}, \frac{1}{10}, \frac{5}{10})$ if $i = 3$.  Furthermore,
let $P_i$ be the CDF of a mixture of uniform measures on $[0,1],
[1,2]$, and $[2,3]$ with weights $p_1^{(i)}, p_2^{(i)}$, and
$p_3^{(i)}$, respectively.  Similarly, let $Q_i$ be the CDF of the
respective mixture with weights $q_1^{(i)},q_2^{(i)}$, and
$q_3^{(i)}$, respectively.

The random CDF for the piecewise uniform forecast in Example
\ref{ex:p-u.tern} can then be written as $F(x) = P_\iota(x-\mu)$,
where the random variables $\iota$ and $\mu$ are independent, and the
integer-valued variable $\iota$ is such that
\[
\left( p_1,p_2,p_3; q_1,q_2,q_3 \right) = 
\left( p_1^{(\iota)},p_2^{(\iota)},p_3^{(\iota)}; q_1^{(\iota)},q_2^{(\iota)},q_3^{(\iota)} \right) \! .
\]

The conditional CDF for the outcome $Y$ given the posited (non)
exceedance probability $F(t)$ at any fixed threshold $t \in \real$ or,
equivalently, given the quantile forecast $F^{-1}(\alpha)$ at any
fixed level $\alpha \in (0,1)$ then is
\begin{align*}
\myQ(Y \leq y \mid F(t) = \alpha) 
&= \myQ(Y \leq y \mid F^{-1}(\alpha) = t)
= \myQ(Y \leq y \mid \mu = t - P_\iota^{-1}(\alpha)) \\
& = \frac{1}{\sum_{i = 1, 2, 3} \varphi \! \left( \frac{t - P_i^{-1}(\alpha)}{c} \right)} 
\sum_{i = 1, 2, 3} \varphi \! \left( \frac{t - P_i^{-1}(\alpha)}{c} \right) Q_i(y - (t - P_i^{-1}(\alpha))),
\end{align*}
where $c$ is the standard deviation of $\mu$, as defined in Example
\ref{ex:p-u.tern}.  The first moment of $F$ is $m_1(F) = \mu + 1 +
\frac{1}{4} \iota$, so that
\begin{align*}
\myQ(Y \leq y \mid m_1(F) = m) 
&= \myQ( {\textstyle Y \leq y \mid \mu = m - 1 - \frac{1}{4} \iota} ) \\
& = \frac{1}{\sum_{i = 1, 2, 3} \varphi \! \left( \frac{m - 1 - \frac{1}{4} i}c \right)} 
\sum_{i = 1, 2, 3} \varphi \! \left( \frac{m - 1 - \frac{1}{4} i}c \right) Q_i({\textstyle y - (m - 1 - \frac{1}{4} i}))
\end{align*}
is a mixture of shifted versions of $Q_1$, $Q_2$, and $Q_3$.  The
associated first moment is the respective mixture of $m +
\frac{3}{20}$, $m$, and $m - \frac{3}{20}$.

Given any integer $k \geq 0$, let $\beta_k = \sum_{j = 1, 2, 3} \left(
j^{k+1} - (j-1)^{k+1} \right) p_j^{(\iota)}$.  The second moment of
$F$ is $m_2(F) = \mu^2 + \beta_1 \mu + \frac{1}{3} \beta_2$, whence
\[
\myQ(Y \leq y \mid m_2(F) = m) 
= \myQ \left( Y \leq y \mid \mu = - \frac{1}{2} \beta_1 \pm \sqrt{\frac{1}{4} \beta_1^2 - \frac{1}{3} \beta_2 + m} \, \right)
\]
also admits an expression in terms of mixtures of shifted versions of
$Q_1$, $Q_2$, and $Q_3$.  Finally, the third moment of $F$ is
$m_3(F) = \mu^3 + \frac{3}{2} \beta_1 \mu^2 + \beta_2 \mu +
\frac{1}{4} \beta_3 = f(\mu;\iota)$, so that the conditional
distribution
$\myQ(Y \leq y \mid m_3(F) = m) = \myQ(Y \leq y \mid f(\mu;\iota) =
m)$ and the associated third moment can be computed analogously.

\subsection{Identification functions, unconditional calibration, and canonical loss}
\label{sec:identification}

In this section, we demonstrate that Definitions \ref{def:uncondT} and
\ref{def:canonical} are unambiguous and do not depend on the choice of
the identification function, which is essentially unique.  To this
end, we first contrast the notions of identification functions in
\cite{Fissler2016} and \cite{Jordan2019}.
\cite{Fissler2016} call $V \colon \real \times \real \to \real$ a
(strict $\cF$-)identification function if $V(x, \cdot)$ is integrable
with respect to all $F \in \cF$ for all $x \in \real$.
\cite{Jordan2019} additionally require $V$ to be increasing and
left-continuous in its first argument.  Furthermore, there is a subtle
difference in the way that the functional is induced.  While
\cite{Fissler2016} define the induced functional as the set
\[
\myT_0(F) = \left\{ x \in \real : \int V(x,y) \dd F(y) = 0 \right\},
\]
\cite{Jordan2019} define it to be the closed interval
$T(F) = [T^-(F),\allowbreak T^+(F)]$, where $\myT^-(F)$ and
$\myT^+(F)$ are defined in \eqref{eq:T}.  The approach by \cite{Jordan2019} allows for
quantiles to be treated in full generality and ensures that the
interval $\myT(F)$ coincides with the closure of $\myT_0(F)$ if the
latter is nonempty.

In the setting of \cite{Fissler2016}, if $V$ is an identification
function, then so is $(x,y) \mapsto h(x) V(x,y)$ whenever
$h(x) \neq 0$ for all $x \in \real$.  If the class $\cF$ is
sufficiently rich, then any two locally bounded identification
functions $V$ and $\tilde{V}$ that induce a functional $\myT_0$ of
singleton type relate to each other in the stated form almost
everywhere on the interior of $\myT_0(\cF) \times \real$
\citep[][Theorem 4]{Dimitriadis2021}, which implies that
increasing identification functions of prediction error form are
unique up to a positive constant.  The following proposition provides
an elementary proof under slightly different conditions that are
tailored to our setting.  Notably, identification functions of
prediction error form induce functionals that are equivariant under
translation by Proposition 4.7 of \citet{Fissler2019}, a result which
can easily be transferred to the setting of \citet{Jordan2019}.

\begin{proposition}
	Let $\cF$ be a convex class of probability measures such that
	$\delta_y \in \cF$ for all $y \in \real$.  If the functional\/
	$\myT$ is induced on $\cF$ by an identification function
	$V(x,y) = v(x-y)$ of prediction error form, where $v$ is increasing
	and left-continuous with $v(-r) < 0$ and $v(r) > 0$ for some
	$r > 0$, then any other identification function of the stated form
	that induces $\myT$ is a positive multiple of\/ $V$.
\end{proposition}

\begin{proof}
	Let $V \colon (x,y) \mapsto v(x-y)$ and
	$\tilde{V} \colon(x,y) \mapsto \tilde{v}(x-y)$ induce the
	functionals $\myT$ and $\tilde{\myT}$, respectively.  We proceed to
	show that $\myT = \tilde{\myT}$ implies $\tilde v = h_0 \cdot v$ for
	some constant $h_0 > 0$.
	
	To this end, suppose that $\myT = \tilde{\myT}$, and let
	\begin{align*}
	r^- & = \sup \{ r : \tilde{v}(r) < 0 \}
	= \tilde{\myT}{}^-(\delta_0)
	= \myT{}^-(\delta_0)
	= \sup \{ r : v(r) < 0 \}
	> - \infty, \\
	r^+ & = \inf \{ r : \tilde{v}(r) > 0 \}
	= \tilde{\myT}{}^+(\delta_0)
	= \myT{}^+(\delta_0)
	= \inf \{ r : v(r) > 0 \}
	< \infty.
	\end{align*}
	By left-continuity and monotonicity of $v$ and $\tilde{v}$, it
	follows that $v(r) = \tilde{v}(r) = 0$ for $r \in (r^-,r^+]$,
	$v(r) < 0$ and $\tilde{v}(r) < 0$ for $r < r^-$, and
	$v(r) > 0$ and $\tilde{v}(r) > 0$ for $r > r^+$.
	
	Let $h(r) = \tilde{v}(r)/v(r) > 0$ for
	$r \in \real \setminus [r^-,r^+]$.  If $r < r^- \leq r^+ < s$, then
	$\tilde{v}(r) = h(r) v(r) < 0$ and $\tilde{v}(s) = h(s) v(s) > 0$.
	Assume $h(r) < h(s)$, and let $p \in (0,1)$ be such that
	\[
	\left( 1 - \frac{h(s)v(s)}{h(r)v(r)} \right)^{-1} < \, p
	< \left( 1 - \frac{v(s)}{v(r)} \right)^{-1}. 
	\]
	Then $(1-p) h(r) v(r) + p h(s) v(s) > 0 > (1-p) v(r) + p v(s)$
	and
	$\tilde{\myT}{}^+ (p \delta_{-s} + (1-p) \delta_{-r}) < 0 \leq \myT{}^- (p
	\delta_{-s} + (1-p) \delta_{-r})$, a contradiction.  An analogous
	argument applies if we assume that $h(r) > h(s)$, and
	we conclude that $h(r) = h(s)$.
	
	If $r, s < r^-$, then $h(r) = h(s) = h(t)$ for any $t > r^+$
	by the above line of reasoning.  An analogous argument yields
	$h(r) = h(s)$ for $r, s > r^+$.  Therefore, the function $h$
	is constant and $v(r) = h_0 \cdot \tilde{v}(r)$ for a constant
	$h_0 > 0$ and all $r \in \real \setminus \{ r^- \}$.  Finally,
	we obtain
	$v(r^-) = \lim_{r \uparrow r^-} v(r) = \lim_{r \uparrow r^-}
	h_0 \cdot \tilde{v}(r) = h_0 \cdot \tilde{v}(r^-)$ by
	left-continuity.
\end{proof}

Hence, if we assume an identification function of type (i) in
Assumption \ref{as:V}, Definitions \ref{def:uncondT} and
\ref{def:canonical} do not depend on the choice of the identification
function, as it is unique up to a positive constant.  Trivially, the
same holds true for type (ii).  To complete the argument that the
definitions are unambiguous, the following technical argument is
needed.

\begin{remark}
	If a functional $\myT$ of singleton type is identified by both an
	identification function $V(x,y) = v(x-y)$ of type (i) and an
	identification function $\tilde{V}(x,y) = x - \myT(\delta_y)$ of
	type (ii), then $\tilde{V}$ is also of type (i).  To see this, let
	$z$ denote the unique value at which the sign of $v$ changes,
	and note that $z = \myT(\delta_y) - y$ for all $y$, since $V$
	induces the functional $\myT$ for each Dirac measure $\delta_y$.
	Hence, $\myT(\delta_y) = y + z$ and $\tilde{V}(x,y) = x - y - z$ is
	of type (i).
\end{remark}

We close this section with comments on the role of the class $\cF$.
As expressed by Assumption \ref{as:V}, we prefer to work with
identification functions that elicit the target functional $\myT$ on a
large, convex class $\cF$ of probability measures, to avoid
unnecessary constraints on forecast(er)s.  Furthermore, when
evaluating stand-alone point forecasts, the underlying predictive
distributions typically are implicit, and assumptions other than the
existence of the functional at hand are unwarranted and contradict the
prequential principle.  Evidently, if the class $\cF$ is sufficiently
restricted, additional identification functions arise.  For example,
the piecewise constant identification function associated with the
median can be used to identify the mean within any class of symmetric
distributions.

\subsection{Probabilistic calibration and unconditional quantile calibration}
\label{sec:PCvsQC}

As noted in the main text, probabilistic calibration implies the
unconditional $\alpha$-quantile calibration condition
\eqref{eq:uncondqcal} at every level $\alpha \in (0,1)$.  To see this,
note that if $F$ is probabilistically calibrated, then $\alpha =
\myQ(Z_F \leq \alpha) \leq \myQ(F(Y-) \leq \alpha) = \myQ(Y \leq
q_\alpha^-(F))$ and $1 - \alpha = \myQ(Z_F > \alpha) \leq \myQ(F(Y) >
\alpha) = \myQ(Y \geq q_\alpha^+(F))$.  As Example
\ref{ex:CEP.q.diff}(b) demonstrates, the reverse implication does not
hold in general.  However, Assumption \ref{as:csi} ensures the
equivalence of probabilistic calibration and unconditional
$\alpha$-quantile calibration at every level $\alpha \in (0,1)$.

\subsection{Counterexample (Proposition \ref{prop:relations}(a))}
\label{sec:STCnotAC}

As pointed out by \citet[p.\ 5]{Sahoo2021}, strong threshold calibration does not imply auto-calibration. Here, we provide a simple example illustrating this, as \cite{Sahoo2021} do not present such. 
The example is similar in spirit to the continuous forecast of Example \ref{ex:CEP.q.diff}(a) (as $c \rightarrow 0$), but with strictly increasing distribution functions satisfying Assumption \ref{as:csi}.

Let $F$ be a
mixture of uniform distributions on the intervals
$[0,1],[1,2],[2,3]$, and $[3,4]$ with
weights $p_1, p_2, p_3$, and $p_4$, respectively, and let $Y$ be
from a mixture with weights $q_1, q_2, q_3$, and $q_4$.
Furthermore, let the tuple $(p_1, p_2, p_3, p_4; q_1, q_2, q_3,
q_4)$ attain each of the values
\begin{align*}
&\textstyle
\left( \frac{4}{10}, \frac{1}{10}, \frac{4}{10}, \frac{1}{10}; \frac{16}{25}, \frac{4}{25}, \frac{4}{25}, \frac{1}{25} \right) \! , \quad
\left( \frac{1}{10}, \frac{4}{10}, \frac{1}{10}, \frac{4}{10}; \frac{4}{25}, \frac{16}{25}, \frac{1}{25}, \frac{4}{25} \right) \! , \\ 
&\textstyle
\left( \frac{4}{10}, \frac{1}{10}, \frac{1}{10}, \frac{4}{10}; \frac{4}{25}, \frac{1}{25}, \frac{4}{25}, \frac{16}{25} \right) \! , \quad
\left( \frac{1}{10}, \frac{4}{10}, \frac{4}{10}, \frac{1}{10}; \frac{1}{25}, \frac{4}{25}, \frac{16}{25}, \frac{4}{25} \right)
\end{align*}
with equal probability. The equal average of the distribution of the PIT conditional on either forecast from the top row, and either forecast from the bottom row, is uniform. As any nontrivial conditioning in terms of a threshold yields a combination of two forecast cases, one from the top row and one from the bottom row, the forecast $F$ is strongly threshold calibrated.

\subsection{Remarks on Figure \ref{fig:reldiag}}  \label{sec:fig3}

\begin{figure}[!t]
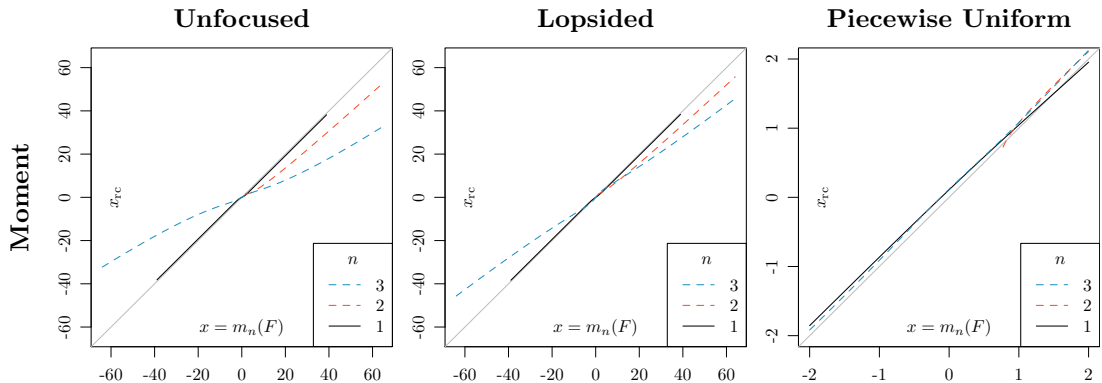
  
	\centering % Please do not change the sizes of the boxes, this will mess up the font size!
	\scalebox{0.65}{\input{figs/tikz/theo_moment_unf_noroot.tex}}
	\scalebox{0.65}{\input{figs/tikz/theo_moment_lop_noroot.tex}}
	\scalebox{0.65}{\input{figs/tikz/theo_moment_pwu_noroot.tex}}
	\caption{Same as the lower row of Figure \ref{fig:reldiag}, but with
		displays on original (rather than root-transformed) scales: Moment
		reliability diagrams for point forecasts induced by (left) the
		unfocused forecast with $\eta_0 = 1.5$ and (middle) the lopsided
		forecast with $\delta_0 = 0.7$ from Example \ref{ex:unf.lop}, and
		(right) the piecewise uniform forecast with $c = 0.5$ from Example
		\ref{ex:p-u.tern}.  \label{fig:reldiag_alt}}
\end{figure}

The root transforms in the moment reliability diagrams in the bottom
row of Figure \ref{fig:reldiag} bring the first, second, and third
moment to the same scale.  The peculiar dent in the reliability curve
for the (third root of the) third moment of the piecewise uniform
forecast results from the transform, which magnifies small deviations
between $x = m_3(F)$ and $x_\rc$ when $x$ is close to zero.  For
comparison, Figure \ref{fig:reldiag_alt} shows moment reliability
diagrams for all three forecast without applying the root transform.

\subsection{Counterexample (Theorem \ref{th:optimal})}  \label{sec:counterexample}

The statement in Theorem \ref{th:optimal} does not hold under
consistent scoring functions in general.  For a counterexample,
consider the empirical distribution of
$(x_1, y_1), \ldots, (x_n, y_n)$, where $x_i = i$ and
$y_i = x_i + \frac{10}{9}$ for $i = 1, \ldots, 9$, and
$x_{10} = y_{10} = - 10$.  The respective mean-forecast $X$ fails to
be unconditionally mean calibrated, whereas the shifted version
$X + 1$ is unconditionally mean calibrated.  Nonetheless, the expected
elementary score \eqref{eq:elementary} for the mean functional (i.e.,
$V(x,y) = x - y$) with index $\eta = - \frac{19}{2}$ increases when
$X$ gets replaced with $X + 1$.

\section{Consistency resamples and calibration tests}  \label{sec:UQ} 

Monte Carlo based consistency bands for $\myT$-reliability diagrams
can be generated from resamples, at any desired nominal level.  The
consistency bands then show the pointwise range of the resampled
calibration curves.  For now, let us assume that we have data
$(x_1,y_1), \ldots, (x_n,y_n)$ of the form \eqref{eq:x.y} along with
$m$ resamples at hand, and defer the critical question of how to
generate the resamples.

\medskip 

\begin{algorithm}[H] 
	\caption{Consistency bands for $\myT$-reliability curves based on resamples}
	\label{alg:bands}  
	\SetAlgoLined 
	\KwIn{resamples $(x_1, \tilde{y}_1^{(j)}), \ldots, (x_n, \tilde{y}_n^{(j)})$ for $j = 1, \ldots, m$} 
	\KwOut{$\alpha \times 100$\% consistency band} 
	\For{$j \in \{ 1, \ldots, m \}$}{apply Algorithm \ref{alg:PAV} to obtain $\hat{x}_1^{(j)}, \ldots, \hat{x}_n^{(j)}$ from 
		$(x_1, \tilde{y}_1^{(j)}), \ldots, (x_n, \tilde{y}_n^{(j)})$} 
	\For{$i \in \{ 1, \ldots, n \}$}{let $l_i$ and $u_i$ be the empirical quantiles of $\hat{x}_i^{(1)}, \ldots, \hat{x}_i^{(m)}$ at level 
		$\frac{\alpha}{2}$ and $1 - \frac{\alpha}{2}$} 
	interpolate the point sets $(x_1, l_1), \ldots, (x_n, l_n)$ and $(x_1,
	u_1), \ldots, (x_n, u_n)$ linearly, to obtain the lower and upper
	bound of the consistency band, respectively
\end{algorithm}

\medskip

Complementary to consistency bands, tests for the assumed type of
calibration, as quantified by the functional $\myT$ and a generic
miscalibration measure $\MCB$, can be performed as usual.
Specifically, we compute $\MCB_j$ for each resample $j = 1, \ldots,
m$, and if $r$ of the resampled measures $\MCB_1, \ldots, \MCB_m$ are
less than or equal to the miscalibration measure computed from the
original data, we declare a Monte Carlo $p$-value of $1 -
\frac{r}{m+1}$.

\subsection{Consistency resamples under the hypothesis of auto-calibration} \label{sec:UQ_auto}

When working with original data of the form \eqref{eq:F.y}, we can
generate resamples under the hypothesis of auto-calibration in the
obvious way, as follows.

\medskip

\begin{algorithm}[H]  
	\caption{Consistency resamples under the hypothesis of auto-calibration}
	\label{alg:auto} 
	\SetAlgoLined
	\KwIn{$(F_1,y_1), \ldots, (F_n,y_n)$}
	\KwOut{resamples $(x_1, \tilde{y}_1^{(j)}), \ldots, (x_n, \tilde{y}_n^{(j)})$ for $j = 1, \ldots, m$} 
	\For{$i \in \{ 1, \ldots, n \}$}{let $x_i = \myT(F_i)$}
	\For{$j \in \{ 1, \ldots, m \}$}{\For{$i = 1, \ldots, n$}{sample $\tilde{y}_i^{(j)}$ from $F_i$}} 
\end{algorithm}

\medskip 

As noted, in the case of threshold calibration, the induced outcome is
binary, whence the assumptions of auto-calibration and
$\myT$-calibration coincide.  For other types of functionals,
auto-calibration is a strictly stronger assumption than
$\myT$-calibration, and it is important to note that the resulting
inferential procedures may be confounded by forecast attributes other
than $\myT$-calibration.  For illustration, let us return to the
setting of Example \ref{ex:perf.clim} and suppose that, conditionally
on a standard normal variate $\mu$, the outcome $Y$ is normal with
mean $\mu$ and variance 1.  Given any fixed $\sigma > 0$, the forecast
$F_\sigma = \cN(\mu,\sigma^2)$ is auto-calibrated if, and only if,
$\sigma = 1$.  However, if $\myT$ is the mean or median functional,
then $F_\sigma$ is $\myT$-calibrated under any $\sigma > 0$.  Clearly,
if we use Algorithm \ref{alg:auto} to generate resamples, then the
consistency bands generated by Algorithm \ref{alg:bands} might be
misleading with regard to the assessment of $\myT$-calibration.  For
example, if $\sigma < 1$ the confidence bands tend to be narrow and
might erroneously suggest a lack of $\myT$-calibration, despite the
forecast being $\myT$-calibrated.

\subsection{Consistency resamples under the hypothesis of\/ $\myT$-calibration} 
\label{sec:UQ_T}

The issues just described call for an alternative to Algorithm
\ref{alg:auto}. Residual-based approaches can be used to generate
resamples under the weaker hypothesis of $\myT$-calibration.  In
developing such a method, we restrict the discussion to single-valued
functionals $\myT$ under which $y_i = \myT(\delta_i)$, which covers
all cases of key interest, such as the mean functional, lower or upper
quantiles, and expectiles.  As is standard in regression diagnostics,
residual-based approaches operate on the basis of tuples
$(x_1,y_1), \ldots, (x_n,y_n)$ of the form \eqref{eq:x.y} under the
assumptions of independence between the point forecast, $x_i$, and the
residual, $y_i - x_i$, and exchangeability of the residuals.  For a
discussion in the context of backtests in banking regulation, see
Example 3 of \cite{Nolde2017}.

Interestingly, Theorem \ref{th:cond=uncond} demonstrates that under
these assumptions a forecast is conditionally $\myT$-calibrated if,
and only if, it is unconditionally $\myT$-calibrated.  Thus, we draw
resamples in a two-stage procedure.  First, we find the constant $c$
from Theorem \ref{th:MCB} such that the empirical distribution of
$(x_1+c,y_1), \ldots, (x_n+c,y_n)$ or, equivalently, $(x_1,y_1-c),
\ldots, (x_n,y_n-c)$, is unconditionally $\myT$-calibrated, and then
we resample from the respective residuals, as follows.

\medskip

\begin{algorithm}[H]
	\caption{Consistency resamples under the joint hypothesis of
		$\myT$-calibration and independence between point forecasts and
		residuals}
	\label{alg:T} 
	\SetAlgoLined
	\KwIn{$(x_1,y_1), \ldots, (x_n,y_n)$}
	\KwOut{resamples $(x_1, \tilde{y}_1^{(j)}), \ldots, (x_n, \tilde{y}_n^{(j)})$ for $j = 1, \ldots, m$} 
	\For{$i = 1, \ldots, n$}{let $r_i = y_i - x_i$}
	find $c$ such that $(x_1+c,y_1), \ldots, (x_n+c,y_n)$ is unconditionally $\myT$-calibrated \\
	\For{$j \in \{ 1, \ldots, m \}$}{
		sample $\tilde{r}_1, \ldots, \tilde{r}_n$ from $\{ r_1, \dots, r_n \}$ with replacement \\
		\For{$i = 1, \ldots, n$}{let $\tilde{y}_i = x_i + \tilde{r}_i - c$}
	}
\end{algorithm}

\medskip

As noted in the main text, the consistency bands for the threshold
reliability diagrams in Figures \ref{fig:reldiag_emp} and \ref{fig:1}
have been generated by Algorithms \ref{alg:bands} and \ref{alg:auto}.
This is nearly the same as the Monte Carlo technique proposed by
\cite{Dimitriadis2020a} that applies in the case of (induced)
binary outcomes (only).  However, unlike \cite{Dimitriadis2020a},
we do not resample the forecasts themselves.  To generate consistency
bands for the mean and quantile reliability diagrams in these figures,
we apply Algorithm \ref{alg:bands} to $m = 1000$ resamples generated
by Algorithm \ref{alg:auto}.  Evidently, this is a crude procedure and
relies on classical assumptions.  Nonetheless, we believe that in many
practical settings, where visual tools for diagnostic checks of
calibration are sought, the consistency bands thus generated provide
useful guidance.

Further methodological development on consistency and confidence bands
needs to be tailored to the specific functional $\myT$ of interest,
and follow-up work on Monte Carlo techniques and large sample theory
is strongly encouraged.  Extant asymptotic theory for nonparametric
isotonic regression, as implemented by Algorithm \ref{alg:PAV}, is
available for quantiles and the mean or expectation functional, as
developed and reviewed by \cite{Barlow1972},
\cite{Casady1976}, \cite{Wright1984},
\cite{Robertson1988}, \cite{ElBarmi2005}, and
\cite{Mosching2020}, and can be leveraged, though with hurdles, as
rates of convergence depend on distributional assumptions and limit
distributions involve nuisance parameters that need to be estimated,
whereas the use of bootstrap methods might be impacted by the issues
described by \cite{Sen2010}.

\subsection{Reliability diagrams and consistency bands for probabilistic and marginal calibration}  
\label{sec:UQ_uncond} 

For the classical notions of unconditional calibration in Section
\ref{sec:unconditional}, the CORP approach does not apply directly,
but its spirit can be retained and adapted.

As for probabilistic calibration, the prevalent practice is to plot
histograms of empirical probability integral transform (PIT) values,
as proposed by \cite{Diebold1998}, \cite{Gneiting2007a} and
\cite{Czado2009}, though this is hindered by the necessity for
binning, as analyzed by \cite{Heinrich2020} in the nearly
equivalent setting of rank histograms.  The population version of our
suggested alternative is the \emph{PIT reliability diagram}, which is
simply the graph of the CDF of the PIT $Z_F$ in \eqref{eq:PIT}.  The
PIT reliability diagram coincides with the diagonal in the unit square
if, and only if, $F$ is probabilistically calibrated.  For tuples of
the form \eqref{eq:F.y} the \emph{empirical PIT reliability diagram}\/
shows the empirical CDF of the (potentially randomized) PIT values.
This approach does not require binning and can be interpreted in much
the same way as a PIT diagram: An inverse S-shape corresponds to a
U-shape in histograms and indicates underdispersion of the forecast,
as typically encountered in practice.  Evidently, this is not a new
idea and extant implementations can be found in work by
\cite{Pinson2012} and \cite{Henzi2019}.

As regards marginal calibration, we define the population version of
the \emph{marginal reliability diagram}\/ as the point set
\[
\{ (\myE_\myQ [F(y)], \myQ(Y \leq y)) \in [0,1]^2 : y \in \real \}.   
\]
The marginal reliability diagram is concentrated on the diagonal in
the unit square if, and only if, $F$ is marginally calibrated.  For
tuples of the form \eqref{eq:F.y} the \emph{empirical marginal
	reliability diagram}\/ is a plot of the empirical non exceedance
probability (NEP) $\hat{F}_0(y) = \frac{1}{n} \sum_{i=1}^n \one \{ y
\geq y_i \}$ against the average forecast NEP $\bar{F}(y) =
\frac{1}{n} \sum_{i=1}^n F_i(y)$ at the unique values $y$ of the
outcomes $y_1, \ldots, y_n$, and interpolated linearly in between.  Of
course, this is not a new idea either and can be interpreted as a P-P
plot.

\begin{figure}[!t]
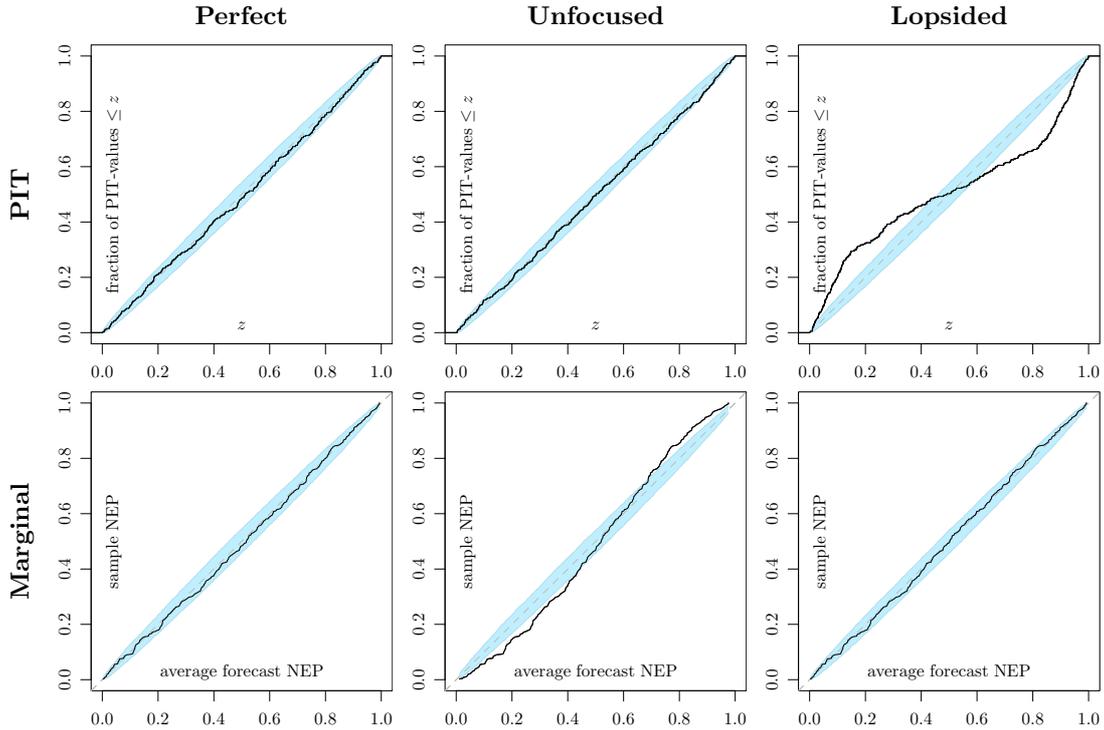
  
	\centering % Please do not change the sizes of the boxes, this will mess up the font size!
	\scalebox{0.65}{\input{figs/tikz/relDiag_PIT_perf400.tex}} 
	\scalebox{0.65}{\input{figs/tikz/relDiag_PIT_unf400.tex}}
	\scalebox{0.65}{\input{figs/tikz/relDiag_PIT_lop400.tex}} \\ \smallskip
	\scalebox{0.65}{\input{figs/tikz/relDiag_marg_perf400.tex}}
	\scalebox{0.65}{\input{figs/tikz/relDiag_marg_unf400.tex}}
	\scalebox{0.65}{\input{figs/tikz/relDiag_marg_lop400.tex}}
	\caption{PIT (top) and marginal (bottom) reliability diagrams for the
		perfect (left), unfocused (middle), and lopsided (right) forecast
		from Examples \ref{ex:perf.clim} and \ref{ex:unf.lop}, along with
		90\% consistency bands based on samples of size
		400.  \label{fig:sim_uncond}}
\end{figure}

For marginal calibration diagrams, we obtain consistency bands under
the assumption of marginal calibration by drawing resamples
$y_1^{(j)}, \ldots, y_n^{(j)}$ from $\bar{F} = \frac{1}{n}
\sum_{i=1}^n F_i$, computing the respective marginal reliability
curve, and repeating over Monte Carlo replicates $j = 1, \ldots, m$.
Then we find consistency bands in the spirit of Algorithm
\ref{alg:bands}.  For PIT reliability diagrams, a trivial technique
applies, as we may obtain consistency bands under the assumption of
probabilistic calibration by (re)sampling $n$ independent standard
uniform variates, computing the respective empirical CDF, and
repeating over Monte Carlo replicates.  Evidently, there are
alternatives based on empirical process theory
\citep{Shorack2009}.

Figure \ref{fig:sim_uncond} illustrates PIT and marginal reliability
diagrams on our customary examples, along with 90\% consistency bands
based on $m = 1000$ Monte Carlo replicates.

\subsection{Testing hypotheses of calibration}  \label{sec:tests}

While the explicit development of calibration tests exceeds the scope
of our paper, we believe that the results and discussion in Section
\ref{sec:population} convey an important general message: It is
critical that the assessed notion of calibration be carefully and
explicitly specified.  Throughout, we consider tests under the
assumption of independent, identically distributed data from a
population.  For extensions to dependent samples, we refer to
\cite{Strahl2017}, who generalized the prediction space concept to
allow for serial dependence, and point at methods introduced by,
e.g., \cite{Corradi2007}, \cite{Knueppel2015} and
\cite{Brocker2020a}.

The most basic case is that of tuples $(x_1,y_1), \ldots, (x_n,y_n)$
of the form \eqref{eq:x.y}, where implicitly or explicitly $x_i =
\myT(F_i)$ for a single-valued functional $\myT$.  We first discuss
tests of unconditional calibration.  If the simplified condition
\eqref{eq:simple} is sufficient, a two-sided $t$-test based on
$\widehat{v} = \frac{1}{n} \sum_{i = 1}^n V(x_i,y_i)$ can be used to
test for unconditional calibration.  In the general case, two
one-sided $t$-tests can be used along with a Bonferroni correction.
In the special case of quantiles, there is no need to resort to the
approximate $t$-tests, and exact binomial tests can be used instead.
Essentially, this is the setting of backtests for value-at-risk
reports in banking regulation, for which we refer to
\cite[Sections 2.1--2.2]{Nolde2017}.

As noted earlier in the section, resamples generated under the
hypothesis of conditional $\myT$-calibration can readily be used to
perform Monte Carlo tests for the respective hypothesis, based on CORP
score components that are computed on the surrogate data.
Alternatively, one might leverage extant large sample theory for
nonparametric isotonic regression \citep{Barlow1972, Casady1976,
	Wright1984, Robertson1988, ElBarmi2005, Mosching2020}.
Independently of the use of resampling or asymptotic theory, CORP
based tests avoid the issues and instabilities incurred by binning
\citep[Section S2]{Dimitriadis2020a} and may simultaneously
improve efficiency and stability.  In passing, we hint at relations to
the null hypothesis of Mincer-Zarnowitz regression
\citep{Krueger2020} and tests of predictive content
\citep{Galbraith2003, Breitung2021}.

We move on to the case of fully specified distributions, where we work
with tuples $(F_1, y_1), \ldots, (F_n, y_n)$ of the form
\eqref{eq:F.y}, where $F_i$ is a posited conditional CDF for $y_i$
($i = 1, \ldots, n$).  Tests for probabilistic calibration then amount
to tests for the uniformity of the (potentially, randomized) PIT
values.  \cite{Wallis2003} and \citet[p.~769]{Wilks2019}
suggest chi-square tests for this purpose, which depend on binning,
and thus are subject to the aforementioned instabilities.  To avoid
binning, we recommend the use of test statistics that operate on the
empirical CDF of the PIT values, such as the classical
Kolmogorov-Smirnov (KS) statistic, as suggested and used to test for
PIT calibration by \cite{Noceti2003} and
\cite{Knueppel2015}, or, more generally, tests based on
distance measures between the empirical CDF of the PIT values, and the
CDF of the standard uniform distribution that arises under the
hypothesis of probabilistic calibration.  Recently proposed
alternatives arise via $e$-values \citep{Henzi2021}.  Similarly,
tests for marginal calibration can be based on resamples and distance
measures between $\bar{F}$ and $\hat{F}_0$, or leverage asymptotic
theory.

In the distributional setting, arbitrarily many types of reliability
can be tested for, and all of the aforementioned tests for
unconditional or conditional $\myT$-calibration apply.  Multiple
testing needs to be accounted for properly, and the development of
simultaneous tests for various types of calibration would be useful.
In this context, let us recall from Theorem \ref{th:TC=QC} that,
subject to technical conditions, CEP, threshold, and quantile
calibration are equivalent and tests for CEP calibration
\citep{Held2010, Strahl2017}, quantile and threshold calibration
assess identical hypotheses.

\section{Time series settings and the Bank of England example}  \label{sec:timeseries}

In typical time series settings, as exemplified by our analysis of
Bank of England forecasts in Section \ref{sec:empirical}, the
assumption of independent replicates of forecasts and observations is
too restrictive.  While the diagnostic methods proposed in our paper
continue to apply, statistical inference requires care, as discussed
by \cite{Corradi2007} and \cite{Knueppel2015}, among other authors.
Here, we elucidate the role of uniform and independent
probability integral transform (PIT) values for calibration in time
series settings, and give further details and results for the Bank of
England example.

\subsection{The role of uniform and independent PITs}  \label{sec:iidPITs}

In a landmark paper, \citet[][p.\ 867]{Diebold1998} showed that a
sequence of continuous predictive distributions $F_t$ for a sequence
$Y_t$ of observations at time $t = 0, 1, \dots$ results in a sequence
of independent, uniformly distributed PITs if $F_t$ is ideal relative
to the $\sigma$-algebra generated by past observations,
$\cA_t = \sigma(Y_0, Y_1, \ldots, Y_{t-1})$.  This property does not
depend on the continuity of $F_t$ and continues to hold under general
predictive CDFs and the randomized definition \eqref{eq:PIT} of the
PIT \citep[][Theorem 3]{Rueschendorf1993}.

In the case of continuous predictive distributions,
\citet[][Section 2]{Tsyplakov2011} noted without proof that if the
forecasts $F_t$ are based only on past observations, i.e., if $F_t$ is
$\cA_t$-measurable, then the converse holds, namely, uniform and
independent PITs arise only if $F_t$ is ideal relative to $\cA_t$.
The following result formalizes Tsyplakov's claim and proves it in the
general setting, without any assumption of continuity.

\begin{theorem} \label{th:PIT} Let $(Y_t)_{t = 0, 1, \ldots}$ be a
	sequence of random variables, and let
	$\cA_t = \sigma(Y_0, \dots, Y_{t-1})$ for $t = 0, 1, \dots$
	Furthermore, let $(F_t)_{t = 0, 1, \dots}$ be a sequence of
	CDFs, such that $F_t$ is $\cA_t$-measurable for
	$t = 0, 1, \dots$, and let $(U_t)_{t = 0, 1, \ldots}$ be a
	sequence of independent, uniformly distributed random variables,
	independent of the sequence $(Y_t)$.  Then the sequence of
	randomized PITs,
	$(Z_t) = (F_t(Y_t-) + U_t(F_t(Y_t) - F_t(Y_t-)))$ is an
	independent sequence of uniform random variables on the
	unit interval if, and only if, $F_t$ is ideal relative to $\cA_t$,
	i.e., $F_t = \cL(Y_t\mid \cA_t)$ almost surely for
	$t = 0, 1, \ldots$
\end{theorem}

The proof utilizes the following simple lemma.

\begin{lemma} \label{lemma:condas} Let $X, Y, Z$ be random variables.
	If $X = Z$ almost surely, then
	$\myE[ Y \mid X] = \myE[Y \mid Z]$ almost surely.
\end{lemma}

\begin{proof}
	Problem 14 of \citet[][Chapter 4]{Breiman1992}, which is proved
	by \citet[][Satz 18.2.10]{Schmidt2011}, states that for random
	variables $X_1$ and $X_2$ such that $\sigma(Y, X_1)$ is independent
	of $\sigma(X_2)$, $\myE[ Y \mid X_1, X_2 ] = \myE[ Y \mid X_1 ]$
	almost surely.  The statement of the lemma follows, as
	$\myE[Y \mid X] = \myE[Y \mid X, X-Z] = \myE[Y\mid Z, X-Z] =
	\myE[Y\mid Z]$ almost surely.
\end{proof}

\begin{proof}[Proof of Theorem \ref{th:PIT}]
	Since $F_t$ is measurable with respect to $\cA_t$, there exists
	a measurable function $f_t \colon \real^t \rightarrow \cF$ such that
	$F_t = f_t(Y_0, \ldots, Y_{t-1})$ for each $t$ by the Doob--Dynkin
	Lemma \citep[][Satz 7.1.16]{Schmidt2011}.\footnote{Note
		that $f_0$ is constant, and $f_t$ is not a random quantity, but a
		fixed function that encodes how the predictive distributions
		are generated from past observations.  The $\sigma$-algebra on
		$\cF$, which is implicitly used throughout, is given by
		\[
		\cA_\cF = \sigma( \{ \{ F \in \cF : F(x) \in B \} : x \in \rational, \, B \in \cB(\real) \} ),
		\]  
		where $\cB(\real)$ denotes the Borel $\sigma$-algebra on $\real$.
		For each $x \in \rational$ there exists a measurable function
		$f_{x,t}$ such that $F_t(x) = f_{x,t}(Y_0, \ldots, Y_{t-1})$ by
		the Doob-Dynkin Lemma, and $f_t$ is essentially the countable
		(and hence measurable) collection
		$(f_{x,t})_{x \in \rational}$.}  We define
	\[
	G_t := f_t(G_0^{-1}(Z_0), \ldots, G_{t-1}^{-1}(Z_{t-1}))
	\]
	recursively for all $t$, and show the ``only if'' assertion by induction.
	
	To this end, let $t \geq 0$ and assume the induction hypothesis that
	$F_i$ is ideal relative to $\cA_i$ for $i = 0, \ldots, t - 1$.  By
	\citet[][Theorem 3(a)]{Rueschendorf1993} and the construction of
	$G_t$, this implies
	\[
	(Y_0, \ldots, Y_{t-1}) = (F_0^{-1}(Z_0), \ldots, F_{t-1}^{-1}(Z_{t-1})) = (G_0^{-1}(Z_0), \ldots, G_{t-1}^{-1}(Z_{t-1}))
	\]
	almost surely, and the last vector is $\sigma(Z_0, \ldots, Z_{t-1})$-measurable.  By Lemma
	\ref{lemma:condas}, it follows that
	\[
	\cL(Z_t \mid \cA_t) = \cL(Z_t \mid \sigma(G_0^{-1}(Z_0), \ldots, G_{t-1}^{-1}(Z_{t-1}))) = U([0,1])
	\]
	almost surely, where the second equality stems from the fact that $Z_t$ is
	independent of $\sigma(G_0^{-1}(Z_0), \ldots, G_{t-1}^{-1}(Z_{t-1})) \subset
	\sigma(Z_0, \ldots, Z_{t-1})$.  This implies that $F_t$ is ideal
	relative to $\cA_t$, because
	\[
	F_t(y) = \myQ(Z_t < F_t(y) \mid \cA_t) \leq \myQ(Y_t \leq y \mid \cA_t)
	\leq \myQ(Z_t \leq F_t(y) \mid \cA_t) = F_t(y)
	\]
	almost surely, and hence
	$F_t(y) = \myQ(Y_t \leq y \mid \cA_t)$ almost surely for all
	$y \in \rational$, thereby completing both the induction step
	and the claim for the base case $t = 0$.
\end{proof}

Evidently, the assumption that no information other than the history
of the time series itself has been utilized to construct the forecasts
is very limiting.  In this light, it is not surprising that, while the
``if'' part of Theorem \ref{th:PIT} is robust, the ``only if'' claim
fails if $F_t$ is allowed to use information beyond the canonical
filtration, even if that information is uninformative.  A simple
counterexample is given by the unfocused forecast from Example
\ref{ex:unf.lop}, which is probabilistically calibrated, but fails to
be auto-calibrated.  Its PIT nevertheless is uniform and independent
even for autoregressive variants \citep[][Section
6]{Tsyplakov2011}.

\subsection{Details and further results for the Bank of England example}  \label{sec:BoE} 

Bank of England forecasts of inflation rates are available within the
data accompanying the quarterly Monetary Policy Report (formerly
Inflation Report), which is available online at
\url{https://www.bankofengland.co.uk/sitemap/monetary-policy-report}.
The forecasts are visualized and communicated in the form of \emph{fan
	charts}\/ that span prediction intervals at increasing forecast
horizons, and derive from two-piece normal forecast distributions.  A
detailed account of the parametrizations for the two-piece normal
distribution used by the Bank of England can be found in
\cite{Julio2006}, and we have implemented the formulas in this
reference.  Historical quarterly CPI inflation rates are published by
the UK Office for National Statistics (ONS) and available online at
\url{https://www.ons.gov.uk/economy/inflationandpriceindices/timeseries/d7g7}.

We consider forecasts of consumer price index (CPI) inflation based on
market expectations for future interest rates at prediction horizons
of zero to six quarters ahead, valid for the third quarter of 2005 up
to the first quarter of 2020, for a total of $n = 59$ quarters.  These
and earlier Bank of England forecasts of inflation rates have been
checked for reliability by \cite{Wallis2003}, who considered
probabilistic calibration, by \cite{Clements2004} in terms of
probabilistic, mean, and threshold calibration, by
\cite{Galbraith2012}, who considered probabilistic and mean
calibration, by \cite{Strahl2017} with focus on conditional
exceedance probability (CEP) calibration, and by \cite{Pohle2020},
who considered quantile calibration.  The 2\% inflation target is
discussed at the Bank of England website at
\url{https://www.bankofengland.co.uk/monetary-policy/inflation}.

Figures \ref{fig:0}--\ref{fig:6} show calibration diagnostics for
inflation forecasts at prediction horizons of $k \in \{ 0, 2, 3, 4, 5,
6 \}$ quarters ahead, in the same format as Figure \ref{fig:1} in the
main text, which concerns forecasts at a lead time of one
quarter.

\def\scalingFactor{0.58} 

\begin{figure}[p]  
	\centering
	\scalebox{\scalingFactor}{\input{figs/tikz/relDiag_PIT_BoE0.tex}}
	\scalebox{\scalingFactor}{% Created by tikzDevice version 0.12.3.1 on 2021-08-06 11:48:11
% !TEX encoding = UTF-8 Unicode
\begin{tikzpicture}[x=1pt,y=1pt]
\definecolor{fillColor}{RGB}{255,255,255}
\path[use as bounding box,fill=fillColor,fill opacity=0.00] (0,0) rectangle (195.13,227.65);
\begin{scope}
\path[clip] ( 21.68, 21.68) rectangle (195.13,195.13);
\definecolor{drawColor}{RGB}{0,0,0}

\path[draw=drawColor,line width= 0.6pt,line join=round,line cap=round] ( 28.11, 58.37) -- ( 28.11,188.70);

\path[draw=drawColor,line width= 0.6pt,line join=round,line cap=round] ( 37.55, 58.37) -- ( 37.55, 28.11);

\path[draw=drawColor,line width= 0.6pt,line join=round,line cap=round] ( 47.00, 58.37) -- ( 47.00, 71.51);

\path[draw=drawColor,line width= 0.6pt,line join=round,line cap=round] ( 56.45, 58.37) -- ( 56.45, 52.72);

\path[draw=drawColor,line width= 0.6pt,line join=round,line cap=round] ( 65.89, 58.37) -- ( 65.89, 54.06);

\path[draw=drawColor,line width= 0.6pt,line join=round,line cap=round] ( 75.34, 58.37) -- ( 75.34,100.70);

\path[draw=drawColor,line width= 0.6pt,line join=round,line cap=round] ( 84.79, 58.37) -- ( 84.79, 57.47);

\path[draw=drawColor,line width= 0.6pt,line join=round,line cap=round] ( 94.23, 58.37) -- ( 94.23, 68.90);

\path[draw=drawColor,line width= 0.6pt,line join=round,line cap=round] (103.68, 58.37) -- (103.68, 40.86);

\path[draw=drawColor,line width= 0.6pt,line join=round,line cap=round] (113.13, 58.37) -- (113.13, 68.22);

\path[draw=drawColor,line width= 0.6pt,line join=round,line cap=round] (122.58, 58.37) -- (122.58, 63.32);

\path[draw=drawColor,line width= 0.6pt,line join=round,line cap=round] (132.02, 58.37) -- (132.02, 75.59);

\path[draw=drawColor,line width= 0.6pt,line join=round,line cap=round] (141.47, 58.37) -- (141.47, 62.20);

\path[draw=drawColor,line width= 0.6pt,line join=round,line cap=round] (150.92, 58.37) -- (150.92, 40.33);

\path[draw=drawColor,line width= 0.6pt,line join=round,line cap=round] (160.36, 58.37) -- (160.36, 72.82);

\path[draw=drawColor,line width= 0.6pt,line join=round,line cap=round] (169.81, 58.37) -- (169.81, 64.24);

\path[draw=drawColor,line width= 0.6pt,line join=round,line cap=round] (179.26, 58.37) -- (179.26, 44.35);

\path[draw=drawColor,line width= 0.6pt,line join=round,line cap=round] (188.71, 58.37) -- (188.71, 69.07);
\end{scope}
\begin{scope}
\path[clip] (  0.00,  0.00) rectangle (195.13,227.65);
\definecolor{drawColor}{RGB}{0,0,0}

\path[draw=drawColor,line width= 0.4pt,line join=round,line cap=round] ( 28.11, 21.68) -- (169.81, 21.68);

\path[draw=drawColor,line width= 0.4pt,line join=round,line cap=round] ( 28.11, 21.68) -- ( 28.11, 15.68);

\path[draw=drawColor,line width= 0.4pt,line join=round,line cap=round] ( 75.34, 21.68) -- ( 75.34, 15.68);

\path[draw=drawColor,line width= 0.4pt,line join=round,line cap=round] (122.58, 21.68) -- (122.58, 15.68);

\path[draw=drawColor,line width= 0.4pt,line join=round,line cap=round] (169.81, 21.68) -- (169.81, 15.68);

\node[text=drawColor,anchor=base,inner sep=0pt, outer sep=0pt, scale=  1.00] at ( 28.11,  2.48) {0};

\node[text=drawColor,anchor=base,inner sep=0pt, outer sep=0pt, scale=  1.00] at ( 75.34,  2.48) {5};

\node[text=drawColor,anchor=base,inner sep=0pt, outer sep=0pt, scale=  1.00] at (122.58,  2.48) {10};

\node[text=drawColor,anchor=base,inner sep=0pt, outer sep=0pt, scale=  1.00] at (169.81,  2.48) {15};

\path[draw=drawColor,line width= 0.4pt,line join=round,line cap=round] ( 21.68, 32.30) -- ( 21.68,188.70);

\path[draw=drawColor,line width= 0.4pt,line join=round,line cap=round] ( 21.68, 32.30) -- ( 15.68, 32.30);

\path[draw=drawColor,line width= 0.4pt,line join=round,line cap=round] ( 21.68, 58.37) -- ( 15.68, 58.37);

\path[draw=drawColor,line width= 0.4pt,line join=round,line cap=round] ( 21.68, 84.44) -- ( 15.68, 84.44);

\path[draw=drawColor,line width= 0.4pt,line join=round,line cap=round] ( 21.68,110.50) -- ( 15.68,110.50);

\path[draw=drawColor,line width= 0.4pt,line join=round,line cap=round] ( 21.68,136.57) -- ( 15.68,136.57);

\path[draw=drawColor,line width= 0.4pt,line join=round,line cap=round] ( 21.68,162.64) -- ( 15.68,162.64);

\path[draw=drawColor,line width= 0.4pt,line join=round,line cap=round] ( 21.68,188.70) -- ( 15.68,188.70);

\node[text=drawColor,rotate= 90.00,anchor=base,inner sep=0pt, outer sep=0pt, scale=  1.00] at (  9.68, 32.30) {-0.2};

\node[text=drawColor,rotate= 90.00,anchor=base,inner sep=0pt, outer sep=0pt, scale=  1.00] at (  9.68, 58.37) {0.0};

\node[text=drawColor,rotate= 90.00,anchor=base,inner sep=0pt, outer sep=0pt, scale=  1.00] at (  9.68, 84.44) {0.2};

\node[text=drawColor,rotate= 90.00,anchor=base,inner sep=0pt, outer sep=0pt, scale=  1.00] at (  9.68,110.50) {0.4};

\node[text=drawColor,rotate= 90.00,anchor=base,inner sep=0pt, outer sep=0pt, scale=  1.00] at (  9.68,136.57) {0.6};

\node[text=drawColor,rotate= 90.00,anchor=base,inner sep=0pt, outer sep=0pt, scale=  1.00] at (  9.68,162.64) {0.8};

\node[text=drawColor,rotate= 90.00,anchor=base,inner sep=0pt, outer sep=0pt, scale=  1.00] at (  9.68,188.70) {1.0};

\path[draw=drawColor,line width= 0.6pt,line join=round,line cap=round] ( 21.68, 21.68) --
	(195.13, 21.68) --
	(195.13,195.13) --
	( 21.68,195.13) --
	( 21.68, 21.68);
\end{scope}
\begin{scope}
\path[clip] (  0.00,  0.00) rectangle (195.13,227.65);
\definecolor{drawColor}{RGB}{0,0,0}

\node[text=drawColor,anchor=base,inner sep=0pt, outer sep=0pt, scale=  1.00] at (108.41, 30.08) {lag};
\end{scope}
\begin{scope}
\path[clip] ( 21.68, 21.68) rectangle (195.13,195.13);
\definecolor{drawColor}{RGB}{0,0,0}

\path[draw=drawColor,line width= 0.6pt,line join=round,line cap=round] ( 21.68, 58.37) -- (195.13, 58.37);
\definecolor{drawColor}{RGB}{0,0,255}

\path[draw=drawColor,line width= 0.6pt,dash pattern=on 4pt off 4pt ,line join=round,line cap=round] ( 21.68, 86.28) -- (195.13, 86.28);

\path[draw=drawColor,line width= 0.6pt,dash pattern=on 4pt off 4pt ,line join=round,line cap=round] ( 21.68, 30.46) -- (195.13, 30.46);
\end{scope}
\begin{scope}
\path[clip] (  0.00,  0.00) rectangle (195.13,227.65);
\definecolor{drawColor}{RGB}{0,0,0}

\node[text=drawColor,anchor=base,inner sep=0pt, outer sep=0pt, scale=  1.50] at (108.41,207.13) {\bfseries (b) ACF of PIT};
\end{scope}
\end{tikzpicture}}
	\scalebox{\scalingFactor}{% Created by tikzDevice version 0.12.3.1 on 2021-08-06 11:48:12
% !TEX encoding = UTF-8 Unicode
\begin{tikzpicture}[x=1pt,y=1pt]
\definecolor{fillColor}{RGB}{255,255,255}
\path[use as bounding box,fill=fillColor,fill opacity=0.00] (0,0) rectangle (195.13,227.65);
\begin{scope}
\path[clip] ( 21.68, 21.68) rectangle (195.13,195.13);
\definecolor{drawColor}{RGB}{0,0,0}

\path[draw=drawColor,line width= 0.6pt,line join=round,line cap=round] ( 28.11, 56.43) -- ( 28.11,188.70);

\path[draw=drawColor,line width= 0.6pt,line join=round,line cap=round] ( 37.55, 56.43) -- ( 37.55, 87.10);

\path[draw=drawColor,line width= 0.6pt,line join=round,line cap=round] ( 47.00, 56.43) -- ( 47.00, 85.86);

\path[draw=drawColor,line width= 0.6pt,line join=round,line cap=round] ( 56.45, 56.43) -- ( 56.45,103.81);

\path[draw=drawColor,line width= 0.6pt,line join=round,line cap=round] ( 65.89, 56.43) -- ( 65.89, 63.75);

\path[draw=drawColor,line width= 0.6pt,line join=round,line cap=round] ( 75.34, 56.43) -- ( 75.34, 78.61);

\path[draw=drawColor,line width= 0.6pt,line join=round,line cap=round] ( 84.79, 56.43) -- ( 84.79, 71.12);

\path[draw=drawColor,line width= 0.6pt,line join=round,line cap=round] ( 94.23, 56.43) -- ( 94.23, 51.21);

\path[draw=drawColor,line width= 0.6pt,line join=round,line cap=round] (103.68, 56.43) -- (103.68, 83.67);

\path[draw=drawColor,line width= 0.6pt,line join=round,line cap=round] (113.13, 56.43) -- (113.13, 71.07);

\path[draw=drawColor,line width= 0.6pt,line join=round,line cap=round] (122.58, 56.43) -- (122.58, 65.68);

\path[draw=drawColor,line width= 0.6pt,line join=round,line cap=round] (132.02, 56.43) -- (132.02, 90.23);

\path[draw=drawColor,line width= 0.6pt,line join=round,line cap=round] (141.47, 56.43) -- (141.47, 54.34);

\path[draw=drawColor,line width= 0.6pt,line join=round,line cap=round] (150.92, 56.43) -- (150.92, 74.26);

\path[draw=drawColor,line width= 0.6pt,line join=round,line cap=round] (160.36, 56.43) -- (160.36, 71.15);

\path[draw=drawColor,line width= 0.6pt,line join=round,line cap=round] (169.81, 56.43) -- (169.81, 49.44);

\path[draw=drawColor,line width= 0.6pt,line join=round,line cap=round] (179.26, 56.43) -- (179.26, 72.18);

\path[draw=drawColor,line width= 0.6pt,line join=round,line cap=round] (188.71, 56.43) -- (188.71, 56.41);
\end{scope}
\begin{scope}
\path[clip] (  0.00,  0.00) rectangle (195.13,227.65);
\definecolor{drawColor}{RGB}{0,0,0}

\path[draw=drawColor,line width= 0.4pt,line join=round,line cap=round] ( 28.11, 21.68) -- (169.81, 21.68);

\path[draw=drawColor,line width= 0.4pt,line join=round,line cap=round] ( 28.11, 21.68) -- ( 28.11, 15.68);

\path[draw=drawColor,line width= 0.4pt,line join=round,line cap=round] ( 75.34, 21.68) -- ( 75.34, 15.68);

\path[draw=drawColor,line width= 0.4pt,line join=round,line cap=round] (122.58, 21.68) -- (122.58, 15.68);

\path[draw=drawColor,line width= 0.4pt,line join=round,line cap=round] (169.81, 21.68) -- (169.81, 15.68);

\node[text=drawColor,anchor=base,inner sep=0pt, outer sep=0pt, scale=  1.00] at ( 28.11,  2.48) {0};

\node[text=drawColor,anchor=base,inner sep=0pt, outer sep=0pt, scale=  1.00] at ( 75.34,  2.48) {5};

\node[text=drawColor,anchor=base,inner sep=0pt, outer sep=0pt, scale=  1.00] at (122.58,  2.48) {10};

\node[text=drawColor,anchor=base,inner sep=0pt, outer sep=0pt, scale=  1.00] at (169.81,  2.48) {15};

\path[draw=drawColor,line width= 0.4pt,line join=round,line cap=round] ( 21.68, 29.98) -- ( 21.68,188.70);

\path[draw=drawColor,line width= 0.4pt,line join=round,line cap=round] ( 21.68, 29.98) -- ( 15.68, 29.98);

\path[draw=drawColor,line width= 0.4pt,line join=round,line cap=round] ( 21.68, 56.43) -- ( 15.68, 56.43);

\path[draw=drawColor,line width= 0.4pt,line join=round,line cap=round] ( 21.68, 82.89) -- ( 15.68, 82.89);

\path[draw=drawColor,line width= 0.4pt,line join=round,line cap=round] ( 21.68,109.34) -- ( 15.68,109.34);

\path[draw=drawColor,line width= 0.4pt,line join=round,line cap=round] ( 21.68,135.80) -- ( 15.68,135.80);

\path[draw=drawColor,line width= 0.4pt,line join=round,line cap=round] ( 21.68,162.25) -- ( 15.68,162.25);

\path[draw=drawColor,line width= 0.4pt,line join=round,line cap=round] ( 21.68,188.70) -- ( 15.68,188.70);

\node[text=drawColor,rotate= 90.00,anchor=base,inner sep=0pt, outer sep=0pt, scale=  1.00] at (  9.68, 29.98) {-0.2};

\node[text=drawColor,rotate= 90.00,anchor=base,inner sep=0pt, outer sep=0pt, scale=  1.00] at (  9.68, 56.43) {0.0};

\node[text=drawColor,rotate= 90.00,anchor=base,inner sep=0pt, outer sep=0pt, scale=  1.00] at (  9.68, 82.89) {0.2};

\node[text=drawColor,rotate= 90.00,anchor=base,inner sep=0pt, outer sep=0pt, scale=  1.00] at (  9.68,109.34) {0.4};

\node[text=drawColor,rotate= 90.00,anchor=base,inner sep=0pt, outer sep=0pt, scale=  1.00] at (  9.68,135.80) {0.6};

\node[text=drawColor,rotate= 90.00,anchor=base,inner sep=0pt, outer sep=0pt, scale=  1.00] at (  9.68,162.25) {0.8};

\node[text=drawColor,rotate= 90.00,anchor=base,inner sep=0pt, outer sep=0pt, scale=  1.00] at (  9.68,188.70) {1.0};

\path[draw=drawColor,line width= 0.6pt,line join=round,line cap=round] ( 21.68, 21.68) --
	(195.13, 21.68) --
	(195.13,195.13) --
	( 21.68,195.13) --
	( 21.68, 21.68);
\end{scope}
\begin{scope}
\path[clip] (  0.00,  0.00) rectangle (195.13,227.65);
\definecolor{drawColor}{RGB}{0,0,0}

\node[text=drawColor,anchor=base,inner sep=0pt, outer sep=0pt, scale=  1.00] at (108.41, 30.08) {lag};
\end{scope}
\begin{scope}
\path[clip] ( 21.68, 21.68) rectangle (195.13,195.13);
\definecolor{drawColor}{RGB}{0,0,0}

\path[draw=drawColor,line width= 0.6pt,line join=round,line cap=round] ( 21.68, 56.43) -- (195.13, 56.43);
\definecolor{drawColor}{RGB}{0,0,255}

\path[draw=drawColor,line width= 0.6pt,dash pattern=on 4pt off 4pt ,line join=round,line cap=round] ( 21.68, 84.76) -- (195.13, 84.76);

\path[draw=drawColor,line width= 0.6pt,dash pattern=on 4pt off 4pt ,line join=round,line cap=round] ( 21.68, 28.11) -- (195.13, 28.11);
\end{scope}
\begin{scope}
\path[clip] (  0.00,  0.00) rectangle (195.13,227.65);
\definecolor{drawColor}{RGB}{0,0,0}

\node[text=drawColor,anchor=base,inner sep=0pt, outer sep=0pt, scale=  1.50] at (108.41,207.13) {\bfseries (c) ACF of $\protect\boldmath{(\text{PIT} - \frac12)^2}$};
\end{scope}
\end{tikzpicture}} \\ \smallskip
	\scalebox{\scalingFactor}{\input{figs/tikz/relDiag_marg_BoE0.tex}}
	\scalebox{\scalingFactor}{\input{figs/tikz/relDiag_thresh_BoE0.tex}}
	\scalebox{\scalingFactor}{\input{figs/tikz/relDiag_quant_BoE0.tex}}
	\caption{Same as Figure \ref{fig:1}, but at
		a prediction horizon of zero quarters.  \label{fig:0}}
	\bigskip
	\centering % Please do not change the sizes of the boxes, this will mess up the font size!
	\scalebox{\scalingFactor}{\input{figs/tikz/relDiag_PIT_BoE2.tex}}
	\scalebox{\scalingFactor}{% Created by tikzDevice version 0.12.3.1 on 2021-08-06 11:48:41
% !TEX encoding = UTF-8 Unicode
\begin{tikzpicture}[x=1pt,y=1pt]
\definecolor{fillColor}{RGB}{255,255,255}
\path[use as bounding box,fill=fillColor,fill opacity=0.00] (0,0) rectangle (195.13,227.65);
\begin{scope}
\path[clip] ( 21.68, 21.68) rectangle (195.13,195.13);
\definecolor{drawColor}{RGB}{0,0,0}

\path[draw=drawColor,line width= 0.6pt,line join=round,line cap=round] ( 28.11, 56.43) -- ( 28.11,188.70);

\path[draw=drawColor,line width= 0.6pt,line join=round,line cap=round] ( 37.55, 56.43) -- ( 37.55,133.91);

\path[draw=drawColor,line width= 0.6pt,line join=round,line cap=round] ( 47.00, 56.43) -- ( 47.00, 77.06);

\path[draw=drawColor,line width= 0.6pt,line join=round,line cap=round] ( 56.45, 56.43) -- ( 56.45, 61.23);

\path[draw=drawColor,line width= 0.6pt,line join=round,line cap=round] ( 65.89, 56.43) -- ( 65.89, 71.65);

\path[draw=drawColor,line width= 0.6pt,line join=round,line cap=round] ( 75.34, 56.43) -- ( 75.34, 84.89);

\path[draw=drawColor,line width= 0.6pt,line join=round,line cap=round] ( 84.79, 56.43) -- ( 84.79, 94.51);

\path[draw=drawColor,line width= 0.6pt,line join=round,line cap=round] ( 94.23, 56.43) -- ( 94.23, 85.82);

\path[draw=drawColor,line width= 0.6pt,line join=round,line cap=round] (103.68, 56.43) -- (103.68, 56.98);

\path[draw=drawColor,line width= 0.6pt,line join=round,line cap=round] (113.13, 56.43) -- (113.13, 52.40);

\path[draw=drawColor,line width= 0.6pt,line join=round,line cap=round] (122.58, 56.43) -- (122.58, 52.85);

\path[draw=drawColor,line width= 0.6pt,line join=round,line cap=round] (132.02, 56.43) -- (132.02, 64.27);

\path[draw=drawColor,line width= 0.6pt,line join=round,line cap=round] (141.47, 56.43) -- (141.47, 77.58);

\path[draw=drawColor,line width= 0.6pt,line join=round,line cap=round] (150.92, 56.43) -- (150.92, 70.06);

\path[draw=drawColor,line width= 0.6pt,line join=round,line cap=round] (160.36, 56.43) -- (160.36, 44.51);

\path[draw=drawColor,line width= 0.6pt,line join=round,line cap=round] (169.81, 56.43) -- (169.81, 31.56);

\path[draw=drawColor,line width= 0.6pt,line join=round,line cap=round] (179.26, 56.43) -- (179.26, 43.88);

\path[draw=drawColor,line width= 0.6pt,line join=round,line cap=round] (188.71, 56.43) -- (188.71, 53.23);
\end{scope}
\begin{scope}
\path[clip] (  0.00,  0.00) rectangle (195.13,227.65);
\definecolor{drawColor}{RGB}{0,0,0}

\path[draw=drawColor,line width= 0.4pt,line join=round,line cap=round] ( 28.11, 21.68) -- (169.81, 21.68);

\path[draw=drawColor,line width= 0.4pt,line join=round,line cap=round] ( 28.11, 21.68) -- ( 28.11, 15.68);

\path[draw=drawColor,line width= 0.4pt,line join=round,line cap=round] ( 75.34, 21.68) -- ( 75.34, 15.68);

\path[draw=drawColor,line width= 0.4pt,line join=round,line cap=round] (122.58, 21.68) -- (122.58, 15.68);

\path[draw=drawColor,line width= 0.4pt,line join=round,line cap=round] (169.81, 21.68) -- (169.81, 15.68);

\node[text=drawColor,anchor=base,inner sep=0pt, outer sep=0pt, scale=  1.00] at ( 28.11,  2.48) {0};

\node[text=drawColor,anchor=base,inner sep=0pt, outer sep=0pt, scale=  1.00] at ( 75.34,  2.48) {5};

\node[text=drawColor,anchor=base,inner sep=0pt, outer sep=0pt, scale=  1.00] at (122.58,  2.48) {10};

\node[text=drawColor,anchor=base,inner sep=0pt, outer sep=0pt, scale=  1.00] at (169.81,  2.48) {15};

\path[draw=drawColor,line width= 0.4pt,line join=round,line cap=round] ( 21.68, 29.98) -- ( 21.68,188.70);

\path[draw=drawColor,line width= 0.4pt,line join=round,line cap=round] ( 21.68, 29.98) -- ( 15.68, 29.98);

\path[draw=drawColor,line width= 0.4pt,line join=round,line cap=round] ( 21.68, 56.43) -- ( 15.68, 56.43);

\path[draw=drawColor,line width= 0.4pt,line join=round,line cap=round] ( 21.68, 82.89) -- ( 15.68, 82.89);

\path[draw=drawColor,line width= 0.4pt,line join=round,line cap=round] ( 21.68,109.34) -- ( 15.68,109.34);

\path[draw=drawColor,line width= 0.4pt,line join=round,line cap=round] ( 21.68,135.80) -- ( 15.68,135.80);

\path[draw=drawColor,line width= 0.4pt,line join=round,line cap=round] ( 21.68,162.25) -- ( 15.68,162.25);

\path[draw=drawColor,line width= 0.4pt,line join=round,line cap=round] ( 21.68,188.70) -- ( 15.68,188.70);

\node[text=drawColor,rotate= 90.00,anchor=base,inner sep=0pt, outer sep=0pt, scale=  1.00] at (  9.68, 29.98) {-0.2};

\node[text=drawColor,rotate= 90.00,anchor=base,inner sep=0pt, outer sep=0pt, scale=  1.00] at (  9.68, 56.43) {0.0};

\node[text=drawColor,rotate= 90.00,anchor=base,inner sep=0pt, outer sep=0pt, scale=  1.00] at (  9.68, 82.89) {0.2};

\node[text=drawColor,rotate= 90.00,anchor=base,inner sep=0pt, outer sep=0pt, scale=  1.00] at (  9.68,109.34) {0.4};

\node[text=drawColor,rotate= 90.00,anchor=base,inner sep=0pt, outer sep=0pt, scale=  1.00] at (  9.68,135.80) {0.6};

\node[text=drawColor,rotate= 90.00,anchor=base,inner sep=0pt, outer sep=0pt, scale=  1.00] at (  9.68,162.25) {0.8};

\node[text=drawColor,rotate= 90.00,anchor=base,inner sep=0pt, outer sep=0pt, scale=  1.00] at (  9.68,188.70) {1.0};

\path[draw=drawColor,line width= 0.6pt,line join=round,line cap=round] ( 21.68, 21.68) --
	(195.13, 21.68) --
	(195.13,195.13) --
	( 21.68,195.13) --
	( 21.68, 21.68);
\end{scope}
\begin{scope}
\path[clip] (  0.00,  0.00) rectangle (195.13,227.65);
\definecolor{drawColor}{RGB}{0,0,0}

\node[text=drawColor,anchor=base,inner sep=0pt, outer sep=0pt, scale=  1.00] at (108.41, 30.08) {lag};
\end{scope}
\begin{scope}
\path[clip] ( 21.68, 21.68) rectangle (195.13,195.13);
\definecolor{drawColor}{RGB}{0,0,0}

\path[draw=drawColor,line width= 0.6pt,line join=round,line cap=round] ( 21.68, 56.43) -- (195.13, 56.43);
\definecolor{drawColor}{RGB}{0,0,255}

\path[draw=drawColor,line width= 0.6pt,dash pattern=on 4pt off 4pt ,line join=round,line cap=round] ( 21.68, 84.76) -- (195.13, 84.76);

\path[draw=drawColor,line width= 0.6pt,dash pattern=on 4pt off 4pt ,line join=round,line cap=round] ( 21.68, 28.11) -- (195.13, 28.11);
\end{scope}
\begin{scope}
\path[clip] (  0.00,  0.00) rectangle (195.13,227.65);
\definecolor{drawColor}{RGB}{0,0,0}

\node[text=drawColor,anchor=base,inner sep=0pt, outer sep=0pt, scale=  1.50] at (108.41,207.13) {\bfseries (b) ACF of PIT};
\end{scope}
\end{tikzpicture}}
	\scalebox{\scalingFactor}{% Created by tikzDevice version 0.12.3.1 on 2021-08-06 11:48:41
% !TEX encoding = UTF-8 Unicode
\begin{tikzpicture}[x=1pt,y=1pt]
\definecolor{fillColor}{RGB}{255,255,255}
\path[use as bounding box,fill=fillColor,fill opacity=0.00] (0,0) rectangle (195.13,227.65);
\begin{scope}
\path[clip] ( 21.68, 21.68) rectangle (195.13,195.13);
\definecolor{drawColor}{RGB}{0,0,0}

\path[draw=drawColor,line width= 0.6pt,line join=round,line cap=round] ( 28.11, 56.43) -- ( 28.11,188.70);

\path[draw=drawColor,line width= 0.6pt,line join=round,line cap=round] ( 37.55, 56.43) -- ( 37.55,112.33);

\path[draw=drawColor,line width= 0.6pt,line join=round,line cap=round] ( 47.00, 56.43) -- ( 47.00, 97.69);

\path[draw=drawColor,line width= 0.6pt,line join=round,line cap=round] ( 56.45, 56.43) -- ( 56.45,110.98);

\path[draw=drawColor,line width= 0.6pt,line join=round,line cap=round] ( 65.89, 56.43) -- ( 65.89,106.23);

\path[draw=drawColor,line width= 0.6pt,line join=round,line cap=round] ( 75.34, 56.43) -- ( 75.34, 82.09);

\path[draw=drawColor,line width= 0.6pt,line join=round,line cap=round] ( 84.79, 56.43) -- ( 84.79, 76.08);

\path[draw=drawColor,line width= 0.6pt,line join=round,line cap=round] ( 94.23, 56.43) -- ( 94.23, 80.93);

\path[draw=drawColor,line width= 0.6pt,line join=round,line cap=round] (103.68, 56.43) -- (103.68, 65.25);

\path[draw=drawColor,line width= 0.6pt,line join=round,line cap=round] (113.13, 56.43) -- (113.13, 52.79);

\path[draw=drawColor,line width= 0.6pt,line join=round,line cap=round] (122.58, 56.43) -- (122.58, 48.20);

\path[draw=drawColor,line width= 0.6pt,line join=round,line cap=round] (132.02, 56.43) -- (132.02, 57.42);

\path[draw=drawColor,line width= 0.6pt,line join=round,line cap=round] (141.47, 56.43) -- (141.47, 60.52);

\path[draw=drawColor,line width= 0.6pt,line join=round,line cap=round] (150.92, 56.43) -- (150.92, 37.05);

\path[draw=drawColor,line width= 0.6pt,line join=round,line cap=round] (160.36, 56.43) -- (160.36, 56.41);

\path[draw=drawColor,line width= 0.6pt,line join=round,line cap=round] (169.81, 56.43) -- (169.81, 61.65);

\path[draw=drawColor,line width= 0.6pt,line join=round,line cap=round] (179.26, 56.43) -- (179.26, 61.79);

\path[draw=drawColor,line width= 0.6pt,line join=round,line cap=round] (188.71, 56.43) -- (188.71, 62.20);
\end{scope}
\begin{scope}
\path[clip] (  0.00,  0.00) rectangle (195.13,227.65);
\definecolor{drawColor}{RGB}{0,0,0}

\path[draw=drawColor,line width= 0.4pt,line join=round,line cap=round] ( 28.11, 21.68) -- (169.81, 21.68);

\path[draw=drawColor,line width= 0.4pt,line join=round,line cap=round] ( 28.11, 21.68) -- ( 28.11, 15.68);

\path[draw=drawColor,line width= 0.4pt,line join=round,line cap=round] ( 75.34, 21.68) -- ( 75.34, 15.68);

\path[draw=drawColor,line width= 0.4pt,line join=round,line cap=round] (122.58, 21.68) -- (122.58, 15.68);

\path[draw=drawColor,line width= 0.4pt,line join=round,line cap=round] (169.81, 21.68) -- (169.81, 15.68);

\node[text=drawColor,anchor=base,inner sep=0pt, outer sep=0pt, scale=  1.00] at ( 28.11,  2.48) {0};

\node[text=drawColor,anchor=base,inner sep=0pt, outer sep=0pt, scale=  1.00] at ( 75.34,  2.48) {5};

\node[text=drawColor,anchor=base,inner sep=0pt, outer sep=0pt, scale=  1.00] at (122.58,  2.48) {10};

\node[text=drawColor,anchor=base,inner sep=0pt, outer sep=0pt, scale=  1.00] at (169.81,  2.48) {15};

\path[draw=drawColor,line width= 0.4pt,line join=round,line cap=round] ( 21.68, 29.98) -- ( 21.68,188.70);

\path[draw=drawColor,line width= 0.4pt,line join=round,line cap=round] ( 21.68, 29.98) -- ( 15.68, 29.98);

\path[draw=drawColor,line width= 0.4pt,line join=round,line cap=round] ( 21.68, 56.43) -- ( 15.68, 56.43);

\path[draw=drawColor,line width= 0.4pt,line join=round,line cap=round] ( 21.68, 82.89) -- ( 15.68, 82.89);

\path[draw=drawColor,line width= 0.4pt,line join=round,line cap=round] ( 21.68,109.34) -- ( 15.68,109.34);

\path[draw=drawColor,line width= 0.4pt,line join=round,line cap=round] ( 21.68,135.80) -- ( 15.68,135.80);

\path[draw=drawColor,line width= 0.4pt,line join=round,line cap=round] ( 21.68,162.25) -- ( 15.68,162.25);

\path[draw=drawColor,line width= 0.4pt,line join=round,line cap=round] ( 21.68,188.70) -- ( 15.68,188.70);

\node[text=drawColor,rotate= 90.00,anchor=base,inner sep=0pt, outer sep=0pt, scale=  1.00] at (  9.68, 29.98) {-0.2};

\node[text=drawColor,rotate= 90.00,anchor=base,inner sep=0pt, outer sep=0pt, scale=  1.00] at (  9.68, 56.43) {0.0};

\node[text=drawColor,rotate= 90.00,anchor=base,inner sep=0pt, outer sep=0pt, scale=  1.00] at (  9.68, 82.89) {0.2};

\node[text=drawColor,rotate= 90.00,anchor=base,inner sep=0pt, outer sep=0pt, scale=  1.00] at (  9.68,109.34) {0.4};

\node[text=drawColor,rotate= 90.00,anchor=base,inner sep=0pt, outer sep=0pt, scale=  1.00] at (  9.68,135.80) {0.6};

\node[text=drawColor,rotate= 90.00,anchor=base,inner sep=0pt, outer sep=0pt, scale=  1.00] at (  9.68,162.25) {0.8};

\node[text=drawColor,rotate= 90.00,anchor=base,inner sep=0pt, outer sep=0pt, scale=  1.00] at (  9.68,188.70) {1.0};

\path[draw=drawColor,line width= 0.6pt,line join=round,line cap=round] ( 21.68, 21.68) --
	(195.13, 21.68) --
	(195.13,195.13) --
	( 21.68,195.13) --
	( 21.68, 21.68);
\end{scope}
\begin{scope}
\path[clip] (  0.00,  0.00) rectangle (195.13,227.65);
\definecolor{drawColor}{RGB}{0,0,0}

\node[text=drawColor,anchor=base,inner sep=0pt, outer sep=0pt, scale=  1.00] at (108.41, 30.08) {lag};
\end{scope}
\begin{scope}
\path[clip] ( 21.68, 21.68) rectangle (195.13,195.13);
\definecolor{drawColor}{RGB}{0,0,0}

\path[draw=drawColor,line width= 0.6pt,line join=round,line cap=round] ( 21.68, 56.43) -- (195.13, 56.43);
\definecolor{drawColor}{RGB}{0,0,255}

\path[draw=drawColor,line width= 0.6pt,dash pattern=on 4pt off 4pt ,line join=round,line cap=round] ( 21.68, 84.76) -- (195.13, 84.76);

\path[draw=drawColor,line width= 0.6pt,dash pattern=on 4pt off 4pt ,line join=round,line cap=round] ( 21.68, 28.11) -- (195.13, 28.11);
\end{scope}
\begin{scope}
\path[clip] (  0.00,  0.00) rectangle (195.13,227.65);
\definecolor{drawColor}{RGB}{0,0,0}

\node[text=drawColor,anchor=base,inner sep=0pt, outer sep=0pt, scale=  1.50] at (108.41,207.13) {\bfseries (c) ACF of $\protect\boldmath{(\text{PIT} - \frac12)^2}$};
\end{scope}
\end{tikzpicture}} \\ \smallskip
	\scalebox{\scalingFactor}{\input{figs/tikz/relDiag_marg_BoE2.tex}}
	\scalebox{\scalingFactor}{\input{figs/tikz/relDiag_thresh_BoE2.tex}}
	\scalebox{\scalingFactor}{\input{figs/tikz/relDiag_quant_BoE2.tex}}
	\caption{Same as Figure \ref{fig:0}, but at a prediction horizon of
		two quarters.  \label{fig:2}}
\end{figure}

\begin{figure}[p]  
	\centering
	\scalebox{\scalingFactor}{\input{figs/tikz/relDiag_PIT_BoE3.tex}}
	\scalebox{\scalingFactor}{% Created by tikzDevice version 0.12.3.1 on 2021-08-06 11:48:54
% !TEX encoding = UTF-8 Unicode
\begin{tikzpicture}[x=1pt,y=1pt]
\definecolor{fillColor}{RGB}{255,255,255}
\path[use as bounding box,fill=fillColor,fill opacity=0.00] (0,0) rectangle (195.13,227.65);
\begin{scope}
\path[clip] ( 21.68, 21.68) rectangle (195.13,195.13);
\definecolor{drawColor}{RGB}{0,0,0}

\path[draw=drawColor,line width= 0.6pt,line join=round,line cap=round] ( 28.11, 56.43) -- ( 28.11,188.70);

\path[draw=drawColor,line width= 0.6pt,line join=round,line cap=round] ( 37.55, 56.43) -- ( 37.55,149.52);

\path[draw=drawColor,line width= 0.6pt,line join=round,line cap=round] ( 47.00, 56.43) -- ( 47.00, 99.73);

\path[draw=drawColor,line width= 0.6pt,line join=round,line cap=round] ( 56.45, 56.43) -- ( 56.45, 74.64);

\path[draw=drawColor,line width= 0.6pt,line join=round,line cap=round] ( 65.89, 56.43) -- ( 65.89, 73.24);

\path[draw=drawColor,line width= 0.6pt,line join=round,line cap=round] ( 75.34, 56.43) -- ( 75.34, 86.49);

\path[draw=drawColor,line width= 0.6pt,line join=round,line cap=round] ( 84.79, 56.43) -- ( 84.79, 96.52);

\path[draw=drawColor,line width= 0.6pt,line join=round,line cap=round] ( 94.23, 56.43) -- ( 94.23, 87.40);

\path[draw=drawColor,line width= 0.6pt,line join=round,line cap=round] (103.68, 56.43) -- (103.68, 68.42);

\path[draw=drawColor,line width= 0.6pt,line join=round,line cap=round] (113.13, 56.43) -- (113.13, 54.59);

\path[draw=drawColor,line width= 0.6pt,line join=round,line cap=round] (122.58, 56.43) -- (122.58, 54.22);

\path[draw=drawColor,line width= 0.6pt,line join=round,line cap=round] (132.02, 56.43) -- (132.02, 66.34);

\path[draw=drawColor,line width= 0.6pt,line join=round,line cap=round] (141.47, 56.43) -- (141.47, 80.34);

\path[draw=drawColor,line width= 0.6pt,line join=round,line cap=round] (150.92, 56.43) -- (150.92, 67.96);

\path[draw=drawColor,line width= 0.6pt,line join=round,line cap=round] (160.36, 56.43) -- (160.36, 43.16);

\path[draw=drawColor,line width= 0.6pt,line join=round,line cap=round] (169.81, 56.43) -- (169.81, 37.10);

\path[draw=drawColor,line width= 0.6pt,line join=round,line cap=round] (179.26, 56.43) -- (179.26, 48.67);

\path[draw=drawColor,line width= 0.6pt,line join=round,line cap=round] (188.71, 56.43) -- (188.71, 61.66);
\end{scope}
\begin{scope}
\path[clip] (  0.00,  0.00) rectangle (195.13,227.65);
\definecolor{drawColor}{RGB}{0,0,0}

\path[draw=drawColor,line width= 0.4pt,line join=round,line cap=round] ( 28.11, 21.68) -- (169.81, 21.68);

\path[draw=drawColor,line width= 0.4pt,line join=round,line cap=round] ( 28.11, 21.68) -- ( 28.11, 15.68);

\path[draw=drawColor,line width= 0.4pt,line join=round,line cap=round] ( 75.34, 21.68) -- ( 75.34, 15.68);

\path[draw=drawColor,line width= 0.4pt,line join=round,line cap=round] (122.58, 21.68) -- (122.58, 15.68);

\path[draw=drawColor,line width= 0.4pt,line join=round,line cap=round] (169.81, 21.68) -- (169.81, 15.68);

\node[text=drawColor,anchor=base,inner sep=0pt, outer sep=0pt, scale=  1.00] at ( 28.11,  2.48) {0};

\node[text=drawColor,anchor=base,inner sep=0pt, outer sep=0pt, scale=  1.00] at ( 75.34,  2.48) {5};

\node[text=drawColor,anchor=base,inner sep=0pt, outer sep=0pt, scale=  1.00] at (122.58,  2.48) {10};

\node[text=drawColor,anchor=base,inner sep=0pt, outer sep=0pt, scale=  1.00] at (169.81,  2.48) {15};

\path[draw=drawColor,line width= 0.4pt,line join=round,line cap=round] ( 21.68, 29.98) -- ( 21.68,188.70);

\path[draw=drawColor,line width= 0.4pt,line join=round,line cap=round] ( 21.68, 29.98) -- ( 15.68, 29.98);

\path[draw=drawColor,line width= 0.4pt,line join=round,line cap=round] ( 21.68, 56.43) -- ( 15.68, 56.43);

\path[draw=drawColor,line width= 0.4pt,line join=round,line cap=round] ( 21.68, 82.89) -- ( 15.68, 82.89);

\path[draw=drawColor,line width= 0.4pt,line join=round,line cap=round] ( 21.68,109.34) -- ( 15.68,109.34);

\path[draw=drawColor,line width= 0.4pt,line join=round,line cap=round] ( 21.68,135.80) -- ( 15.68,135.80);

\path[draw=drawColor,line width= 0.4pt,line join=round,line cap=round] ( 21.68,162.25) -- ( 15.68,162.25);

\path[draw=drawColor,line width= 0.4pt,line join=round,line cap=round] ( 21.68,188.70) -- ( 15.68,188.70);

\node[text=drawColor,rotate= 90.00,anchor=base,inner sep=0pt, outer sep=0pt, scale=  1.00] at (  9.68, 29.98) {-0.2};

\node[text=drawColor,rotate= 90.00,anchor=base,inner sep=0pt, outer sep=0pt, scale=  1.00] at (  9.68, 56.43) {0.0};

\node[text=drawColor,rotate= 90.00,anchor=base,inner sep=0pt, outer sep=0pt, scale=  1.00] at (  9.68, 82.89) {0.2};

\node[text=drawColor,rotate= 90.00,anchor=base,inner sep=0pt, outer sep=0pt, scale=  1.00] at (  9.68,109.34) {0.4};

\node[text=drawColor,rotate= 90.00,anchor=base,inner sep=0pt, outer sep=0pt, scale=  1.00] at (  9.68,135.80) {0.6};

\node[text=drawColor,rotate= 90.00,anchor=base,inner sep=0pt, outer sep=0pt, scale=  1.00] at (  9.68,162.25) {0.8};

\node[text=drawColor,rotate= 90.00,anchor=base,inner sep=0pt, outer sep=0pt, scale=  1.00] at (  9.68,188.70) {1.0};

\path[draw=drawColor,line width= 0.6pt,line join=round,line cap=round] ( 21.68, 21.68) --
	(195.13, 21.68) --
	(195.13,195.13) --
	( 21.68,195.13) --
	( 21.68, 21.68);
\end{scope}
\begin{scope}
\path[clip] (  0.00,  0.00) rectangle (195.13,227.65);
\definecolor{drawColor}{RGB}{0,0,0}

\node[text=drawColor,anchor=base,inner sep=0pt, outer sep=0pt, scale=  1.00] at (108.41, 30.08) {lag};
\end{scope}
\begin{scope}
\path[clip] ( 21.68, 21.68) rectangle (195.13,195.13);
\definecolor{drawColor}{RGB}{0,0,0}

\path[draw=drawColor,line width= 0.6pt,line join=round,line cap=round] ( 21.68, 56.43) -- (195.13, 56.43);
\definecolor{drawColor}{RGB}{0,0,255}

\path[draw=drawColor,line width= 0.6pt,dash pattern=on 4pt off 4pt ,line join=round,line cap=round] ( 21.68, 84.76) -- (195.13, 84.76);

\path[draw=drawColor,line width= 0.6pt,dash pattern=on 4pt off 4pt ,line join=round,line cap=round] ( 21.68, 28.11) -- (195.13, 28.11);
\end{scope}
\begin{scope}
\path[clip] (  0.00,  0.00) rectangle (195.13,227.65);
\definecolor{drawColor}{RGB}{0,0,0}

\node[text=drawColor,anchor=base,inner sep=0pt, outer sep=0pt, scale=  1.50] at (108.41,207.13) {\bfseries (b) ACF of PIT};
\end{scope}
\end{tikzpicture}}
	\scalebox{\scalingFactor}{% Created by tikzDevice version 0.12.3.1 on 2021-08-06 11:48:54
% !TEX encoding = UTF-8 Unicode
\begin{tikzpicture}[x=1pt,y=1pt]
\definecolor{fillColor}{RGB}{255,255,255}
\path[use as bounding box,fill=fillColor,fill opacity=0.00] (0,0) rectangle (195.13,227.65);
\begin{scope}
\path[clip] ( 21.68, 21.68) rectangle (195.13,195.13);
\definecolor{drawColor}{RGB}{0,0,0}

\path[draw=drawColor,line width= 0.6pt,line join=round,line cap=round] ( 28.11, 56.43) -- ( 28.11,188.70);

\path[draw=drawColor,line width= 0.6pt,line join=round,line cap=round] ( 37.55, 56.43) -- ( 37.55,135.25);

\path[draw=drawColor,line width= 0.6pt,line join=round,line cap=round] ( 47.00, 56.43) -- ( 47.00,107.17);

\path[draw=drawColor,line width= 0.6pt,line join=round,line cap=round] ( 56.45, 56.43) -- ( 56.45, 89.62);

\path[draw=drawColor,line width= 0.6pt,line join=round,line cap=round] ( 65.89, 56.43) -- ( 65.89, 93.77);

\path[draw=drawColor,line width= 0.6pt,line join=round,line cap=round] ( 75.34, 56.43) -- ( 75.34, 96.62);

\path[draw=drawColor,line width= 0.6pt,line join=round,line cap=round] ( 84.79, 56.43) -- ( 84.79, 92.99);

\path[draw=drawColor,line width= 0.6pt,line join=round,line cap=round] ( 94.23, 56.43) -- ( 94.23, 89.00);

\path[draw=drawColor,line width= 0.6pt,line join=round,line cap=round] (103.68, 56.43) -- (103.68, 67.67);

\path[draw=drawColor,line width= 0.6pt,line join=round,line cap=round] (113.13, 56.43) -- (113.13, 67.22);

\path[draw=drawColor,line width= 0.6pt,line join=round,line cap=round] (122.58, 56.43) -- (122.58, 57.72);

\path[draw=drawColor,line width= 0.6pt,line join=round,line cap=round] (132.02, 56.43) -- (132.02, 61.70);

\path[draw=drawColor,line width= 0.6pt,line join=round,line cap=round] (141.47, 56.43) -- (141.47, 59.72);

\path[draw=drawColor,line width= 0.6pt,line join=round,line cap=round] (150.92, 56.43) -- (150.92, 51.42);

\path[draw=drawColor,line width= 0.6pt,line join=round,line cap=round] (160.36, 56.43) -- (160.36, 54.44);

\path[draw=drawColor,line width= 0.6pt,line join=round,line cap=round] (169.81, 56.43) -- (169.81, 53.82);

\path[draw=drawColor,line width= 0.6pt,line join=round,line cap=round] (179.26, 56.43) -- (179.26, 60.77);

\path[draw=drawColor,line width= 0.6pt,line join=round,line cap=round] (188.71, 56.43) -- (188.71, 70.42);
\end{scope}
\begin{scope}
\path[clip] (  0.00,  0.00) rectangle (195.13,227.65);
\definecolor{drawColor}{RGB}{0,0,0}

\path[draw=drawColor,line width= 0.4pt,line join=round,line cap=round] ( 28.11, 21.68) -- (169.81, 21.68);

\path[draw=drawColor,line width= 0.4pt,line join=round,line cap=round] ( 28.11, 21.68) -- ( 28.11, 15.68);

\path[draw=drawColor,line width= 0.4pt,line join=round,line cap=round] ( 75.34, 21.68) -- ( 75.34, 15.68);

\path[draw=drawColor,line width= 0.4pt,line join=round,line cap=round] (122.58, 21.68) -- (122.58, 15.68);

\path[draw=drawColor,line width= 0.4pt,line join=round,line cap=round] (169.81, 21.68) -- (169.81, 15.68);

\node[text=drawColor,anchor=base,inner sep=0pt, outer sep=0pt, scale=  1.00] at ( 28.11,  2.48) {0};

\node[text=drawColor,anchor=base,inner sep=0pt, outer sep=0pt, scale=  1.00] at ( 75.34,  2.48) {5};

\node[text=drawColor,anchor=base,inner sep=0pt, outer sep=0pt, scale=  1.00] at (122.58,  2.48) {10};

\node[text=drawColor,anchor=base,inner sep=0pt, outer sep=0pt, scale=  1.00] at (169.81,  2.48) {15};

\path[draw=drawColor,line width= 0.4pt,line join=round,line cap=round] ( 21.68, 29.98) -- ( 21.68,188.70);

\path[draw=drawColor,line width= 0.4pt,line join=round,line cap=round] ( 21.68, 29.98) -- ( 15.68, 29.98);

\path[draw=drawColor,line width= 0.4pt,line join=round,line cap=round] ( 21.68, 56.43) -- ( 15.68, 56.43);

\path[draw=drawColor,line width= 0.4pt,line join=round,line cap=round] ( 21.68, 82.89) -- ( 15.68, 82.89);

\path[draw=drawColor,line width= 0.4pt,line join=round,line cap=round] ( 21.68,109.34) -- ( 15.68,109.34);

\path[draw=drawColor,line width= 0.4pt,line join=round,line cap=round] ( 21.68,135.80) -- ( 15.68,135.80);

\path[draw=drawColor,line width= 0.4pt,line join=round,line cap=round] ( 21.68,162.25) -- ( 15.68,162.25);

\path[draw=drawColor,line width= 0.4pt,line join=round,line cap=round] ( 21.68,188.70) -- ( 15.68,188.70);

\node[text=drawColor,rotate= 90.00,anchor=base,inner sep=0pt, outer sep=0pt, scale=  1.00] at (  9.68, 29.98) {-0.2};

\node[text=drawColor,rotate= 90.00,anchor=base,inner sep=0pt, outer sep=0pt, scale=  1.00] at (  9.68, 56.43) {0.0};

\node[text=drawColor,rotate= 90.00,anchor=base,inner sep=0pt, outer sep=0pt, scale=  1.00] at (  9.68, 82.89) {0.2};

\node[text=drawColor,rotate= 90.00,anchor=base,inner sep=0pt, outer sep=0pt, scale=  1.00] at (  9.68,109.34) {0.4};

\node[text=drawColor,rotate= 90.00,anchor=base,inner sep=0pt, outer sep=0pt, scale=  1.00] at (  9.68,135.80) {0.6};

\node[text=drawColor,rotate= 90.00,anchor=base,inner sep=0pt, outer sep=0pt, scale=  1.00] at (  9.68,162.25) {0.8};

\node[text=drawColor,rotate= 90.00,anchor=base,inner sep=0pt, outer sep=0pt, scale=  1.00] at (  9.68,188.70) {1.0};

\path[draw=drawColor,line width= 0.6pt,line join=round,line cap=round] ( 21.68, 21.68) --
	(195.13, 21.68) --
	(195.13,195.13) --
	( 21.68,195.13) --
	( 21.68, 21.68);
\end{scope}
\begin{scope}
\path[clip] (  0.00,  0.00) rectangle (195.13,227.65);
\definecolor{drawColor}{RGB}{0,0,0}

\node[text=drawColor,anchor=base,inner sep=0pt, outer sep=0pt, scale=  1.00] at (108.41, 30.08) {lag};
\end{scope}
\begin{scope}
\path[clip] ( 21.68, 21.68) rectangle (195.13,195.13);
\definecolor{drawColor}{RGB}{0,0,0}

\path[draw=drawColor,line width= 0.6pt,line join=round,line cap=round] ( 21.68, 56.43) -- (195.13, 56.43);
\definecolor{drawColor}{RGB}{0,0,255}

\path[draw=drawColor,line width= 0.6pt,dash pattern=on 4pt off 4pt ,line join=round,line cap=round] ( 21.68, 84.76) -- (195.13, 84.76);

\path[draw=drawColor,line width= 0.6pt,dash pattern=on 4pt off 4pt ,line join=round,line cap=round] ( 21.68, 28.11) -- (195.13, 28.11);
\end{scope}
\begin{scope}
\path[clip] (  0.00,  0.00) rectangle (195.13,227.65);
\definecolor{drawColor}{RGB}{0,0,0}

\node[text=drawColor,anchor=base,inner sep=0pt, outer sep=0pt, scale=  1.50] at (108.41,207.13) {\bfseries (c) ACF of $\protect\boldmath{(\text{PIT} - \frac12)^2}$};
\end{scope}
\end{tikzpicture}} \\ \smallskip
	\scalebox{\scalingFactor}{\input{figs/tikz/relDiag_marg_BoE3.tex}}
	\scalebox{\scalingFactor}{\input{figs/tikz/relDiag_thresh_BoE3.tex}}
	\scalebox{\scalingFactor}{\input{figs/tikz/relDiag_quant_BoE3.tex}}
	\caption{Same as Figure \ref{fig:0}, but at a prediction horizon of
		three quarters.  \label{fig:3}}
	\bigskip
	\scalebox{\scalingFactor}{\input{figs/tikz/relDiag_PIT_BoE4.tex}}
	\scalebox{\scalingFactor}{% Created by tikzDevice version 0.12.3.1 on 2021-08-06 11:49:08
% !TEX encoding = UTF-8 Unicode
\begin{tikzpicture}[x=1pt,y=1pt]
\definecolor{fillColor}{RGB}{255,255,255}
\path[use as bounding box,fill=fillColor,fill opacity=0.00] (0,0) rectangle (195.13,227.65);
\begin{scope}
\path[clip] ( 21.68, 21.68) rectangle (195.13,195.13);
\definecolor{drawColor}{RGB}{0,0,0}

\path[draw=drawColor,line width= 0.6pt,line join=round,line cap=round] ( 28.11, 56.43) -- ( 28.11,188.70);

\path[draw=drawColor,line width= 0.6pt,line join=round,line cap=round] ( 37.55, 56.43) -- ( 37.55,153.72);

\path[draw=drawColor,line width= 0.6pt,line join=round,line cap=round] ( 47.00, 56.43) -- ( 47.00,109.72);

\path[draw=drawColor,line width= 0.6pt,line join=round,line cap=round] ( 56.45, 56.43) -- ( 56.45, 89.32);

\path[draw=drawColor,line width= 0.6pt,line join=round,line cap=round] ( 65.89, 56.43) -- ( 65.89, 81.20);

\path[draw=drawColor,line width= 0.6pt,line join=round,line cap=round] ( 75.34, 56.43) -- ( 75.34, 80.11);

\path[draw=drawColor,line width= 0.6pt,line join=round,line cap=round] ( 84.79, 56.43) -- ( 84.79, 89.60);

\path[draw=drawColor,line width= 0.6pt,line join=round,line cap=round] ( 94.23, 56.43) -- ( 94.23, 92.17);

\path[draw=drawColor,line width= 0.6pt,line join=round,line cap=round] (103.68, 56.43) -- (103.68, 75.73);

\path[draw=drawColor,line width= 0.6pt,line join=round,line cap=round] (113.13, 56.43) -- (113.13, 61.53);

\path[draw=drawColor,line width= 0.6pt,line join=round,line cap=round] (122.58, 56.43) -- (122.58, 58.39);

\path[draw=drawColor,line width= 0.6pt,line join=round,line cap=round] (132.02, 56.43) -- (132.02, 66.23);

\path[draw=drawColor,line width= 0.6pt,line join=round,line cap=round] (141.47, 56.43) -- (141.47, 71.48);

\path[draw=drawColor,line width= 0.6pt,line join=round,line cap=round] (150.92, 56.43) -- (150.92, 65.24);

\path[draw=drawColor,line width= 0.6pt,line join=round,line cap=round] (160.36, 56.43) -- (160.36, 49.91);

\path[draw=drawColor,line width= 0.6pt,line join=round,line cap=round] (169.81, 56.43) -- (169.81, 43.80);

\path[draw=drawColor,line width= 0.6pt,line join=round,line cap=round] (179.26, 56.43) -- (179.26, 50.62);

\path[draw=drawColor,line width= 0.6pt,line join=round,line cap=round] (188.71, 56.43) -- (188.71, 61.77);
\end{scope}
\begin{scope}
\path[clip] (  0.00,  0.00) rectangle (195.13,227.65);
\definecolor{drawColor}{RGB}{0,0,0}

\path[draw=drawColor,line width= 0.4pt,line join=round,line cap=round] ( 28.11, 21.68) -- (169.81, 21.68);

\path[draw=drawColor,line width= 0.4pt,line join=round,line cap=round] ( 28.11, 21.68) -- ( 28.11, 15.68);

\path[draw=drawColor,line width= 0.4pt,line join=round,line cap=round] ( 75.34, 21.68) -- ( 75.34, 15.68);

\path[draw=drawColor,line width= 0.4pt,line join=round,line cap=round] (122.58, 21.68) -- (122.58, 15.68);

\path[draw=drawColor,line width= 0.4pt,line join=round,line cap=round] (169.81, 21.68) -- (169.81, 15.68);

\node[text=drawColor,anchor=base,inner sep=0pt, outer sep=0pt, scale=  1.00] at ( 28.11,  2.48) {0};

\node[text=drawColor,anchor=base,inner sep=0pt, outer sep=0pt, scale=  1.00] at ( 75.34,  2.48) {5};

\node[text=drawColor,anchor=base,inner sep=0pt, outer sep=0pt, scale=  1.00] at (122.58,  2.48) {10};

\node[text=drawColor,anchor=base,inner sep=0pt, outer sep=0pt, scale=  1.00] at (169.81,  2.48) {15};

\path[draw=drawColor,line width= 0.4pt,line join=round,line cap=round] ( 21.68, 29.98) -- ( 21.68,188.70);

\path[draw=drawColor,line width= 0.4pt,line join=round,line cap=round] ( 21.68, 29.98) -- ( 15.68, 29.98);

\path[draw=drawColor,line width= 0.4pt,line join=round,line cap=round] ( 21.68, 56.43) -- ( 15.68, 56.43);

\path[draw=drawColor,line width= 0.4pt,line join=round,line cap=round] ( 21.68, 82.89) -- ( 15.68, 82.89);

\path[draw=drawColor,line width= 0.4pt,line join=round,line cap=round] ( 21.68,109.34) -- ( 15.68,109.34);

\path[draw=drawColor,line width= 0.4pt,line join=round,line cap=round] ( 21.68,135.80) -- ( 15.68,135.80);

\path[draw=drawColor,line width= 0.4pt,line join=round,line cap=round] ( 21.68,162.25) -- ( 15.68,162.25);

\path[draw=drawColor,line width= 0.4pt,line join=round,line cap=round] ( 21.68,188.70) -- ( 15.68,188.70);

\node[text=drawColor,rotate= 90.00,anchor=base,inner sep=0pt, outer sep=0pt, scale=  1.00] at (  9.68, 29.98) {-0.2};

\node[text=drawColor,rotate= 90.00,anchor=base,inner sep=0pt, outer sep=0pt, scale=  1.00] at (  9.68, 56.43) {0.0};

\node[text=drawColor,rotate= 90.00,anchor=base,inner sep=0pt, outer sep=0pt, scale=  1.00] at (  9.68, 82.89) {0.2};

\node[text=drawColor,rotate= 90.00,anchor=base,inner sep=0pt, outer sep=0pt, scale=  1.00] at (  9.68,109.34) {0.4};

\node[text=drawColor,rotate= 90.00,anchor=base,inner sep=0pt, outer sep=0pt, scale=  1.00] at (  9.68,135.80) {0.6};

\node[text=drawColor,rotate= 90.00,anchor=base,inner sep=0pt, outer sep=0pt, scale=  1.00] at (  9.68,162.25) {0.8};

\node[text=drawColor,rotate= 90.00,anchor=base,inner sep=0pt, outer sep=0pt, scale=  1.00] at (  9.68,188.70) {1.0};

\path[draw=drawColor,line width= 0.6pt,line join=round,line cap=round] ( 21.68, 21.68) --
	(195.13, 21.68) --
	(195.13,195.13) --
	( 21.68,195.13) --
	( 21.68, 21.68);
\end{scope}
\begin{scope}
\path[clip] (  0.00,  0.00) rectangle (195.13,227.65);
\definecolor{drawColor}{RGB}{0,0,0}

\node[text=drawColor,anchor=base,inner sep=0pt, outer sep=0pt, scale=  1.00] at (108.41, 30.08) {lag};
\end{scope}
\begin{scope}
\path[clip] ( 21.68, 21.68) rectangle (195.13,195.13);
\definecolor{drawColor}{RGB}{0,0,0}

\path[draw=drawColor,line width= 0.6pt,line join=round,line cap=round] ( 21.68, 56.43) -- (195.13, 56.43);
\definecolor{drawColor}{RGB}{0,0,255}

\path[draw=drawColor,line width= 0.6pt,dash pattern=on 4pt off 4pt ,line join=round,line cap=round] ( 21.68, 84.76) -- (195.13, 84.76);

\path[draw=drawColor,line width= 0.6pt,dash pattern=on 4pt off 4pt ,line join=round,line cap=round] ( 21.68, 28.11) -- (195.13, 28.11);
\end{scope}
\begin{scope}
\path[clip] (  0.00,  0.00) rectangle (195.13,227.65);
\definecolor{drawColor}{RGB}{0,0,0}

\node[text=drawColor,anchor=base,inner sep=0pt, outer sep=0pt, scale=  1.50] at (108.41,207.13) {\bfseries (b) ACF of PIT};
\end{scope}
\end{tikzpicture}}
	\scalebox{\scalingFactor}{% Created by tikzDevice version 0.12.3.1 on 2021-08-06 11:49:08
% !TEX encoding = UTF-8 Unicode
\begin{tikzpicture}[x=1pt,y=1pt]
\definecolor{fillColor}{RGB}{255,255,255}
\path[use as bounding box,fill=fillColor,fill opacity=0.00] (0,0) rectangle (195.13,227.65);
\begin{scope}
\path[clip] ( 21.68, 21.68) rectangle (195.13,195.13);
\definecolor{drawColor}{RGB}{0,0,0}

\path[draw=drawColor,line width= 0.6pt,line join=round,line cap=round] ( 28.11, 56.43) -- ( 28.11,188.70);

\path[draw=drawColor,line width= 0.6pt,line join=round,line cap=round] ( 37.55, 56.43) -- ( 37.55,142.48);

\path[draw=drawColor,line width= 0.6pt,line join=round,line cap=round] ( 47.00, 56.43) -- ( 47.00,116.67);

\path[draw=drawColor,line width= 0.6pt,line join=round,line cap=round] ( 56.45, 56.43) -- ( 56.45,108.58);

\path[draw=drawColor,line width= 0.6pt,line join=round,line cap=round] ( 65.89, 56.43) -- ( 65.89, 98.92);

\path[draw=drawColor,line width= 0.6pt,line join=round,line cap=round] ( 75.34, 56.43) -- ( 75.34,106.70);

\path[draw=drawColor,line width= 0.6pt,line join=round,line cap=round] ( 84.79, 56.43) -- ( 84.79, 98.05);

\path[draw=drawColor,line width= 0.6pt,line join=round,line cap=round] ( 94.23, 56.43) -- ( 94.23, 98.42);

\path[draw=drawColor,line width= 0.6pt,line join=round,line cap=round] (103.68, 56.43) -- (103.68, 85.43);

\path[draw=drawColor,line width= 0.6pt,line join=round,line cap=round] (113.13, 56.43) -- (113.13, 70.10);

\path[draw=drawColor,line width= 0.6pt,line join=round,line cap=round] (122.58, 56.43) -- (122.58, 75.42);

\path[draw=drawColor,line width= 0.6pt,line join=round,line cap=round] (132.02, 56.43) -- (132.02, 69.50);

\path[draw=drawColor,line width= 0.6pt,line join=round,line cap=round] (141.47, 56.43) -- (141.47, 71.84);

\path[draw=drawColor,line width= 0.6pt,line join=round,line cap=round] (150.92, 56.43) -- (150.92, 61.44);

\path[draw=drawColor,line width= 0.6pt,line join=round,line cap=round] (160.36, 56.43) -- (160.36, 56.36);

\path[draw=drawColor,line width= 0.6pt,line join=round,line cap=round] (169.81, 56.43) -- (169.81, 66.35);

\path[draw=drawColor,line width= 0.6pt,line join=round,line cap=round] (179.26, 56.43) -- (179.26, 67.20);

\path[draw=drawColor,line width= 0.6pt,line join=round,line cap=round] (188.71, 56.43) -- (188.71, 70.90);
\end{scope}
\begin{scope}
\path[clip] (  0.00,  0.00) rectangle (195.13,227.65);
\definecolor{drawColor}{RGB}{0,0,0}

\path[draw=drawColor,line width= 0.4pt,line join=round,line cap=round] ( 28.11, 21.68) -- (169.81, 21.68);

\path[draw=drawColor,line width= 0.4pt,line join=round,line cap=round] ( 28.11, 21.68) -- ( 28.11, 15.68);

\path[draw=drawColor,line width= 0.4pt,line join=round,line cap=round] ( 75.34, 21.68) -- ( 75.34, 15.68);

\path[draw=drawColor,line width= 0.4pt,line join=round,line cap=round] (122.58, 21.68) -- (122.58, 15.68);

\path[draw=drawColor,line width= 0.4pt,line join=round,line cap=round] (169.81, 21.68) -- (169.81, 15.68);

\node[text=drawColor,anchor=base,inner sep=0pt, outer sep=0pt, scale=  1.00] at ( 28.11,  2.48) {0};

\node[text=drawColor,anchor=base,inner sep=0pt, outer sep=0pt, scale=  1.00] at ( 75.34,  2.48) {5};

\node[text=drawColor,anchor=base,inner sep=0pt, outer sep=0pt, scale=  1.00] at (122.58,  2.48) {10};

\node[text=drawColor,anchor=base,inner sep=0pt, outer sep=0pt, scale=  1.00] at (169.81,  2.48) {15};

\path[draw=drawColor,line width= 0.4pt,line join=round,line cap=round] ( 21.68, 29.98) -- ( 21.68,188.70);

\path[draw=drawColor,line width= 0.4pt,line join=round,line cap=round] ( 21.68, 29.98) -- ( 15.68, 29.98);

\path[draw=drawColor,line width= 0.4pt,line join=round,line cap=round] ( 21.68, 56.43) -- ( 15.68, 56.43);

\path[draw=drawColor,line width= 0.4pt,line join=round,line cap=round] ( 21.68, 82.89) -- ( 15.68, 82.89);

\path[draw=drawColor,line width= 0.4pt,line join=round,line cap=round] ( 21.68,109.34) -- ( 15.68,109.34);

\path[draw=drawColor,line width= 0.4pt,line join=round,line cap=round] ( 21.68,135.80) -- ( 15.68,135.80);

\path[draw=drawColor,line width= 0.4pt,line join=round,line cap=round] ( 21.68,162.25) -- ( 15.68,162.25);

\path[draw=drawColor,line width= 0.4pt,line join=round,line cap=round] ( 21.68,188.70) -- ( 15.68,188.70);

\node[text=drawColor,rotate= 90.00,anchor=base,inner sep=0pt, outer sep=0pt, scale=  1.00] at (  9.68, 29.98) {-0.2};

\node[text=drawColor,rotate= 90.00,anchor=base,inner sep=0pt, outer sep=0pt, scale=  1.00] at (  9.68, 56.43) {0.0};

\node[text=drawColor,rotate= 90.00,anchor=base,inner sep=0pt, outer sep=0pt, scale=  1.00] at (  9.68, 82.89) {0.2};

\node[text=drawColor,rotate= 90.00,anchor=base,inner sep=0pt, outer sep=0pt, scale=  1.00] at (  9.68,109.34) {0.4};

\node[text=drawColor,rotate= 90.00,anchor=base,inner sep=0pt, outer sep=0pt, scale=  1.00] at (  9.68,135.80) {0.6};

\node[text=drawColor,rotate= 90.00,anchor=base,inner sep=0pt, outer sep=0pt, scale=  1.00] at (  9.68,162.25) {0.8};

\node[text=drawColor,rotate= 90.00,anchor=base,inner sep=0pt, outer sep=0pt, scale=  1.00] at (  9.68,188.70) {1.0};

\path[draw=drawColor,line width= 0.6pt,line join=round,line cap=round] ( 21.68, 21.68) --
	(195.13, 21.68) --
	(195.13,195.13) --
	( 21.68,195.13) --
	( 21.68, 21.68);
\end{scope}
\begin{scope}
\path[clip] (  0.00,  0.00) rectangle (195.13,227.65);
\definecolor{drawColor}{RGB}{0,0,0}

\node[text=drawColor,anchor=base,inner sep=0pt, outer sep=0pt, scale=  1.00] at (108.41, 30.08) {lag};
\end{scope}
\begin{scope}
\path[clip] ( 21.68, 21.68) rectangle (195.13,195.13);
\definecolor{drawColor}{RGB}{0,0,0}

\path[draw=drawColor,line width= 0.6pt,line join=round,line cap=round] ( 21.68, 56.43) -- (195.13, 56.43);
\definecolor{drawColor}{RGB}{0,0,255}

\path[draw=drawColor,line width= 0.6pt,dash pattern=on 4pt off 4pt ,line join=round,line cap=round] ( 21.68, 84.76) -- (195.13, 84.76);

\path[draw=drawColor,line width= 0.6pt,dash pattern=on 4pt off 4pt ,line join=round,line cap=round] ( 21.68, 28.11) -- (195.13, 28.11);
\end{scope}
\begin{scope}
\path[clip] (  0.00,  0.00) rectangle (195.13,227.65);
\definecolor{drawColor}{RGB}{0,0,0}

\node[text=drawColor,anchor=base,inner sep=0pt, outer sep=0pt, scale=  1.50] at (108.41,207.13) {\bfseries (c) ACF of $\protect\boldmath{(\text{PIT} - \frac12)^2}$};
\end{scope}
\end{tikzpicture}} \\ \smallskip
	\scalebox{\scalingFactor}{\input{figs/tikz/relDiag_marg_BoE4.tex}}
	\scalebox{\scalingFactor}{\input{figs/tikz/relDiag_thresh_BoE4.tex}}
	\scalebox{\scalingFactor}{\input{figs/tikz/relDiag_quant_BoE4.tex}}
	\caption{Same as Figure \ref{fig:0}, but at a prediction horizon of
		four quarters.  \label{fig:4}}
\end{figure}

\begin{figure}[p]  
	\centering
	\scalebox{\scalingFactor}{\input{figs/tikz/relDiag_PIT_BoE5.tex}}
	\scalebox{\scalingFactor}{% Created by tikzDevice version 0.12.3.1 on 2021-08-06 11:49:22
% !TEX encoding = UTF-8 Unicode
\begin{tikzpicture}[x=1pt,y=1pt]
\definecolor{fillColor}{RGB}{255,255,255}
\path[use as bounding box,fill=fillColor,fill opacity=0.00] (0,0) rectangle (195.13,227.65);
\begin{scope}
\path[clip] ( 21.68, 21.68) rectangle (195.13,195.13);
\definecolor{drawColor}{RGB}{0,0,0}

\path[draw=drawColor,line width= 0.6pt,line join=round,line cap=round] ( 28.11, 56.43) -- ( 28.11,188.70);

\path[draw=drawColor,line width= 0.6pt,line join=round,line cap=round] ( 37.55, 56.43) -- ( 37.55,160.52);

\path[draw=drawColor,line width= 0.6pt,line join=round,line cap=round] ( 47.00, 56.43) -- ( 47.00,118.72);

\path[draw=drawColor,line width= 0.6pt,line join=round,line cap=round] ( 56.45, 56.43) -- ( 56.45, 93.10);

\path[draw=drawColor,line width= 0.6pt,line join=round,line cap=round] ( 65.89, 56.43) -- ( 65.89, 82.34);

\path[draw=drawColor,line width= 0.6pt,line join=round,line cap=round] ( 75.34, 56.43) -- ( 75.34, 78.71);

\path[draw=drawColor,line width= 0.6pt,line join=round,line cap=round] ( 84.79, 56.43) -- ( 84.79, 83.83);

\path[draw=drawColor,line width= 0.6pt,line join=round,line cap=round] ( 94.23, 56.43) -- ( 94.23, 86.62);

\path[draw=drawColor,line width= 0.6pt,line join=round,line cap=round] (103.68, 56.43) -- (103.68, 76.84);

\path[draw=drawColor,line width= 0.6pt,line join=round,line cap=round] (113.13, 56.43) -- (113.13, 62.47);

\path[draw=drawColor,line width= 0.6pt,line join=round,line cap=round] (122.58, 56.43) -- (122.58, 54.52);

\path[draw=drawColor,line width= 0.6pt,line join=round,line cap=round] (132.02, 56.43) -- (132.02, 61.26);

\path[draw=drawColor,line width= 0.6pt,line join=round,line cap=round] (141.47, 56.43) -- (141.47, 64.97);

\path[draw=drawColor,line width= 0.6pt,line join=round,line cap=round] (150.92, 56.43) -- (150.92, 59.63);

\path[draw=drawColor,line width= 0.6pt,line join=round,line cap=round] (160.36, 56.43) -- (160.36, 51.13);

\path[draw=drawColor,line width= 0.6pt,line join=round,line cap=round] (169.81, 56.43) -- (169.81, 47.10);

\path[draw=drawColor,line width= 0.6pt,line join=round,line cap=round] (179.26, 56.43) -- (179.26, 48.82);

\path[draw=drawColor,line width= 0.6pt,line join=round,line cap=round] (188.71, 56.43) -- (188.71, 58.66);
\end{scope}
\begin{scope}
\path[clip] (  0.00,  0.00) rectangle (195.13,227.65);
\definecolor{drawColor}{RGB}{0,0,0}

\path[draw=drawColor,line width= 0.4pt,line join=round,line cap=round] ( 28.11, 21.68) -- (169.81, 21.68);

\path[draw=drawColor,line width= 0.4pt,line join=round,line cap=round] ( 28.11, 21.68) -- ( 28.11, 15.68);

\path[draw=drawColor,line width= 0.4pt,line join=round,line cap=round] ( 75.34, 21.68) -- ( 75.34, 15.68);

\path[draw=drawColor,line width= 0.4pt,line join=round,line cap=round] (122.58, 21.68) -- (122.58, 15.68);

\path[draw=drawColor,line width= 0.4pt,line join=round,line cap=round] (169.81, 21.68) -- (169.81, 15.68);

\node[text=drawColor,anchor=base,inner sep=0pt, outer sep=0pt, scale=  1.00] at ( 28.11,  2.48) {0};

\node[text=drawColor,anchor=base,inner sep=0pt, outer sep=0pt, scale=  1.00] at ( 75.34,  2.48) {5};

\node[text=drawColor,anchor=base,inner sep=0pt, outer sep=0pt, scale=  1.00] at (122.58,  2.48) {10};

\node[text=drawColor,anchor=base,inner sep=0pt, outer sep=0pt, scale=  1.00] at (169.81,  2.48) {15};

\path[draw=drawColor,line width= 0.4pt,line join=round,line cap=round] ( 21.68, 29.98) -- ( 21.68,188.70);

\path[draw=drawColor,line width= 0.4pt,line join=round,line cap=round] ( 21.68, 29.98) -- ( 15.68, 29.98);

\path[draw=drawColor,line width= 0.4pt,line join=round,line cap=round] ( 21.68, 56.43) -- ( 15.68, 56.43);

\path[draw=drawColor,line width= 0.4pt,line join=round,line cap=round] ( 21.68, 82.89) -- ( 15.68, 82.89);

\path[draw=drawColor,line width= 0.4pt,line join=round,line cap=round] ( 21.68,109.34) -- ( 15.68,109.34);

\path[draw=drawColor,line width= 0.4pt,line join=round,line cap=round] ( 21.68,135.80) -- ( 15.68,135.80);

\path[draw=drawColor,line width= 0.4pt,line join=round,line cap=round] ( 21.68,162.25) -- ( 15.68,162.25);

\path[draw=drawColor,line width= 0.4pt,line join=round,line cap=round] ( 21.68,188.70) -- ( 15.68,188.70);

\node[text=drawColor,rotate= 90.00,anchor=base,inner sep=0pt, outer sep=0pt, scale=  1.00] at (  9.68, 29.98) {-0.2};

\node[text=drawColor,rotate= 90.00,anchor=base,inner sep=0pt, outer sep=0pt, scale=  1.00] at (  9.68, 56.43) {0.0};

\node[text=drawColor,rotate= 90.00,anchor=base,inner sep=0pt, outer sep=0pt, scale=  1.00] at (  9.68, 82.89) {0.2};

\node[text=drawColor,rotate= 90.00,anchor=base,inner sep=0pt, outer sep=0pt, scale=  1.00] at (  9.68,109.34) {0.4};

\node[text=drawColor,rotate= 90.00,anchor=base,inner sep=0pt, outer sep=0pt, scale=  1.00] at (  9.68,135.80) {0.6};

\node[text=drawColor,rotate= 90.00,anchor=base,inner sep=0pt, outer sep=0pt, scale=  1.00] at (  9.68,162.25) {0.8};

\node[text=drawColor,rotate= 90.00,anchor=base,inner sep=0pt, outer sep=0pt, scale=  1.00] at (  9.68,188.70) {1.0};

\path[draw=drawColor,line width= 0.6pt,line join=round,line cap=round] ( 21.68, 21.68) --
	(195.13, 21.68) --
	(195.13,195.13) --
	( 21.68,195.13) --
	( 21.68, 21.68);
\end{scope}
\begin{scope}
\path[clip] (  0.00,  0.00) rectangle (195.13,227.65);
\definecolor{drawColor}{RGB}{0,0,0}

\node[text=drawColor,anchor=base,inner sep=0pt, outer sep=0pt, scale=  1.00] at (108.41, 30.08) {lag};
\end{scope}
\begin{scope}
\path[clip] ( 21.68, 21.68) rectangle (195.13,195.13);
\definecolor{drawColor}{RGB}{0,0,0}

\path[draw=drawColor,line width= 0.6pt,line join=round,line cap=round] ( 21.68, 56.43) -- (195.13, 56.43);
\definecolor{drawColor}{RGB}{0,0,255}

\path[draw=drawColor,line width= 0.6pt,dash pattern=on 4pt off 4pt ,line join=round,line cap=round] ( 21.68, 84.76) -- (195.13, 84.76);

\path[draw=drawColor,line width= 0.6pt,dash pattern=on 4pt off 4pt ,line join=round,line cap=round] ( 21.68, 28.11) -- (195.13, 28.11);
\end{scope}
\begin{scope}
\path[clip] (  0.00,  0.00) rectangle (195.13,227.65);
\definecolor{drawColor}{RGB}{0,0,0}

\node[text=drawColor,anchor=base,inner sep=0pt, outer sep=0pt, scale=  1.50] at (108.41,207.13) {\bfseries (b) ACF of PIT};
\end{scope}
\end{tikzpicture}}
	\scalebox{\scalingFactor}{% Created by tikzDevice version 0.12.3.1 on 2021-08-06 11:49:22
% !TEX encoding = UTF-8 Unicode
\begin{tikzpicture}[x=1pt,y=1pt]
\definecolor{fillColor}{RGB}{255,255,255}
\path[use as bounding box,fill=fillColor,fill opacity=0.00] (0,0) rectangle (195.13,227.65);
\begin{scope}
\path[clip] ( 21.68, 21.68) rectangle (195.13,195.13);
\definecolor{drawColor}{RGB}{0,0,0}

\path[draw=drawColor,line width= 0.6pt,line join=round,line cap=round] ( 28.11, 56.43) -- ( 28.11,188.70);

\path[draw=drawColor,line width= 0.6pt,line join=round,line cap=round] ( 37.55, 56.43) -- ( 37.55,145.09);

\path[draw=drawColor,line width= 0.6pt,line join=round,line cap=round] ( 47.00, 56.43) -- ( 47.00,113.56);

\path[draw=drawColor,line width= 0.6pt,line join=round,line cap=round] ( 56.45, 56.43) -- ( 56.45, 84.02);

\path[draw=drawColor,line width= 0.6pt,line join=round,line cap=round] ( 65.89, 56.43) -- ( 65.89, 72.24);

\path[draw=drawColor,line width= 0.6pt,line join=round,line cap=round] ( 75.34, 56.43) -- ( 75.34, 94.39);

\path[draw=drawColor,line width= 0.6pt,line join=round,line cap=round] ( 84.79, 56.43) -- ( 84.79, 97.08);

\path[draw=drawColor,line width= 0.6pt,line join=round,line cap=round] ( 94.23, 56.43) -- ( 94.23, 88.46);

\path[draw=drawColor,line width= 0.6pt,line join=round,line cap=round] (103.68, 56.43) -- (103.68, 72.23);

\path[draw=drawColor,line width= 0.6pt,line join=round,line cap=round] (113.13, 56.43) -- (113.13, 52.43);

\path[draw=drawColor,line width= 0.6pt,line join=round,line cap=round] (122.58, 56.43) -- (122.58, 54.48);

\path[draw=drawColor,line width= 0.6pt,line join=round,line cap=round] (132.02, 56.43) -- (132.02, 62.84);

\path[draw=drawColor,line width= 0.6pt,line join=round,line cap=round] (141.47, 56.43) -- (141.47, 67.09);

\path[draw=drawColor,line width= 0.6pt,line join=round,line cap=round] (150.92, 56.43) -- (150.92, 66.10);

\path[draw=drawColor,line width= 0.6pt,line join=round,line cap=round] (160.36, 56.43) -- (160.36, 51.94);

\path[draw=drawColor,line width= 0.6pt,line join=round,line cap=round] (169.81, 56.43) -- (169.81, 48.85);

\path[draw=drawColor,line width= 0.6pt,line join=round,line cap=round] (179.26, 56.43) -- (179.26, 62.67);

\path[draw=drawColor,line width= 0.6pt,line join=round,line cap=round] (188.71, 56.43) -- (188.71, 76.93);
\end{scope}
\begin{scope}
\path[clip] (  0.00,  0.00) rectangle (195.13,227.65);
\definecolor{drawColor}{RGB}{0,0,0}

\path[draw=drawColor,line width= 0.4pt,line join=round,line cap=round] ( 28.11, 21.68) -- (169.81, 21.68);

\path[draw=drawColor,line width= 0.4pt,line join=round,line cap=round] ( 28.11, 21.68) -- ( 28.11, 15.68);

\path[draw=drawColor,line width= 0.4pt,line join=round,line cap=round] ( 75.34, 21.68) -- ( 75.34, 15.68);

\path[draw=drawColor,line width= 0.4pt,line join=round,line cap=round] (122.58, 21.68) -- (122.58, 15.68);

\path[draw=drawColor,line width= 0.4pt,line join=round,line cap=round] (169.81, 21.68) -- (169.81, 15.68);

\node[text=drawColor,anchor=base,inner sep=0pt, outer sep=0pt, scale=  1.00] at ( 28.11,  2.48) {0};

\node[text=drawColor,anchor=base,inner sep=0pt, outer sep=0pt, scale=  1.00] at ( 75.34,  2.48) {5};

\node[text=drawColor,anchor=base,inner sep=0pt, outer sep=0pt, scale=  1.00] at (122.58,  2.48) {10};

\node[text=drawColor,anchor=base,inner sep=0pt, outer sep=0pt, scale=  1.00] at (169.81,  2.48) {15};

\path[draw=drawColor,line width= 0.4pt,line join=round,line cap=round] ( 21.68, 29.98) -- ( 21.68,188.70);

\path[draw=drawColor,line width= 0.4pt,line join=round,line cap=round] ( 21.68, 29.98) -- ( 15.68, 29.98);

\path[draw=drawColor,line width= 0.4pt,line join=round,line cap=round] ( 21.68, 56.43) -- ( 15.68, 56.43);

\path[draw=drawColor,line width= 0.4pt,line join=round,line cap=round] ( 21.68, 82.89) -- ( 15.68, 82.89);

\path[draw=drawColor,line width= 0.4pt,line join=round,line cap=round] ( 21.68,109.34) -- ( 15.68,109.34);

\path[draw=drawColor,line width= 0.4pt,line join=round,line cap=round] ( 21.68,135.80) -- ( 15.68,135.80);

\path[draw=drawColor,line width= 0.4pt,line join=round,line cap=round] ( 21.68,162.25) -- ( 15.68,162.25);

\path[draw=drawColor,line width= 0.4pt,line join=round,line cap=round] ( 21.68,188.70) -- ( 15.68,188.70);

\node[text=drawColor,rotate= 90.00,anchor=base,inner sep=0pt, outer sep=0pt, scale=  1.00] at (  9.68, 29.98) {-0.2};

\node[text=drawColor,rotate= 90.00,anchor=base,inner sep=0pt, outer sep=0pt, scale=  1.00] at (  9.68, 56.43) {0.0};

\node[text=drawColor,rotate= 90.00,anchor=base,inner sep=0pt, outer sep=0pt, scale=  1.00] at (  9.68, 82.89) {0.2};

\node[text=drawColor,rotate= 90.00,anchor=base,inner sep=0pt, outer sep=0pt, scale=  1.00] at (  9.68,109.34) {0.4};

\node[text=drawColor,rotate= 90.00,anchor=base,inner sep=0pt, outer sep=0pt, scale=  1.00] at (  9.68,135.80) {0.6};

\node[text=drawColor,rotate= 90.00,anchor=base,inner sep=0pt, outer sep=0pt, scale=  1.00] at (  9.68,162.25) {0.8};

\node[text=drawColor,rotate= 90.00,anchor=base,inner sep=0pt, outer sep=0pt, scale=  1.00] at (  9.68,188.70) {1.0};

\path[draw=drawColor,line width= 0.6pt,line join=round,line cap=round] ( 21.68, 21.68) --
	(195.13, 21.68) --
	(195.13,195.13) --
	( 21.68,195.13) --
	( 21.68, 21.68);
\end{scope}
\begin{scope}
\path[clip] (  0.00,  0.00) rectangle (195.13,227.65);
\definecolor{drawColor}{RGB}{0,0,0}

\node[text=drawColor,anchor=base,inner sep=0pt, outer sep=0pt, scale=  1.00] at (108.41, 30.08) {lag};
\end{scope}
\begin{scope}
\path[clip] ( 21.68, 21.68) rectangle (195.13,195.13);
\definecolor{drawColor}{RGB}{0,0,0}

\path[draw=drawColor,line width= 0.6pt,line join=round,line cap=round] ( 21.68, 56.43) -- (195.13, 56.43);
\definecolor{drawColor}{RGB}{0,0,255}

\path[draw=drawColor,line width= 0.6pt,dash pattern=on 4pt off 4pt ,line join=round,line cap=round] ( 21.68, 84.76) -- (195.13, 84.76);

\path[draw=drawColor,line width= 0.6pt,dash pattern=on 4pt off 4pt ,line join=round,line cap=round] ( 21.68, 28.11) -- (195.13, 28.11);
\end{scope}
\begin{scope}
\path[clip] (  0.00,  0.00) rectangle (195.13,227.65);
\definecolor{drawColor}{RGB}{0,0,0}

\node[text=drawColor,anchor=base,inner sep=0pt, outer sep=0pt, scale=  1.50] at (108.41,207.13) {\bfseries (c) ACF of $\protect\boldmath{(\text{PIT} - \frac12)^2}$};
\end{scope}
\end{tikzpicture}} \\ \smallskip
	\scalebox{\scalingFactor}{\input{figs/tikz/relDiag_marg_BoE5.tex}}
	\scalebox{\scalingFactor}{\input{figs/tikz/relDiag_thresh_BoE5.tex}} 
	\scalebox{\scalingFactor}{\input{figs/tikz/relDiag_quant_BoE5.tex}}
	\caption{Same as Figure \ref{fig:0}, but at a prediction horizon of
		five quarters.  \label{fig:5}}
	\bigskip
	\scalebox{\scalingFactor}{\input{figs/tikz/relDiag_PIT_BoE6.tex}}
	\scalebox{\scalingFactor}{% Created by tikzDevice version 0.12.3.1 on 2021-08-06 11:49:36
% !TEX encoding = UTF-8 Unicode
\begin{tikzpicture}[x=1pt,y=1pt]
\definecolor{fillColor}{RGB}{255,255,255}
\path[use as bounding box,fill=fillColor,fill opacity=0.00] (0,0) rectangle (195.13,227.65);
\begin{scope}
\path[clip] ( 21.68, 21.68) rectangle (195.13,195.13);
\definecolor{drawColor}{RGB}{0,0,0}

\path[draw=drawColor,line width= 0.6pt,line join=round,line cap=round] ( 28.11, 56.43) -- ( 28.11,188.70);

\path[draw=drawColor,line width= 0.6pt,line join=round,line cap=round] ( 37.55, 56.43) -- ( 37.55,165.87);

\path[draw=drawColor,line width= 0.6pt,line join=round,line cap=round] ( 47.00, 56.43) -- ( 47.00,132.55);

\path[draw=drawColor,line width= 0.6pt,line join=round,line cap=round] ( 56.45, 56.43) -- ( 56.45,104.89);

\path[draw=drawColor,line width= 0.6pt,line join=round,line cap=round] ( 65.89, 56.43) -- ( 65.89, 90.29);

\path[draw=drawColor,line width= 0.6pt,line join=round,line cap=round] ( 75.34, 56.43) -- ( 75.34, 84.83);

\path[draw=drawColor,line width= 0.6pt,line join=round,line cap=round] ( 84.79, 56.43) -- ( 84.79, 88.20);

\path[draw=drawColor,line width= 0.6pt,line join=round,line cap=round] ( 94.23, 56.43) -- ( 94.23, 87.92);

\path[draw=drawColor,line width= 0.6pt,line join=round,line cap=round] (103.68, 56.43) -- (103.68, 77.51);

\path[draw=drawColor,line width= 0.6pt,line join=round,line cap=round] (113.13, 56.43) -- (113.13, 64.95);

\path[draw=drawColor,line width= 0.6pt,line join=round,line cap=round] (122.58, 56.43) -- (122.58, 57.03);

\path[draw=drawColor,line width= 0.6pt,line join=round,line cap=round] (132.02, 56.43) -- (132.02, 60.76);

\path[draw=drawColor,line width= 0.6pt,line join=round,line cap=round] (141.47, 56.43) -- (141.47, 65.00);

\path[draw=drawColor,line width= 0.6pt,line join=round,line cap=round] (150.92, 56.43) -- (150.92, 64.12);

\path[draw=drawColor,line width= 0.6pt,line join=round,line cap=round] (160.36, 56.43) -- (160.36, 58.23);

\path[draw=drawColor,line width= 0.6pt,line join=round,line cap=round] (169.81, 56.43) -- (169.81, 52.54);

\path[draw=drawColor,line width= 0.6pt,line join=round,line cap=round] (179.26, 56.43) -- (179.26, 49.92);

\path[draw=drawColor,line width= 0.6pt,line join=round,line cap=round] (188.71, 56.43) -- (188.71, 53.55);
\end{scope}
\begin{scope}
\path[clip] (  0.00,  0.00) rectangle (195.13,227.65);
\definecolor{drawColor}{RGB}{0,0,0}

\path[draw=drawColor,line width= 0.4pt,line join=round,line cap=round] ( 28.11, 21.68) -- (169.81, 21.68);

\path[draw=drawColor,line width= 0.4pt,line join=round,line cap=round] ( 28.11, 21.68) -- ( 28.11, 15.68);

\path[draw=drawColor,line width= 0.4pt,line join=round,line cap=round] ( 75.34, 21.68) -- ( 75.34, 15.68);

\path[draw=drawColor,line width= 0.4pt,line join=round,line cap=round] (122.58, 21.68) -- (122.58, 15.68);

\path[draw=drawColor,line width= 0.4pt,line join=round,line cap=round] (169.81, 21.68) -- (169.81, 15.68);

\node[text=drawColor,anchor=base,inner sep=0pt, outer sep=0pt, scale=  1.00] at ( 28.11,  2.48) {0};

\node[text=drawColor,anchor=base,inner sep=0pt, outer sep=0pt, scale=  1.00] at ( 75.34,  2.48) {5};

\node[text=drawColor,anchor=base,inner sep=0pt, outer sep=0pt, scale=  1.00] at (122.58,  2.48) {10};

\node[text=drawColor,anchor=base,inner sep=0pt, outer sep=0pt, scale=  1.00] at (169.81,  2.48) {15};

\path[draw=drawColor,line width= 0.4pt,line join=round,line cap=round] ( 21.68, 29.98) -- ( 21.68,188.70);

\path[draw=drawColor,line width= 0.4pt,line join=round,line cap=round] ( 21.68, 29.98) -- ( 15.68, 29.98);

\path[draw=drawColor,line width= 0.4pt,line join=round,line cap=round] ( 21.68, 56.43) -- ( 15.68, 56.43);

\path[draw=drawColor,line width= 0.4pt,line join=round,line cap=round] ( 21.68, 82.89) -- ( 15.68, 82.89);

\path[draw=drawColor,line width= 0.4pt,line join=round,line cap=round] ( 21.68,109.34) -- ( 15.68,109.34);

\path[draw=drawColor,line width= 0.4pt,line join=round,line cap=round] ( 21.68,135.80) -- ( 15.68,135.80);

\path[draw=drawColor,line width= 0.4pt,line join=round,line cap=round] ( 21.68,162.25) -- ( 15.68,162.25);

\path[draw=drawColor,line width= 0.4pt,line join=round,line cap=round] ( 21.68,188.70) -- ( 15.68,188.70);

\node[text=drawColor,rotate= 90.00,anchor=base,inner sep=0pt, outer sep=0pt, scale=  1.00] at (  9.68, 29.98) {-0.2};

\node[text=drawColor,rotate= 90.00,anchor=base,inner sep=0pt, outer sep=0pt, scale=  1.00] at (  9.68, 56.43) {0.0};

\node[text=drawColor,rotate= 90.00,anchor=base,inner sep=0pt, outer sep=0pt, scale=  1.00] at (  9.68, 82.89) {0.2};

\node[text=drawColor,rotate= 90.00,anchor=base,inner sep=0pt, outer sep=0pt, scale=  1.00] at (  9.68,109.34) {0.4};

\node[text=drawColor,rotate= 90.00,anchor=base,inner sep=0pt, outer sep=0pt, scale=  1.00] at (  9.68,135.80) {0.6};

\node[text=drawColor,rotate= 90.00,anchor=base,inner sep=0pt, outer sep=0pt, scale=  1.00] at (  9.68,162.25) {0.8};

\node[text=drawColor,rotate= 90.00,anchor=base,inner sep=0pt, outer sep=0pt, scale=  1.00] at (  9.68,188.70) {1.0};

\path[draw=drawColor,line width= 0.6pt,line join=round,line cap=round] ( 21.68, 21.68) --
	(195.13, 21.68) --
	(195.13,195.13) --
	( 21.68,195.13) --
	( 21.68, 21.68);
\end{scope}
\begin{scope}
\path[clip] (  0.00,  0.00) rectangle (195.13,227.65);
\definecolor{drawColor}{RGB}{0,0,0}

\node[text=drawColor,anchor=base,inner sep=0pt, outer sep=0pt, scale=  1.00] at (108.41, 30.08) {lag};
\end{scope}
\begin{scope}
\path[clip] ( 21.68, 21.68) rectangle (195.13,195.13);
\definecolor{drawColor}{RGB}{0,0,0}

\path[draw=drawColor,line width= 0.6pt,line join=round,line cap=round] ( 21.68, 56.43) -- (195.13, 56.43);
\definecolor{drawColor}{RGB}{0,0,255}

\path[draw=drawColor,line width= 0.6pt,dash pattern=on 4pt off 4pt ,line join=round,line cap=round] ( 21.68, 84.76) -- (195.13, 84.76);

\path[draw=drawColor,line width= 0.6pt,dash pattern=on 4pt off 4pt ,line join=round,line cap=round] ( 21.68, 28.11) -- (195.13, 28.11);
\end{scope}
\begin{scope}
\path[clip] (  0.00,  0.00) rectangle (195.13,227.65);
\definecolor{drawColor}{RGB}{0,0,0}

\node[text=drawColor,anchor=base,inner sep=0pt, outer sep=0pt, scale=  1.50] at (108.41,207.13) {\bfseries (b) ACF of PIT};
\end{scope}
\end{tikzpicture}}
	\scalebox{\scalingFactor}{% Created by tikzDevice version 0.12.3.1 on 2021-08-06 11:49:36
% !TEX encoding = UTF-8 Unicode
\begin{tikzpicture}[x=1pt,y=1pt]
\definecolor{fillColor}{RGB}{255,255,255}
\path[use as bounding box,fill=fillColor,fill opacity=0.00] (0,0) rectangle (195.13,227.65);
\begin{scope}
\path[clip] ( 21.68, 21.68) rectangle (195.13,195.13);
\definecolor{drawColor}{RGB}{0,0,0}

\path[draw=drawColor,line width= 0.6pt,line join=round,line cap=round] ( 28.11, 56.43) -- ( 28.11,188.70);

\path[draw=drawColor,line width= 0.6pt,line join=round,line cap=round] ( 37.55, 56.43) -- ( 37.55,148.78);

\path[draw=drawColor,line width= 0.6pt,line join=round,line cap=round] ( 47.00, 56.43) -- ( 47.00,103.69);

\path[draw=drawColor,line width= 0.6pt,line join=round,line cap=round] ( 56.45, 56.43) -- ( 56.45, 67.38);

\path[draw=drawColor,line width= 0.6pt,line join=round,line cap=round] ( 65.89, 56.43) -- ( 65.89, 48.81);

\path[draw=drawColor,line width= 0.6pt,line join=round,line cap=round] ( 75.34, 56.43) -- ( 75.34, 63.59);

\path[draw=drawColor,line width= 0.6pt,line join=round,line cap=round] ( 84.79, 56.43) -- ( 84.79, 77.72);

\path[draw=drawColor,line width= 0.6pt,line join=round,line cap=round] ( 94.23, 56.43) -- ( 94.23, 79.62);

\path[draw=drawColor,line width= 0.6pt,line join=round,line cap=round] (103.68, 56.43) -- (103.68, 68.89);

\path[draw=drawColor,line width= 0.6pt,line join=round,line cap=round] (113.13, 56.43) -- (113.13, 51.37);

\path[draw=drawColor,line width= 0.6pt,line join=round,line cap=round] (122.58, 56.43) -- (122.58, 42.61);

\path[draw=drawColor,line width= 0.6pt,line join=round,line cap=round] (132.02, 56.43) -- (132.02, 49.89);

\path[draw=drawColor,line width= 0.6pt,line join=round,line cap=round] (141.47, 56.43) -- (141.47, 57.27);

\path[draw=drawColor,line width= 0.6pt,line join=round,line cap=round] (150.92, 56.43) -- (150.92, 63.48);

\path[draw=drawColor,line width= 0.6pt,line join=round,line cap=round] (160.36, 56.43) -- (160.36, 55.29);

\path[draw=drawColor,line width= 0.6pt,line join=round,line cap=round] (169.81, 56.43) -- (169.81, 53.93);

\path[draw=drawColor,line width= 0.6pt,line join=round,line cap=round] (179.26, 56.43) -- (179.26, 65.34);

\path[draw=drawColor,line width= 0.6pt,line join=round,line cap=round] (188.71, 56.43) -- (188.71, 76.55);
\end{scope}
\begin{scope}
\path[clip] (  0.00,  0.00) rectangle (195.13,227.65);
\definecolor{drawColor}{RGB}{0,0,0}

\path[draw=drawColor,line width= 0.4pt,line join=round,line cap=round] ( 28.11, 21.68) -- (169.81, 21.68);

\path[draw=drawColor,line width= 0.4pt,line join=round,line cap=round] ( 28.11, 21.68) -- ( 28.11, 15.68);

\path[draw=drawColor,line width= 0.4pt,line join=round,line cap=round] ( 75.34, 21.68) -- ( 75.34, 15.68);

\path[draw=drawColor,line width= 0.4pt,line join=round,line cap=round] (122.58, 21.68) -- (122.58, 15.68);

\path[draw=drawColor,line width= 0.4pt,line join=round,line cap=round] (169.81, 21.68) -- (169.81, 15.68);

\node[text=drawColor,anchor=base,inner sep=0pt, outer sep=0pt, scale=  1.00] at ( 28.11,  2.48) {0};

\node[text=drawColor,anchor=base,inner sep=0pt, outer sep=0pt, scale=  1.00] at ( 75.34,  2.48) {5};

\node[text=drawColor,anchor=base,inner sep=0pt, outer sep=0pt, scale=  1.00] at (122.58,  2.48) {10};

\node[text=drawColor,anchor=base,inner sep=0pt, outer sep=0pt, scale=  1.00] at (169.81,  2.48) {15};

\path[draw=drawColor,line width= 0.4pt,line join=round,line cap=round] ( 21.68, 29.98) -- ( 21.68,188.70);

\path[draw=drawColor,line width= 0.4pt,line join=round,line cap=round] ( 21.68, 29.98) -- ( 15.68, 29.98);

\path[draw=drawColor,line width= 0.4pt,line join=round,line cap=round] ( 21.68, 56.43) -- ( 15.68, 56.43);

\path[draw=drawColor,line width= 0.4pt,line join=round,line cap=round] ( 21.68, 82.89) -- ( 15.68, 82.89);

\path[draw=drawColor,line width= 0.4pt,line join=round,line cap=round] ( 21.68,109.34) -- ( 15.68,109.34);

\path[draw=drawColor,line width= 0.4pt,line join=round,line cap=round] ( 21.68,135.80) -- ( 15.68,135.80);

\path[draw=drawColor,line width= 0.4pt,line join=round,line cap=round] ( 21.68,162.25) -- ( 15.68,162.25);

\path[draw=drawColor,line width= 0.4pt,line join=round,line cap=round] ( 21.68,188.70) -- ( 15.68,188.70);

\node[text=drawColor,rotate= 90.00,anchor=base,inner sep=0pt, outer sep=0pt, scale=  1.00] at (  9.68, 29.98) {-0.2};

\node[text=drawColor,rotate= 90.00,anchor=base,inner sep=0pt, outer sep=0pt, scale=  1.00] at (  9.68, 56.43) {0.0};

\node[text=drawColor,rotate= 90.00,anchor=base,inner sep=0pt, outer sep=0pt, scale=  1.00] at (  9.68, 82.89) {0.2};

\node[text=drawColor,rotate= 90.00,anchor=base,inner sep=0pt, outer sep=0pt, scale=  1.00] at (  9.68,109.34) {0.4};

\node[text=drawColor,rotate= 90.00,anchor=base,inner sep=0pt, outer sep=0pt, scale=  1.00] at (  9.68,135.80) {0.6};

\node[text=drawColor,rotate= 90.00,anchor=base,inner sep=0pt, outer sep=0pt, scale=  1.00] at (  9.68,162.25) {0.8};

\node[text=drawColor,rotate= 90.00,anchor=base,inner sep=0pt, outer sep=0pt, scale=  1.00] at (  9.68,188.70) {1.0};

\path[draw=drawColor,line width= 0.6pt,line join=round,line cap=round] ( 21.68, 21.68) --
	(195.13, 21.68) --
	(195.13,195.13) --
	( 21.68,195.13) --
	( 21.68, 21.68);
\end{scope}
\begin{scope}
\path[clip] (  0.00,  0.00) rectangle (195.13,227.65);
\definecolor{drawColor}{RGB}{0,0,0}

\node[text=drawColor,anchor=base,inner sep=0pt, outer sep=0pt, scale=  1.00] at (108.41, 30.08) {lag};
\end{scope}
\begin{scope}
\path[clip] ( 21.68, 21.68) rectangle (195.13,195.13);
\definecolor{drawColor}{RGB}{0,0,0}

\path[draw=drawColor,line width= 0.6pt,line join=round,line cap=round] ( 21.68, 56.43) -- (195.13, 56.43);
\definecolor{drawColor}{RGB}{0,0,255}

\path[draw=drawColor,line width= 0.6pt,dash pattern=on 4pt off 4pt ,line join=round,line cap=round] ( 21.68, 84.76) -- (195.13, 84.76);

\path[draw=drawColor,line width= 0.6pt,dash pattern=on 4pt off 4pt ,line join=round,line cap=round] ( 21.68, 28.11) -- (195.13, 28.11);
\end{scope}
\begin{scope}
\path[clip] (  0.00,  0.00) rectangle (195.13,227.65);
\definecolor{drawColor}{RGB}{0,0,0}

\node[text=drawColor,anchor=base,inner sep=0pt, outer sep=0pt, scale=  1.50] at (108.41,207.13) {\bfseries (c) ACF of $\protect\boldmath{(\text{PIT} - \frac12)^2}$};
\end{scope}
\end{tikzpicture}} \\ \smallskip
	\scalebox{\scalingFactor}{\input{figs/tikz/relDiag_marg_BoE6.tex}}
	\scalebox{\scalingFactor}{\input{figs/tikz/relDiag_thresh_BoE6.tex}}
	\scalebox{\scalingFactor}{\input{figs/tikz/relDiag_quant_BoE6.tex}}
	\caption{Same as Figure \ref{fig:0}, but at a prediction horizon of
		six quarters.  \label{fig:6}}
\end{figure}

\end{appendices}

\end{document}